\documentclass[12pt, preprint]{aastex}

\shorttitle{Local Group LBVs}
\shortauthors{Massey et al}

\begin{document}

\title{A Survey of Local Group Galaxies Currently Forming Stars:  III. A Search for 
Luminous Blue Variables and Other H$\alpha$ Emission-Lined Stars\altaffilmark{1}}

\author{
Philip Massey\altaffilmark{2,3}, 
Reagin T. McNeill\altaffilmark{2,4}, 
K. A. G. Olsen\altaffilmark{3,5}, 
Paul W. Hodge\altaffilmark{6}, 
Cynthia Blaha\altaffilmark{7}, 
George H. Jacoby\altaffilmark{8}, 
R. C. Smith\altaffilmark{9},
Shay B. Strong\altaffilmark{2,10}
}

\altaffiltext{1}{Observations reported
here were obtained at the MMT Observatory, a joint facility of the Smithsonian Institution
and the University of Arizona.}
\altaffiltext{2}{Lowell Observatory, 1400 W Mars Hill Road, Flagstaff,
AZ 86001; Phil.Massey@lowell.edu}
\altaffiltext{3}{Visiting Astronomer, Kitt Peak National Observatory (KPNO)
and Cerro Tololo
Inter-American Observatory (CTIO),
National Optical Astronomy Observatory (NOAO), which is
operated by AURA, Inc., under cooperative agreement with the National
Science Foundation (NSF).}
\altaffiltext{4} {Participant in the Research Experiences for Undergraduates (REU)
program at Lowell Observatory, 2006.  Present address: Five College Astronomy Dept., Smith College, McConnell Hall, Northampton, MA 01063; rmcneill@email.smith.edu.}
\altaffiltext{5}{Gemini Science Center, National Optical Astronomy Observatory,
950 N. Cherry Ave., Tucson, AZ 85726-6732; kolsen@noao.edu.}
\altaffiltext{6}{Department of Astronomy, University of Washington, Seattle, WA 98195;
hodge@astro.washington.edu.}
\altaffiltext{7}{Department of Physics and Astronomy, Carleton College, One North College Street, Northfield, MN 55057; cblaha@carleton.edu}
\altaffiltext{8}{WIYN Observatory, P. O. Box 26732, Tucson, AZ 85726-6732;
 gjacoby@noao.edu. The WIYN Observatory is a joint facility of the University of Wisconsin-Madison, Indiana University, Yale University, and the National Optical Astronomy Observatory.}
 \altaffiltext{9}{CTIO, NOAO, Casilla 603, La Serena, Chile; rsmith@noao.edu.}
 \altaffiltext{10}{Research Experiences for Undergraduates at CTIO, 2001.  Current
 address: Department of Astronomy, University of Texas, RLM 16318, Austin,
 TX 78712-1083; sholmes@astro.as.utexas.edu.}

\begin{abstract}
We describe a search for H$\alpha$ emission-lined stars in M31, M33, and 
seven dwarfs in or near the Local Group (IC 10, NGC 6822, WLM,
Sextans B, Sextans A, Pegasus and the Phoenix dwarf)
using interference filter imaging with the KPNO and CTIO 4-m telescope and Mosaic cameras.  The survey is aimed primarily
at identifying new Luminous Blue Variables (LBVs) from their spectroscopic
similarity to known LBVs, avoiding the bias towards photometric variability,
which may require centuries to manifest itself if LBVs go through long
quiescent periods.  Followup spectroscopy with WIYN confirms that our survey detected a wealth of stars whose spectra
are similar to the known LBVs.  We ``classify" the spectra of known LBVs, and
compare these to the spectra of the new LBV candidates.  We demonstrate 
spectacular spectral variability for several of the new LBV candidates, such
as AM2, previously classified as a Wolf-Rayet star, which now shows
Fe I, Fe II and Balmer emission lines but neither the N~III $\lambda$$\lambda$4634,42 nor
He~II $\lambda$4686 emission  that it did in 1982.  Profound spectral changes
are also noted for other suspected and known LBVs.  Several of the LBV candidates also
show $>0.5$ mag changes in $V$ over the past 10-20 years.  The number of known or
suspected LBVs is now 24 in M31, 37 in M33, 1 in NGC 6822, and 3 in IC 10.
We estimate that the total number of LBVs in M31 and M33 may be several
hundred, in contrast to the 8  known historically through large-scale photometric
variability.  This has significant implications for the time scale of the LBV phase.
We also identify a few new WRs and peculiar emission-lined objects. 
\end{abstract}


\section{Introduction}
\label{Sec-intro}

The nearby galaxies of the Local Group serve as our astrophysical
laboratories for understanding the effect that metallicity has on
the evolution of massive stars.  For instance, the relative number
of red supergiants (RSGs) and Wolf-Rayet stars (WRs) is known to
vary by two orders of magnitude with just a 0.8~dex change in
oxygen abundance (Massey 2003 and references therein), in
keeping with the general predictions of massive star evolutionary
theory (Maeder et al.\ 1980).  Although such studies are made more
difficult by the stars' faintness and crowding, they have the
advantages over Galactic studies that the distances of these systems
are relatively well known (usually to a few percent; see van den Bergh
2000), and that in general the correction for interstellar extinction
is low and relatively uniform within these galaxies.

In previous papers of this series, we presented  {\it UBVRI} catalogs
for the two Local Group spiral galaxies M31 and M33, (Massey et al.
2006, hereafter Paper I) and for the seven dwarf irregular galaxies
IC 10, NGC 6822, WLM, Sextans B, Sextans A, Pegasus, and Phoenix
(Massey et al. 2007, hereafter Paper II).  This paper now discusses
the analysis of narrow band (50 \AA, FWHM) images centered on
H$\alpha$, [SII] $\lambda \lambda 6717,31$ , and [OIII] $\lambda 5007$.  The data are used to identify $H\alpha$
emission-lined stars in these nine galaxies, and in particular stars which
are spectroscopically similar to Luminous Blue Variables (LBVs).

LBVs are a rare class of luminous stars
which undergo episodic mass-loss, and represent a transitional phase
between the most massive O stars and the WR stage.  The archetype
LBVs in the Milky Way are the stars $\eta$ Carinae and P Cygni.
Observations of nebulae around these stars reveal very high ejecta
masses, of order 10$M_\odot$, and evidence of multiple ejections
on the time-scales of order $10^3$ years.  (For two recent provocative
reviews, see Smith \& Owocki 2006 and Smith 2007.) Massey (2003)
has argued that if $\eta$ Car or P Cyg were located in M31 or
M33 we would not know of them today, since their spectacular
photometric outbursts were hundreds of years ago.  Indeed, Hubble
\& Sandage (1953) identified only 5 such stars in all of M31 and M33
looking for photometric variability on archival plates dating back
40 years.  (Subsequent work brought this number to 8; see Parker
1997.)   In the absence of large photometric variability (which
might take a thousand years to manifest itself), better statistics
concerning the number of LBVs (and hence their lifetimes) might be
gathered by finding stars which are spectroscopically indistinguishable
from the known LBVs.  In the recent past, such {\it LBV candidates}
have been identified in nearby galaxies by spectroscopic surveys
of UV-bright objects (Massey et al.\ 1996), He-emission objects
(Corral \& Herrero 2003), $H\alpha$ emission stars (Neese et al.\ 1991;
Corral 1996; King et al.\
1998), or just blundering across them spectroscopically (Massey et
al.\ 2000; Massey 2006) by spectroscopic surveys of bright blue
stars.  

While we are mainly searching for LBVs, our selection has the
potential to uncover many other types of interesting $H\alpha$
emission-lined stars,  such as the brightest Of-type stars, WRs,
and Be and  B[e] stars.  
We expect that if there exists an equivalent to SS~433 (a
``mini-quasar" with a pair of jets ejecting material at 0.26 times the speed
of light; see Margon 1984) in the Local Group, our survey is likely
to pick it out, as such an object would also have strong $H\alpha$
emission with colors similar to LBVs and other early-type stars.

In \S\ref{Sec-obs} we describe our observations and how we did our
photometry.  The basis of selecting H$\alpha$ emission-lined stars
is given in \S\ref{Sec-pick}, along with the resulting catalogs.
Spectroscopy of the initial sample for M31, M33, NGC 6822, and IC 10 is given in
\S\ref{Sec-WIYN}.  We discuss our results in \S\ref{Sec-discuss}.

\section{Data and Photometry}
\label{Sec-obs}

The narrow-band observations were made with the Mosaic cameras
mounted at the prime foci of the KPNO and CTIO 4-m telescopes.  The
instruments and basic reduction procedures are given in detail in
Paper I.  Here we note that each camera consists of a 2x4 array of
2048x4096 SITe CCDs.  The field of view is 36'x36', and the plate-scale
of the final reduced images is 0.27" pixel$^{-1}$.  Exposures were
obtained through 50 \AA\ wide filters centered on H$\alpha$, [SII]
$\lambda \lambda 6717, 31$, and [OIII] $\lambda$ 5007.  A typical
sequence consisted of 5 exposures of 300s each with the telescope
offset slightly between adjacent exposures in order to compensate
for the small gaps between adjacent CCDs.  The journal of observations
is given in Table~\ref{tab:journal}, along with the delivered image
quality (DIQ) measured on each frame.  Conditions were generally 
good, but not necessarily photometric.  

For the seven
dwarfs in our sample 
we restricted our analysis to the regions centered on the galaxy, as the FOV
is much larger than the size of the galaxies. 
The coordinate ranges can be found in the captions of Figs.~14 to 20 of Paper II,
except that there is misprint for the coordinates used for Sextans A (Fig.~18 of Paper II),
where the correct range is $\alpha_{\rm 2000}=10^h10^m48^s$ to $10^h11^m15^s$
and $\delta_{\rm 2000}=-4^\circ 45'$ to $-4^\circ 39'$, an area of 0.011 deg$^2$.
We also extended the analysis region of NGC~6822 by 2' to the south of the region
listed in Paper II, as we found it had been too conservative; the range we used
was $\alpha_{\rm 2000}=19^h44^m34^s$ to $19^h45^m22^s$
and $\delta_{\rm 2000}=-14^\circ 58'$ to $-14^\circ 40'$, an area of 0.058~deg$^2$.

For the {\it UBVRI} photometry in Papers~I and II we found that we
needed to treat each CCD as a separate detector due to the slight
differences in color-terms.  However, this effect is negligible for
the narrow-band imaging discussed here, and so we did the photometry
of these images on the average of the registered (``stacked") images.
Our photometry relied upon the broad-band source lists given in Papers~I
and II.  We began with the 0.1" coordinates given in those catalogs,
and performed aperture photometry at those positions on the H$\alpha$,
[SII], and [OIII] stacked images.  We chose a 5-pixel (1.35") radius measuring
aperture as a reasonable compromise between seeing and crowding.
Sky was determined from the modal value within an annulus extending
from 10 to 20 pixels from each object.  Only objects that had
positive fluxes in all three apertures were retained.  For the
galaxies consisting of multiple, overlapping fields (M31 and M33)
the results were averaged for stars in common.

In order to convert counts into approximate flux, we used observations
of spectrophotometric standard stars.  Our goal was to achieve an
absolute calibration good to 5-10\%.  Although this is fairly crude, it provides
a sufficiently accurate measurement of the absolute measures 
of the fluxes, while our selection criteria
(described below) will rely principally on the 
relative measures between different filters, and not the absolute fluxes.
 For the standard stars, 
we performed
aperture photometry using a 15 pixel radius aperture, with a sky
annulus extending from 20 to 25 pixels.  We corrected for airmass
(both with the standards and program photometry)
using the ``standard'' KPNO and CTIO coefficients.  The KPNO data
were collected over four observing runs (2000 October, 2001 February,
2001 September, 2002 September), but the zero-points were found to
be quite similar (to within 0.05~mag), and a single average was
adopted. The differences between the northern and southern calibrations
are also small, as shown in Table~\ref{tab:calib}, and we adopted an
average value for each filter.  In general,
the standards were observed on a single chip (``im2"), but tests
conducted one night moving a spectrophotometric standard between
the eight chips confirmed that the flat-fielding was sufficiently
good to keep chip-to-chip differences in the zero-point to less
than one percent.  We determined a mean aperture correction (from 15
pixels used for the standards to the 5 pixels used on our program
frames) of -0.15~mag, with a scatter of 0.06~mag, and applied the -0.15~mag
correction to all of our H$\alpha$ magnitudes.  

We will refer to our fluxes in terms of  ``AB" magnitudes
(Oke 1974), with the reminder that this is defined in terms of the flux $f_\nu$ [cgs]
as simply $-2.5\log f_\nu - 48.60$.  Since our H$\alpha$ images are publicly 
available\footnote{http://www.noao.edu/nsa and ftp://ftp.lowell.edu/pub/massey/lgsurvey/datarelease/} 
we also give in Table~\ref{tab:calib} the conversion from
counts s$^{-1}$ to $f_\lambda$ [cgs], as these may be useful to others wishing to use
our images.  Note that while
this calibration allows us to determine the fluxes of continuum
sources, purely emission-lined objects (such as HII regions and
PNe) will have their lines ``diluted'' by the bandpass of our filter.
We determined a correction factor by following Jacoby et al. (1987),
where we determined the effective bandpass using the filter response
curves simulated in an f/3 beam, similar to the situation at prime
focus at each of the 4-m telescopes.  We then divided by the
normalized transmission of the filter at the wavelength of the line.
These are also provided in Table \ref{tab:calib} for the
benefit of others.  Although our sources are continuum and emission,
the emission is a minor component of the flux.

We give in Table~\ref{tab:errors} the photometric errors as a function of
magnitude for our three narrow-band filters. These errors reflect only
the {\it internal} errors (i.e., photon statistics and read-noise) and not the
external (calibration) errors, but since we will rely upon photometric indices
to select our objects, it is the internal errors which matter.  We see that our
errors were negligible $<0.02$ to roughly 20th mag.  

\section{Analysis: Separating the Wheat from the Chaff}
\label{Sec-pick}

Determining what criteria to apply to select H$\alpha$ emission-lined stars, while
rejecting HII regions and planetary nebulae (PNe), required careful consideration and
experimentation.  We constructed a number of photometric indices, using our
H$\alpha$, [SII], and [OIII] magnitudes, in combination with the broad-band
values from Papers I and II.   We determined what was effective by examining
the distribution of  known interesting
objects  in M31 and M33 (LBVs, LBV candidates, and WRs from Tables 8 and 9 of
Paper~I) to the general stellar populations of these two galaxies. 
We illustrate our selection in Figs.~\ref{fig:m31cuts}-\ref{fig:phxcuts}, and 
summarize our selection criteria in
Table~\ref{tab:cuts}.

First, we imposed a flux limit. This was necessary in order to not include objects whose
photometry was so poor that they appeared to be emission-lined objects when
they were not, and it also included the practical consideration that we wanted stars
that were bright enough to be observable on 6.5-m  telescopes.  We used
 the magnitude in the H$\alpha$ image itself as a general measure
of the brightness of the object (continuum plus emission).  Based upon the numbers in
Table~\ref{tab:errors}, we decided to impose a cut-off at an H$\alpha$ magnitude of 20th.
 In Table~\ref{tab:galaxies} we list our adopted distance moduli and reddenings
(from Table~1 of Paper II and references therein),
along with the apparent distance modulus at H$\alpha$, using a correction for
interstellar reddening of $A_{H\alpha}=2.54 E(B-V)$ Cardelli et al.\ 1989). 
We see that 20th magnitude corresponds to an absolute magnitude (at H$\alpha$) of
roughly $-5$ in M31 and M33.  For the galaxy with the largest apparent distance modulus,
IC10, it corresponds roughly to -6.  Although our source catalogs are based upon 
how good our photometry is (i.e., flux-limited, to a certain apparent magnitude), we still wanted to 
compare the number
of objects to the same absolute magnitude, and so we have adjusted the flux limit for IC10
slightly to reach $M_{H\alpha}=-6$ (i.e., Table~\ref{tab:cuts}).  We show the effects
of this cut on the full sample in Figs.~\ref{fig:m31cuts}-\ref{fig:phxcuts}{\it a}.

Secondly, for a measure of the actual H$\alpha$ 
emission, we chose to use [SII] as our continuum filter, as we found
that this provided cleaner separation in the various diagnostic two-color plots 
than did (say) broad-band $R$.  (King et al.\ 1998 also used H$\alpha -$[SII]
in their search for LBVs in M31.)   Based upon our examination of the known LBVs
in M31 and M33 we chose H$\alpha$-[SII]$\leq -0.15$ to separate emission-lined
stars from the general stellar population\footnote{For all of our indices, we applied
small shifts (of order a few tenths of a magnitude) in order for the general
population of stars to have a zero value.  This was partially an artifact that the
emission-lined indices, such as H$\alpha$-[SII] were constructed without 
separate aperture corrections.  We list these shifts in Table~\ref{tab:shifts}.}.
We illustrate the effect of this selection in Fig.~\ref{fig:m31cuts}-\ref{fig:phxcuts}{\it b}.

Thirdly, we constructed an index that would measure the amount of [OIII] emission.
For this, we found that a continuum comprised of the average of $V$ and $B$
worked well; i.e., [OIII]$-$C, where C=$(V+B)/2$.  This was necessary in order to
eliminate compact HII regions and PNe.  Our examination of the locations 
of the known LBVs showed that some showed emission in the [OIII] bandpass; this
is likely due to emission in the He~I $\lambda 5016$ line.   Therefore
we adopted a fairly conservative criteria of keeping only objects with [OIII]$-$C$>-0.75$.
However, even this will eliminate legitimate H$\alpha$ emission-lined stars which happen
to excite small HII regions.  The most extreme example of this problem is in IC~10, where
the [OIII]$-$C cut resulted in eliminating the vast majority of WRs (Fig.~\ref{fig:ic10cuts}).
In most cases, though, we expect that this cut eliminated unwanted gaseous emission
regions (Figs.~\ref{fig:m31cuts}-\ref{fig:phxcuts}{\it c}).

After applying all of these criteria, we were surprised to find that some of
the remaining objects were actually very red stars.  We attribute this to a molecular 
absorption band located at 6715 \AA. (Mostly these would be
foreground red dwarfs, although a few might be bona-fide red supergiants in these
galaxies.) 
This absorption band falls within the [SII] filter and would create the appearance of 
$H\alpha$ emission.  To eliminate these unwanted objects a color criteria was imposed 
using $B-V< 1.2$ for all galaxies (Fig.~\ref{fig:m31cuts}-\ref{fig:phxcuts}{\it d}), save one.  
In IC10, the colors were shifted a fair amount due to reddening.  From careful examination 
of color plots for IC10, we determined that a 
value of 1.5 would be equivalent for the $B-V$ criteria. (Based purely on
the reddening we would have picked a much higher cut-off, about 1.9, but we
infer from our plots that the vast majority of the stars are nearer foreground
stars, and that just a small increase in our cut-off was sufficient.) 

 In addition, we decided to use the reddening-free color index $Q=U-B-0.72(B-V)$  
 to keep only the intrinsically bluest stars, using  Q$\le-0.3$ (Fig.~\ref{fig:m31cuts}-\ref{fig:phxcuts}{\it e})
 (O-type stars typically have $Q<-0.9$, while a $-0.4$ corresponds to a B5 dwarf or giant, or an A0 supergiant;
 see Table~3 of Massey 1998.)
 Not all of our stars had $U$ band photometry, and the result of this cut meant that 
 any star without good $U$ band photometry would be eliminated.
This was judged to be a particular problem for IC10, and so we decided to 
 include stars without $U$ but which met the other criteria.

We show our final selections in Figs.~\ref{fig:m31cuts}-\ref{fig:pegcuts}{\it f}, and
 list these stars in Tables~\ref{tab:m31}
through \ref{tab:peg}.  (No objects were found in the Phoenix dwarf.) We have included the 
relevant broad-band photometry
from Papers I and II, along with spectral types, both those that were previously
known (from Papers I and II and references therein) and those which are
newly determined here (\S~\ref{Sec-WIYN}).  

We include stars in these tables as faint as an AB magnitude of 20.0 in H$\alpha$ (20.2
in the case of IC~10).  For M31 we find 2,334 potential H$\alpha$ emission-lined stars.
In our complete catalog of M31 there are 32,802 sources this bright or brighter
in H$\alpha$, so the fraction is about 7\%.  In M33 we find 3,707 potential 
H$\alpha$ emission-lined stars, while the catalog contains 18,867 stars this bright
or brighter, about 20\%: a much larger fraction.  

The full lists in Tables~\ref{tab:m31}-\ref{tab:peg} are based upon a given apparent
magnitude in H$\alpha$.  A more meaningful comparison is to use the absolute
luminosity.  In Table~\ref{tab:galaxies} we list the apparent H$\alpha$ magnitude
corresponding to $M_{H\alpha}=-6$.  If we count only stars that bright or brighter,
then we arrive at the number of H$\alpha$ sources listed in Table~\ref{tab:nums}.
We include in that table the absolute visual magnitude of each galaxy, along with the
current SFR.  There is clearly a much better correlation with SFR than with $M_V$:
for instance, we find about twice as many sources in M33 as in M31, although the
latter is 8 times more luminous.  The current SFR in M33 (integrated over the 
galaxy) is twice as large as in M31.    It is interesting that the number of sources
we detect in IC~10 is much lower than the current SFR would indicate.  However,
our [OIII]-C cut eliminated most of the WRs, and we believe that therefore our
potential list is spuriously low for that galaxy. The results for NGC 6822 are harder to explain, as only 9 
objects were found, while we might expect 85-90 by scaling from M31 or M33.
Perhaps in the dwarfs we are seeing the effects of low metallicity: that 
lower mass-loss rates result in disproportionately fewer H$\alpha$ emission stars.
The paucity of potential
H$\alpha$ emission-lined stars in NGC 6822 {\it is} consistent with its scant number (4)
of WRs, compared to (say) M33, where a deep surveys of about half of the galaxy
have confirmed roughly 160 WRs (Massey \& Johnson 1998).   We also include
in Table~\ref{tab:nums} the number of stars with $M_{H\alpha}\leq-6$ for which
we know the spectral types, either from previous work or from the present study.
We can see that despite the efforts reported in the next section, we have only
just begun to investigate the interesting emission-lined stars in these nearby galaxies.

\section{A Spectroscopic Reconnaissance in M31, M33, IC10, and NGC 6822}
\label{Sec-WIYN}

The stars listed in Tables~\ref{tab:m31} to \ref{tab:peg} are {\it potentially} interesting
objects, likely---but not certain---to have H$\alpha$ emission.  Although our detection
criteria were chosen to be fairly conservative (by necessity, so as to not to include
too many non-emission lined objects), the inherent uncertainties of such photometry
in crowded fields necessitates spectroscopy to see what it is we actually have.

We observed on four nights (19-22 September 2006) with the Hydra fiber positioner
at the WIYN 3.5-m telescope. 
On the first two nights we used the blue fiber cable
(consisting of $\sim 100$ fibers of 3.1" diameter) with a 790 line mm$^{-1}$ grating (``KPC-18C") used in second order
with a BG-39 blocking filter.  The spectral coverage was 3970-5030 \AA, with a spectral
resolution of 1.5 \AA; the setup was identical to that described in Paper I.  The first night was clear, with
good conditions (seeing $\approx$1"), while there were intermittent clouds on the second.
 On the third and fourth nights we used the red fiber cable
($\sim 100$ fibers of 2.0" diameter) with a the 600 line mm$^{-1}$ grating (``600\@10.1") in first
order with a GG-420 blocking filter.  The spectral coverage was 4550-7400 \AA, with a
spectral resolution of 3.4 \AA.  Conditions were marginal on the first red night (21 September 2006) and
we were only able to observe for the first hour or so of the night due to humidity; conditions were again good for 22 September.
For both setups we used the Bench Spectrograph Camera with the T2KA
CCD, a 2048$\times$2048 device with 24$\mu$m pixels and excellent cosmetics.  

The FOV of Hydra is 1$^\circ$, and we assigned fibers to two overlapping fields in M31, two
overlapping fields in M33, and one field each in NGC 6822 and IC10.  Each field was observed
for 2.3-3.0 hrs in the blue, and 1.0-2.0 hrs in the red, as summarized in Table~\ref{tab:spectexp}.
The IC10 field was not observed in the blue, and the second M31 field was not observed in the
red, due to the variable conditions.

In addition to our Hydra spectra, we also obtained a new high signal-to-noise spectrum with the 6.5-m
MMT of two of our P Cygni
like LBV candidates: the star J004341.84+411112.0, previously described by
Massey (2006), and J013416.07+303642.1, newly described here.  We used
the Blue Channel spectrograph on 28 October 2006 with
the 832 line mm$^{-1}$ grating in second order (CuSO$_4$ blocking filter)
with a 1.25" slit, for a resolution of 1.1 \AA, and covering 4075-5020 \AA.  These
spectra were obtained principally for the purposes of modeling, but we will use these
spectra here for illustration.

Finally, we include here three spectra  contributed by N. Caldwell, who contacted us
during the course of our writing this paper with questions about several M31 objects that
he thought resembled LBVs.  
Three of these turned out to be in our list of M31 potential H$\alpha$ sources
(Table~\ref{tab:m31}), J004229.87+410551.8, J004322.50+413940.9, and
J004442.28+415823.1.  These had
not been observed at WIYN, and he kindly suggested we include their
spectra here. He obtained these spectra on the
MMT 6.5-m with the Hectospect fiber positioner on 15 November 2006 using the 270 line mm$^{-1}$
grating, which provided wavelength coverage from 3650\AA\ to 9200\AA\ and a resolution of
roughly 5 \AA.

Nearly every object
had H$\alpha$ emission, as shown by the red spectra. Many of our objects (60\%)
were simply stars
in low excitation HII regions (usually with minimal [OIII] $\lambda 5007$) but in general our selection
criteria worked very well, and our spectroscopy revealed a wealth of interesting objects.
We list the results of spectroscopy in Table~\ref{tab:newsp} and 
discuss the newly discovered objects below. 

\subsection{New LBV Candidates}

The vast majority of our discoveries were stars whose spectra are 
extremely similar to that of known LBVs.  Of course,
these stars cover a significant range in spectral properties, and, in addition, some of these
stars are known to have drastic changes in their spectra.  We illustrate some current, and
past, spectra in Fig.~\ref{fig:lbv1}-\ref{fig:pcyg}, where we draw primarily
from the previously known 8 LBVs in M31 and M33.

\subsubsection{Hot LBV Candidates}
Generally, during the visual minimum (``quiescent") phase,
the optical spectra of high-luminosity LBVs are marked by strong emission in the
lower hydrogen lines (H$\alpha$, H$\beta$, and H$\gamma$), plus emission
of singly ionized metals, primarily [FeII] (Massey 2000 and references therein).  
We consider the prototype of this spectrum to be the current spectral state of Var C,
one of the original Hubble \& Sandage (1953) variables, although many other LBVs in M31 and
M33 share this spectral characteristic, such as A-1 and Var 15, as shown in Fig.~\ref{fig:lbv1}.
A 1983 spectrum of AE And shown in Kenyon \& Gallagher (1985) is very similar, and
we used this star's line list to identify lines shown in Fig.~\ref{fig:lbv1}\footnote{Our 
September 2006 spectrum of this star shows some changes in the intervening 23 years, with the Balmer lines and He~I lines now showing P Cygni profiles, which were
missing in the 1983 spectrum.
We show our
spectrum in Fig.~\ref{fig:AEnow} for comparison to Fig.~4 in Kenyon \& Gallagher
(1985).}.
S Doradus itself, the star whose name used to refer to the class of objects now known as
LBVs, has shown similar spectra (see Fig.~2 of Wolf \& Kaufer 1997).
It should be noted that at our dispersion, these spectra are sometimes indistinguishable
from those of
the high luminosity B[e] stars, which Conti (1997a) has argued are not LBVs.  The emission
in B[e] stars is believed to originate in a disk, as demonstrated by the study of R136 (in the LMC)
by Zickgraf et al.\ (1985).  In contrast, the emission in the quiescent LBVs shows P Cygni profiles
at  high dispersion  (cf.\ Kenyon \& Gallagher 1985).  Here we will call stars
that resemble Var C, AE And, and Var A-1 ``hot LBV candidates", but note that more detailed
studies might reclassify some of  them as B[e]. In the nomenclature of Bohannan (1989) such stars
would be called ``A extr".  
Massey et al.\ (1996) discovered four such stars in M33 by 
observing the brightest UV sources, and King et al.\ (1998) discovered five similar objects
in M31 based upon observing H$\alpha$-bright sources.  It is unsurprising that many more 
remained to be discovered.

We begin by showing the spectra of 10 similarly hot LBV candidates,  five in M31 and five in M33, compared to that of Var A-1 in Fig.~\ref{fig:lbvhot}.
We include in this figure one previously identified LBV candidate, J004417.10+411928.0
(``k350" in the study of King et al.\ 1998).
For ease of displaying these spectra we have (roughly) divided these into stars whose 
[FeII] and FeII lines are stronger than those in Var A-1 (Fig.~\ref{fig:lbvhot} {\it top}), and
those with weaker lines (Fig.~\ref{fig:lbvhot} {\it bottom}).

Two additional
  such stars were found in M31 by N. Caldwell, 
  J004229.87+410551.8 and J004322.50+413940.9.  Both spectra are somewhat 
  peculiar (Fig.~\ref{fig:nelson}).
  The spectrum of J004229.87+410551.8  may be a composite.
Although the Balmer lines are in emission and [FeII] is present, the strength of the H and K Ca II
lines and the presence of the G-band would indicate a much cooler absorption spectrum
than we see in the other hot stars. 
The spectrum of the other star, J004322.50+413940.9, looks at first blush to be that
of a P Cygni type LBV, with strong P Cygni profiles in the lower Balmer lines
(H$\alpha$, H$\beta$, and H$\gamma$).  However, closer inspection reveals
P Cygni lines in (permitted) Fe II, notably $\lambda 4924$, $\lambda 5018$, and
$\lambda 5169$\footnote{Given the dispersion of the Hectospect spectra, we could
not be certain of the line identifications. We considered the possibility that the first
two lines were He~I $\lambda 4922$ and $\lambda 5016$.  However, the absence
of any other He~I lines in the spectra ruled this out.}.  The spectrum of Var~C shown
by Kenyon \& Gallagher (1985) also showed strong P Cygni profiles in many of the
permitted Fe II lines, and is also lacking in He~I.  However, J004322.50+413940.9
is somewhat peculiar compared to the Kenyon \& Gallagher (1985) Var~C exposure, as
the many other strong Fe II emission lines in the blue aren't evident.  Possibly this
is due to the lower resolution and modest signal-to-noise ratio of the spectrum.
There is also an absorption line near 4643 \AA\ that we were unable to identify
unambiguously. Both of these stars show significant photometric variability over the time
scale of a decade, as shown below.

There are another five stars (one in M31 and four in M33) for which we have blue
spectra similar to those of the hot LBV candidates.  All have H$\beta$ and H$\gamma$
in emission, and all also show He I $\lambda 4922$ and $\lambda 5016$ in emission. 
However, any [Fe II] or Fe II emission is so weak as to be considered  either incipient or uncertain.
We show their spectra in Fig.~\ref{fig:hotmaybe}, where we have again included the
spectrum of Var A-1 for comparison.  One of these stars, J004411.36+413257.2,
was previously described as an LBV candidate by King et al.\ (1998),  where
it was listed as ``k315a".

We have only a red spectrum of the M33 star J013332.64+304127.2 
(Fig.~\ref{fig:am2} {\it top} and {\it middle}), but it too
reveals Balmer emission plus [FeII] and FeII emission.  This star would simply serve
as another example of a ``hot LBV candidate" were it not for the fact that this
star is M33WR41 (AM2), whose spectrum was called WNL by Massey \& Johnson (1997), 
based on an observation made in September 1982 and shown 
in Fig.~2b of Massey \& Conti (1983), where it
is identified as number 28.  We have checked the cross identification carefully
and there appears to be nothing amiss.  The star J013332.64+304127.2 is certainly the star 
labeled 2 in Armandroff \& Massey (1985) where it showed up in on-band,
off-band imaging  in
He~II $\lambda 4686$.  Furthermore, it was independently identified as being
bright in He~II $\lambda 4686$ in the {\it HST} imaging of Drissen et al.\ (1993), where they identified it as NGC~595-WR6.  In Fig.~\ref{fig:am2}  we compare the two.
Although the 1982 spectrum is quite noisy, there is no question that it is of
a WN-type WR.   The resolution of the 1982 data (taken with the Intensified Image Dissector Scanner on the KPNO 4-m) is 6\AA, and it is possible that the star was
actually a Ofpe/WN9, with the P Cygni absorption components washed out by the relatively low resolution. Unfortunately,  it was not in the sample of M33 WRs recently re-observed
by Abbott et al.\  (2004). The star HDE 269858 (Radcliffe 127) is (or rather, was) an Ofpe/WN9
star that underwent an LBV-like outburst in 1980 (Stahl et al.\ 1983).  We believe that J013332.64+304127.2 (AM2)
is another such example.

Spectra of two stars in IC 10 are very similar, although they do not
have the signal-to-noise ratio needed to reveal weak emission (Fig.~\ref{fig:ic10}).  One of
these stars, J002020.35+591837.6, is located just 1.6" from 
RSMV8 (J002020.56+591837.3),
classified as ``WN10" by Crowther et al.\ (2003).  
Their classification is equivalent to the Ofpe/WN9 designation.  We do not
think, though, that our spectrum of J002020.35+591837.6
 has been contaminated by this object, as we used the 2" diameter red fibers
for the observation, and RSMV8 is at least a magnitude fainter than
J002020.35+591837.6.

A third IC 10 star, and the only NGC 6822 star to prove of interest, are
even less certain examples.
J002016.48+591906.9 (IC 10) and J194503.77-145619.1 (NGC 6822) show
only Balmer emission.   J013442.14+303216.0 (M33) is very similar.
We show their spectra in Fig.~\ref{fig:n6822}. It is a stretch
to call these  LBV candidates based on these spectra, so we inlcude
a ``?" in their classification.
Higher S/N data is clearly warranted, as they {\it may} reveal [FeII] emission.
We will note, however, that J002016.48+591906.9 (IC 10) showed photometric variability at the
0.7~mag level in a 10 year period (see below), bolstering its case.

\subsubsection{Cool LBV Candidates}
At outburst (visual maximum), the spectra of LBVs are often said to
resemble that of an extreme F-type supergiant, with
the absorption arising in a ``pseudo-photosphere" (see, for example, Humpheys \& Davidson 1994).  The spectra of all five of the original Hubble-Sandage variables 
(Var 2, Var A, Var B, Var C, all of which are in M33; Var 19 in M31) 
were in this state when observed by
Hubble \& Sandage (1953)\footnote{Note that Var A is now considered to be
an LBV {\it candidate} by some; see Humphreys \& Davidson (1994) and Parker (1998).}.
We use Var B during its 1992-1993 
outburst (Szeifert et al.\ 1996) as the archetype (Fig.~\ref{fig:lbv2}).   We would estimate the spectral type of this
photosphere to be late F or even early G-type.  The figure shows that
S Doradus exhibited
a very similar spectrum in October 1999, although the star was clearly {\it not} undergoing a
(photometric) outburst at the time (Massey 2000). This, we believe, underscores how complex and poorly
understood the LBV phenomenon really is, and why additional examples of LBVs can only
help improve our understanding of the LBV phenomenon.
  We will refer to LBVs showing such spectra as being in their ``cool state" rather than in outburst.

We did not observe any star that was quite as extreme as Var B during its outburst.
However, there are six stars that show 
underlying absorption spectra which are not that late, more like B8 to early F, along with very
strong emission at H$\alpha$ and P Cygni profiles at H$\beta$.   We compare the spectrum
of one of these, J004507.65+413740.8, to that of the outburst spectrum of Var~B in 
Fig.~\ref{fig:coolguys} ({\it top}).  J004507.65+413740.8 is clearly of earlier type (compare, for
example, the strength of the G-band).  We estimate the absorption spectral type as F2~Ia.
We show all six spectra in Fig.~\ref{fig:coolguys} ({\it bottom}), where we have scaled the figure
to emphasize the absorption spectra.  Note that the two lines to the red of H$\beta$ are clearly
FeII $\lambda \lambda 4924, 5018$, and not the He~I $\lambda \lambda 4922, 5016$ lines visible
in the spectra of the hotter stars discussed above.  This is evident not only from the wavelengths, but
also from the lack of other, usually stronger, He~I lines, such as $\lambda 4471$.
The emission profiles of H$\beta$ and H$\alpha$ can
be found in Fig.~\ref{fig:coolprofiles}.  This shows that all six have strong P Cygni profiles in the
Balmer lines.
  
It is interesting to note that our September 2006
observation of Var B shows a fairly 
boringly weak emission-line spectrum with the Balmer lines
in emission, slight P Cygni emission in the He I lines, and a few {\it doubly} ionized forbidden Fe
lines, i.e., [FeIII] $\lambda \lambda 4658, 4702$.)   We compare this to the December 1993
spectrum in Fig.~\ref{fig:varb}.  It is hard to believe  we are looking at the same star!   Less dramatic spectral changes for Var B
on the time-scale of months were discussed and illustrated by
Szeifert et al.\ (1996).

We do have spectra of two other M33 stars which may be ``cool" LBV candidates: 
J013416.44+303120.8 and J013429.64+303732.1.  Both of these show an absorption
spectrum typical of a B8~I.  However, H$\beta$ shows a very narrow emission component
superposed on a very broad component.  We show the blue spectra in Fig.~\ref{fig:weird} ({\it top}).
We also show the H$\alpha$ profiles in Fig.~\ref{fig:weird} ({\it bottom}).  Such a broad profile
{\it could} be indicative of a rapidly rotating disk, but it could also be indicative of a more
optically thick wind, reminiscent of Wolf-Rayet stars.  The narrow emission component
could be nebular, although the region around the H$\alpha$ profile seems to suggest otherwise,
as there is no sign of the [NII] $\lambda \lambda 6548, 6584$ lines nor is there [OIII] $\lambda 5007$
emission characteristic of a nebula.  The broad components extend to $\pm 600-1000$ km s$^{-1}$.

\subsubsection{P Cygni LBV Candidates}

Of course, not all LBV spectra fall into these two extremes.  The most notable exception is
the Galactic star P Cygni.   This star contains no Fe~II or [Fe II] features, but rather the
He I and Balmer lines show characteristic line
profiles that bear the name of this famous star: there
is a blue-shifted absorption component,
with a strong emission component extending redwards from line center.  A good 
signal-to-noise spectrum
also reveals lines of NII, indicative of enriched material at the stellar surface.
 We show the spectrum of P Cygni, AF And, and
J004341.84+411112.0 in Fig.~\ref{fig:pcyg}.   Although the latter has not shown the same sort of
spectacular photometric outbursts that characterize LBVs, its spectrum is uniquely similar to 
that of P Cygni itself, and like P Cygni it may be surrounded by nebulosity indicative of a 
past eruption---about 2 millennia ago in the case of J004341.84+411112.0 (Massey 2006).
The spectrum of J004341.84+411112.0 shown here is new, and was acquired with the MMT 6.5-m
on 28 October 2006, and will be discussed elsewhere; here we just note that the NII features
only hinted at in the lower S/N spectra shown by Massey (2006) are quite obvious in this higher
S/N data.

We identify six similar stars here, and compare their spectra to that of P Cygni in Fig.~\ref{fig:pcygs}.
Although all six show strong P Cygni components in the Balmer lines, and some in the He I lines
(i.e., J004242.33+413922.7, J013351.46+304057.0, and J013416.07+303642.1)
none show the startling close resemblance to P Cygni itself as does J004341.84+411112.0
(Fig.~\ref{fig:pcyg}).   Still, the same three that show He~I P Cygni also show evidence of the
NII emission bands indicative of enriched material at the surface.  In the case of J013416.07+303642.1
we show our MMT spectrum of the star, as our Hydra spectrum has significant nebular
contamination.

One of these P Cygni-like stars, J013339.52+304540.5, had previously been classified
as ``B0.5I+WNE" (Massey et al.\ 1996, where it was called UIT154; the star is listed
in Massey \& Johnson 1998 as ``M33-WR57")  Its spectrum was also described
by Crowther et al.\  (1997), where it is referred to by its Humphreys \& Sandage
(1980) designation, ``B517".  We compare the 1993 spectrum
with that from 2006 in Fig.~\ref{fig:pcygprob1}.  The broad He~II $\lambda 4686$ 
feature has clearly disappeared, while the Balmer emission has gotten
stronger.  (In 1993 H$\delta$ was primarily absorption while in 2006 it shows strong
P Cygni emission.)  Nevertheless, the P Cygni lines were quite evident in the 1993 spectrum, and the classification of this star as ``B0.5I+WNE" appears, in retrospect,
to have been overly simplistic.  Crowther et al.\ (1997) described it as ``WN11h"
based upon a 1995 optical spectrum; the star showed only weak He~II
$\lambda 4686$ at that time.
Finally, we note that another of these P Cygni-like LBV candidates, J013341.28+302237.2,
had previously been called a B1~Ia by Monteverde et al.\ (1996), who refer to the
star by its designation in Humphreys \& Sandage (1980), ``110-A".  Its spectrum is
shown in Monteverde et al.\ (1996)'s Figs.~1 and 2.  Although H$\alpha$ is in
emission, there is no sign of P Cygni emission at H$\gamma$, although it may
be partially filled in by emission.  Thus the spectrum has clearly changed, as
H$\gamma$ has roughly equal absorption and emission components in our
spectrum (Fig.~\ref{fig:pcyg}) of this star.  Monteverde et al.\ (1996) did not include
H$\beta$. 

\subsubsection{Ofpe/WN9 LBV Candidates}

Further complicating the LBV issue, R 127, AG Car,
and HDE 269582 are three examples of stars whose spectra at minimum light are
Ofpe/WN9 stars (Walborn 1977; Bohannan \& Walborn 1989; Humphreys \& Davidson 1994 and references therein; see also Smith et al.\ 1995).   
Therefore we have chosen to discuss newly found
``slash stars" with the other LBV candidates rather than with Wolf-Rayets.  We also find below
that the Ofpe/WN9 stars tend to be photometrically variable.

We show the blue spectra of J013509.73+304157.3 (M33) and J004334.50+410951.7
(M31)
in Fig.~\ref{fig:ofpe}, where we have used the LMC star BE 381 as a reference.
BE 381 was 
one of the stars that originally defined the class (see Bohannan \& Walborn 1989), 
and its spectrum
was kindly made available by B. Bohannan.  The M33 star J013509.73+304157.3,
also known as Romano's star, known to be photometrically variable by more than
a magnitude (Romano 1978), and is usually included in lists of LBV candidates
(see, for example, Parker 1997).  Its spectrum was recently discovered independently
to have
changed to that of Ofpe/WN9 (Viotti et al.\ 2007).   The Ofpe/WN9 nature of the M31
star J004334.50+410951.7 is newly discovered here.

A third star, J013432.76+304717.2, located in M33,   we also call Ofpe/WN9.   We have
only a ``red" spectrum ($>4550$\AA) of this star.  The spectrum is shown in Fig.~\ref{fig:ofpeRED},
where we compare it to that of J013509.73+304157.3.  They are clearly quite similar.

With the addition of these newly found Ofpe/WN9 stars, the number
of such stars known in M31 is now two, and in M33 is eight.  Such stars are
quite rare.

\subsection{Wolf-Rayet and Of Stars}

We expected to find few WRs: we saw in \S~\ref{Sec-pick} that our
selection criteria eliminated the vast majority of known WRs, 
primarily at the stage of eliminating HII regions via the [OIII]-C restriction.
This is simply because most WRs are found in HII regions.  The exceptions tended
to be the previously known 
Ofpe/WN9 stars, which 
generally were retained in our H$\alpha$ catalogs.  
Nevertheless, we found 3 new WRs in M31, 3 or 4 new WRs in M33 (one of which
may simply be an Of star), and identified
a classification problem with a previously identified WR. 

Of the other  stars with WR-like emission lines, one is a WC-type, three are
WN-type, and one is likely to be an Of star rather than a WN. We show their spectra in 
Fig.~\ref{fig:wrguys}.  For the Wolf-Rayets, the low S/N of the data, plus the presence of nebular
emission, precludes a more precise classification than WNL (based on the
strength of NIII $\lambda 4634,42$ for the WNs) and WC.  

The star M33 star
J013406.72+304154.5 by far has the weakest emission of any of these.
That star had been classified by Massey et al.\ (1996) as an ``early O star",
which we would indeed expect to have the NIII $\lambda 43634,42$ and
He II $\lambda 4686$ emission that characterizes Of stars.  We measure
its equivalent width to be $-1.4$ \AA.  This is significantly weaker than the
-10 \AA\ usually taken as the cut-off between WNs and Ofs, and so we
call it an Of star.  We note with some amusement that this star is located just
5" to the north of the much brighter LBV candidate J013406.63+304147.8.

The M33 star J013334.27+304136.7 is very close (1.7") to another known WNL star,
J013334.31+304138.3 (M33WR49=AM6).  It is possible that the light from
latter contaminated our spectrum of the former, although the region around
He~II $\lambda 4686$ looks identical on both our blue exposure, taken with a
3.1" diameter fiber,  and red, taken with a 2.0" diameter fiber.  It is also conceivable
that the cross-identification of M33WR49 with J013334.31+304138.3 is wrong,
and that the correct identification should have been with J013334.27+304136.7.
Future observations will have to resolve that issue.  
The WNL star J013355.87+304528.4 (M33) is 7.4" from J013355.60+304534.9 (M33WR107=MJ B17), another WN star, but this separation is large enough so there
is likely to be no confusion.

{\subsubsection{J013307.50+304258.5=M33-WR19}
The star J013307.50+304258.5 was classified by Massey et al.\ (1996) as
WNE+B, where it was designated UIT041.  
The star was recognized as being blue by Humphreys \& Sandage (1980),
where it is called ``B52".  Massey \& Johnson (1998) repeat the WNE+B spectral
type and re-christen it as M33-WR19.  It is quite bright, with a continuum magnitude
of $\sim 17.2$.

Our blue spectrum of this star reveals a very flat-topped He~II $\lambda 4686$ and
no  absorption lines present.
We rechecked the spectra used by Massey et al.\ (1996), and
found that the spectra of UIT041 and UIT177 were interchanged in their Fig.~6.
The classification ``WNE+B" really belongs to UIT 177 
(M33-WR75 in Massey \& Johnson
1998, or J013343.34+303534.1), which was also called ``WN4.5+O6-9" on the basis of a better spectrum.  
The real classification of UIT041 (=J013307.50+304258.5=B 52=M33-WR19)
was never made.

The spectrum we show in Fig.~\ref{fig:UIT} shows a  
very broad He~II $\lambda 4686$ feature. The width (60 \AA) would suggest
it is an early-type WN, but  the equivalent width of about -20\AA is quite weak
for a WNE; see Fig.~5 in Armandroff \& Massey (1991).   The star is 1.5" to the west
of a bright late-type star (J013307.60+304259.0) which compromised our red
spectrum of the object.

\subsection{Supergiants in HII Regions}

There are six stars which we observed whose presence in the H$\alpha$ list
is due to their lying in an HII region (as evidenced by [OIII] 
$\lambda \lambda 4959, 5007$
[NII] $\lambda \lambda 6548, 6584$
and [S II] $\lambda \lambda 6717, 31$ emission) but whose blue spectra
reveal them to be interesting supergiants in their own right.  We show their
spectra in Fig.~\ref{fig:nice}.  In the top panel are two late-A or early-F type
supergiants.  Such stars are very rare.  In the bottom two panels we show the spectra
of four B supergiants, ranging from B1~I to B8~I: J013300.86+303504.9 (B1~I),
J013339.42+303124.8 (B3~I), J004434.65+412503.6 (B3~I), and J013359.01+303353.9 (B8~I).
We have identified the strongest lines, using
Coluzzi (1993) and Walborn \& Fitzpatrick (1990). Three of the stars
had been previously classified.
 The star J004434.65+412503.6 was
classified as ``B1:" from spectra obtained in Paper~I; our spectrum here is
superior, and we adopt the B3~I type.  J013300.86+303504.9 was called
B1.5~Ia+ by Massey et al.\ (1995), in essential agreement with the B1~I type assigned
here, which we prefer.
 The star J013339.42+303124.8 was called B1~Ia by Massey et al.\ (1995),
but our current spectra shows it is somewhat later, and we adopt the B3~I type here.

\subsection{A Very Broad-lined H$\alpha$ Emission Object}

As in any such survey, there are a few truly peculiar objects which are not easily categorized. 
Our most interesting such object is J004057.03+405238.6,
shown in Fig.~\ref{fig:bizzare}.  We see very broad emission
at H$\beta$ (as well as at H$\gamma$ and H$\delta$, not shown).  Are the two broad emission
to the redwards of H$\beta$ He~I $\lambda \lambda 4922, 5016$ or could they be
Fe II $\lambda \lambda 4924, 5018$?  We lack a red spectrum of the star, and the rest of
the blue is too noisy to permit identification of weak lines.
  The full width at half maximum of the H$\beta$ profile is
20.5\AA, or nearly 1270 km s$^{-1}$.   Such broad lines might be seen in WR stars, but there
is no He~II $\lambda 4686$.  The heliocentric
radial velocity based on H$\beta$ is $-225$ km s$^{-1}$.  However,
at its location at $X/R=-0.973$ (in the notation of Rubin \& Ford 1970)
we would expect a radial velocity of $-530$ km s$^{-1}$, significantly more negative.
The object might be part of M31's halo population and not partake of the disk's rotation. 
Alternatively, the object might be variable in radial velocity.  The broadness of the lines
is a bit reminiscent of those of SS 433 (Margon 1984), although without showing the
characteristic three components.  Still, one is reminded of one of the early spectra of  SS 433, 
caught when all three components had similar velocities (Margon et al.\ 1979).
The object is coincident with a long-period variable 15043 with a 330 day period
(Mould et al.\ 2004), but does not appear to be an x-ray source.  A similar object in M33
{\it may} be described by Neese et al.\ (1991).  Whatever the object is, it is not one of
the familiar emission-line objects (LBVs, WRs) that we have so far discussed.

\section{Discussion}
\label{Sec-discuss}

Clearly our survey has been very effective in identifying interesting objects.
Our spectroscopy has identified 12 new LBV candidates in M31 (excluding the
 3 previously suggested by King et al.\ 1998 that we re-observed here) and  
 18 new LBV candidates in M33.  This brings the number of known or suspected
LBVs to 24 in M31 and 37 in M33\footnote{Parker (1998) lists 4 known 
and 5 candidate LBVs in M31, plus 4 known and 12 candidate LBVs in M33.
However, among the latter group, he lists two of the stars twice, under separate
names, not recognizing that Corral (1996)'s S193 is the same star as
Massey et al.\ (1996)'s UIT 301, and that Corral (1996)'s S95 is the same star
as Massey et al.\ (1996)'s UIT 212.  The fact that Corral (1996) and Massey
et al.\ (1996) were published at about the same time meant that these works did
not cross-reference each other, resulting in some subsequent confusion.}.
In addition, our spectroscopy identified 3 possible LBVs in IC 10, and 1 in 
NGC 6822; none had previously been reported in either galaxy.  We list
the full list of known and candidate LBVs in Table~\ref{tab:LBVs}.

To what extent are we certain that these newly found objects are truly LBVs?
In some cases, spectral and photometric variability may require decades---if
not centuries, or even millennia, to manifest itself.  Still, it is worth recalling
that in this paper we found evidence of spectral variability for several of the
LBV candidates.  In some cases this spectral variability is spectacular, as we
found for the (former) WR star AM2, whose spectrum now lacks either He~II or
N~III.  

We can also ask to what extent we know that these stars
vary photometrically.  Answering that is somewhat more difficult than is 
sometimes assumed, as many of the past photometric surveys, particularly
the photographic ones, have large magnitude-dependent errors associated
with them when compared to modern CCD data (see discussion in Paper~I).
For IC 10, we used some August 1992 CCD images obtained on the 4-m by
PM and GHJ and performed differential photometry of the LBV candidates relative to
nearby stars, which we calibrated using the IC 10 LGGS photometry from Paper II.
These data are previously unpublished, and were obtained as part of testing
``T2KB", a (then) brand new 2048x2048 device with 24$\mu$m pixels.  The seeing was
quite poor, about 2.0", which is why the data were not previously used. However, they do
suffice to check for variability of two of the IC 10 LBV candidates; the third was too crowded.
 For M31 we compare the LGGS 2000-2001 $V$-band photometry of 
Paper I with the CCD photometry of Magnier et al.\ (1992), obtained from images
taken in 1990.
For M33 we use the photographic  photometry of Ivanov et al.\ (1993).  
Some of the plates used in that study were obtained at the CFHT in the early
1980s; the rest were from plates obtained from the Rozhen 2-m telescope 
at an unspecified time.  In Paper I we showed that the Magnier et al.\ (1992) data
agreed well, on average, with the LGGS data, while the Ivanov et al.\ (1993) data
agreed moderately well only for stars brighter than about $V=19$, fainter than which
the Ivanov et al.\ (1993) values are too bright compared to the LGGS data.

We list the photometry and $\Delta V$ (in the sense of LGGS {\it minus} other) in
Table~\ref{tab:LBVs}.  We see that the known LBVs show (absolute)
differences of 0.6~mag to 1.5~mag, with some exceptions (Magnier et al.\ 1992
did not include photometry of AE And, and Var~B shows only a $-0.19$ mag difference.)
In general, the (absolute) differences in the photometry of the LBV candidates is 
smaller, but perhaps this just means that these stars are not variable {\it at present}:
after all, the archetypical P Cyg has shown little photometric variability over many
decades of monitoring; see Israelian \& de Groot (1999).  Nevertheless, {\it some} of
the LBV {\it candidates} listed in Table~\ref{tab:LBVs} show quite significant variations, 
of 0.6-1.3~mag.  This includes some of the spectroscopically questionable LBV candidates, such as the
IC~10 star J002016.48+591906.9.

We also checked to see which stars appear in the list of variables discovered by
the DIRECT project, using a merged list of stars kindly prepared by A. Bonanos,
and based upon the nine separate lists published by 
Kaluzny et al.\ (1998, DIRECT I;  1999, DIRECT IV);
Stanek et al.\ (1998, DIRECT II; 1999, DIRECT III);
Mochejska et al.\ (1999, DIRECT V; 2001a, DIRECT VII; 2001b, DIRECT VIII);
Macri et al.\ (2001, DIRECT VI); and
Bonanos et al.\ (2003, DIRECT IX). Their fields do not cover all of M31 and M33,
so the lack of designation of a variable is not particularly telling, but several
of these stars are listed.  Two of them are identified as Cepheids from their
light curves, which is not consistent with the spectra we show here.  For M33
we also checked two additional resources: (1) the
updated version of the Hartman et al.\ (2006) variable list\footnote{http://www.astro.livjm.ac.uk/~dfb/M33/} and the
Wise Observatory M33 Variability study (Shporer \& Mazeh 2006)
\footnote{http://wise-obs.tau.ac.il/~shporer/m33/}. We are indebted to K. Stanek 
for calling both of these
to our attention.  These lists reveal several additional variables, the majority
of which turned out to be the Ofpe/WN9 stars!  Such long-term monitoring programs as the DIRECT, Hartman et al.\ (2006), and the Wise Observatory project are quite valuable, and their continuation into the future
is quite useful.

Our NGC 6822 LBV candidate is slightly too south to be included in the
photometry of Bianchi et al.\ (2001).  However, the star was found to be
a low-amplitude, non-periodic  variable in the $I$ band 
by Mennickent et al.\ (2006).

Only about 40\% of the sources we observed spectroscopically proved to
be interesting. It is not clear how the percentage will be affected going to 
fainter limits.  Of the non-interesting objects, most proved to be HII regions,
with a small smattering of stars without emission, principally in the two dwarf
galaxies.

With our additional knowledge, could we further refine our selection criteria?
In Fig.~\ref{fig:winlose} we compare our selection criteria to the ``winners"
(LBVs, LBV candidates, and Ofpe/WN9 stars) and ``losers" (HII regions and
the occasional star with no emission).  We see that there is no adjustment we
could have made in our selection criteria that would have favored the ``winners"
over the ``losers".  Still, our approach has been one dimensional, and it is possible
that with the new data one could devise more effective selection criteria by
a multidimensional approach.

It is worth noting that we cut off the selection of objects
for spectroscopic observation at about $B=19.5$; 
fainter than this, we expected to obtain no useful spectroscopic data with
the WIYN telescope and Hydra.  We obtained spectra
of 42 stars in M31 and 79 in M33 (Table~\ref{tab:newsp}),
 but the complete catalogs (Tables~\ref{tab:m31} and \ref{tab:m33})
contain 307 and 820 stars, respectively, brighter than this limit.
Thus our spectroscopy, even at this magnitude limit, was only about 10\% complete.
We can therefore expect {\it many} more LBV candidates and other interesting objects
to be found, with potentially profound effects on the statistics of LBVs in nearby
galaxies.  We expect that the actual number of LBVs, rather than being 4 each
in M31
and M33 (as the generally recognized LBVs number), is probably more on order
several hundred.  This clearly has implications for the duration of the LBV phase.

Throughout this paper we have taken the conservative approach of referring to 
stars with spectroscopic similarities to known LBVs as LBV {\it candidates}.
In attempting to ``define" an LBV at the 1996 Kona meeting, 
Bohannan (1997) argued that ``A star should not
be considered an LBV because its current spectroscopic character is similar to that of a known
LBV.  Remember what is said about ducks: it may look like a duck, walk like a duck, but
it is not a duck until it quacks."    
In this case, the ``quacking" involves photometric variability, or possibly an ``outburst".
In his ``redefinition" of an LBV at the same meeting, 
Conti (1997b) concurred, arguing that an outburst is
needed to ``promote" a candidate to an LBV.
Here, we would make two points.  First, many of the stars in this sample in fact do have
demonstrated photometric variability, and at a level comparable to the known LBVs, at least
over a twenty year time span.  We leave it to others to decide if this variability is sufficient to
``promote" any of these stars from ``candidate" status to true LBV.  But, our more important
point is that we would argue that the ``requirement" of variability may be misguided.  Major
outbursts may occur only on the time scales of centuries, or milleninum, and so we would argue
that the {\it lack} of variability should not preclude a star from being considered an LBV.
Perhaps instead spectral resemblance to known LBVs should be considered a sufficient criterion,
as what other objects have such spectra? In a raft of ducks, at any one time, some will be quacking and some will be not.
Momentary silence does not transform a duck into a goose, nor would it confuse most bird spotters.

\acknowledgments
We are grateful for correspondence with A. Bonanos, S. Kenyon, and K. Stanek. 
We acknowledge the suggestion by N. King that we include the [SII] filter for use in
our study, and the excellent support received at Kitt Peak and Cerro Tololo where
the observations were made.  N. Caldwell graciously donated three of his spectra
for this paper.  D. Hunter kindly offered useful comments on an early version of this
manuscript.  Comments from an anonymous referee improved the paper.
RTM's involvement in this project was supported through an NSF
grant, AST-0453611.

\clearpage

\begin{figure}
\plotone{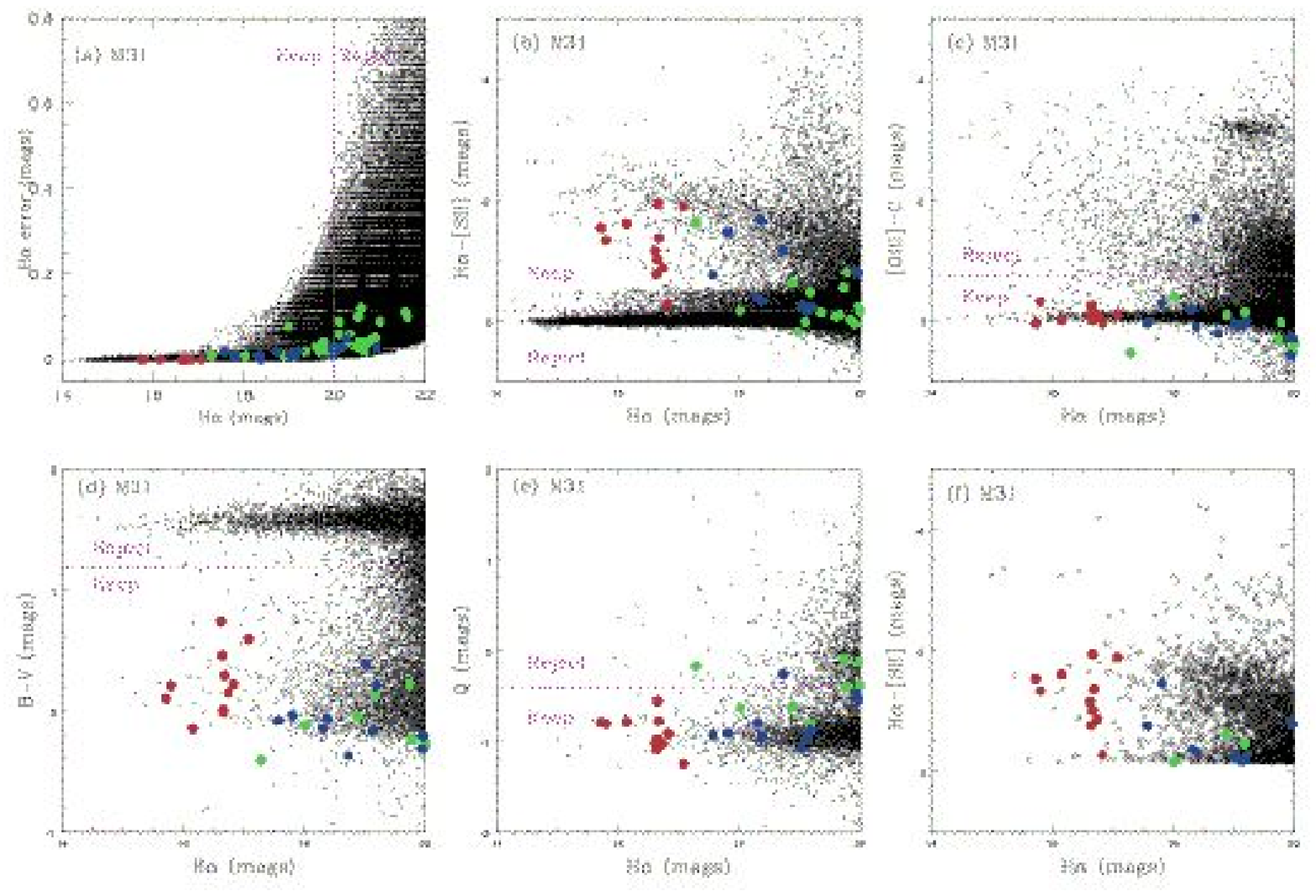}
\vskip -200pt
\caption{\label{fig:m31cuts} Selection criteria for H$\alpha$ emission-lined stars applied to M31.
Red points denote the
previously known LBVs and LBV candidates, while blue and green points denote
WRs of WN- and WC-type, respectively. 
(a) We demonstrate that the errors in the H$\alpha$ fluxes begin to increase
past a (spectrophotometric) magnitude of $\sim$ 20th.  Imposing 20th mag
as a cutoff  keeps all of the LBVs but only the brightest of
the WRs.  (b) Rejecting the stars that show little or no H$\alpha$ emission is done
by eliminating stars with H$\alpha$-[S II]$\geq-0.15$.  (c) We attempt to eliminate HII regions and PNe by requiring little [O III] emission, i.e., [OIII]-C$\geq -0.75$.  (d) Our sample is still dominated by red foreground stars, which we eliminate by $B-V<1.2$. (e) Finally, in order to keep stars which are only intrinsically quite blue, we impose a cutoff $Q\leq -0.4$, where $Q=(U-B)-0.72(B-V)$, the Johnson reddening-free index. (f) Our final
sample contains 2,334 stars. }
\end{figure}

\begin{figure}
\plotone{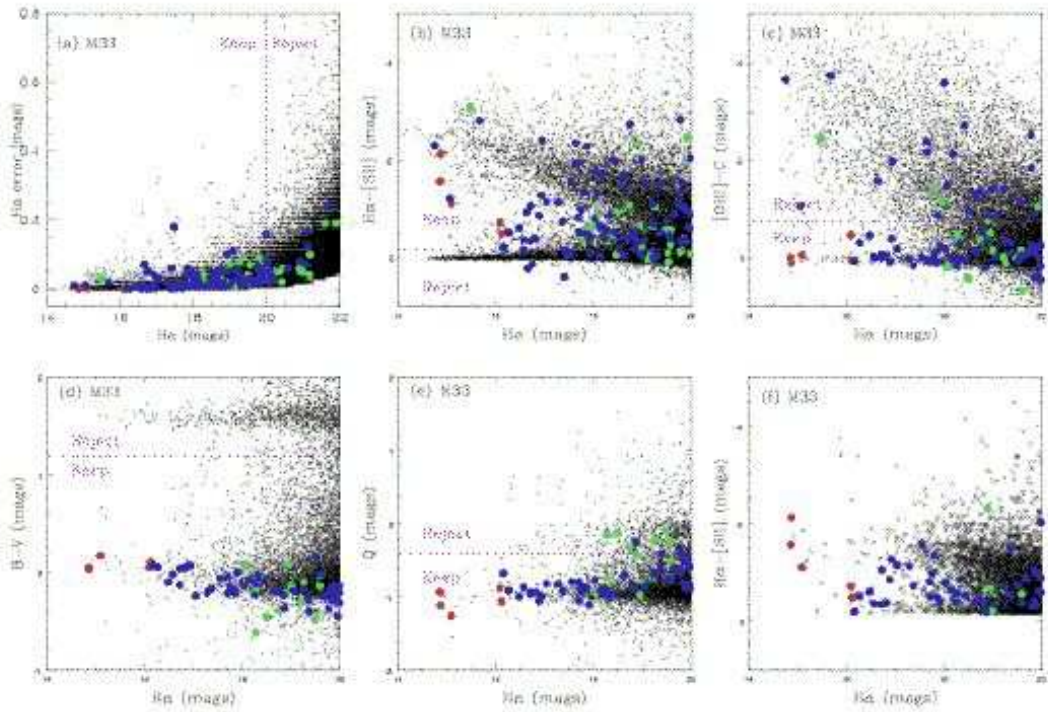}
\vskip -200pt
\caption{\label{fig:M33cuts} Selection criteria for H$\alpha$ emission-lined stars applied to M33.
Same as Fig.~\ref{fig:m31cuts}, except (f) our final sample contains 3,707 stars.
}
\end{figure}

\begin{figure}
\plotone{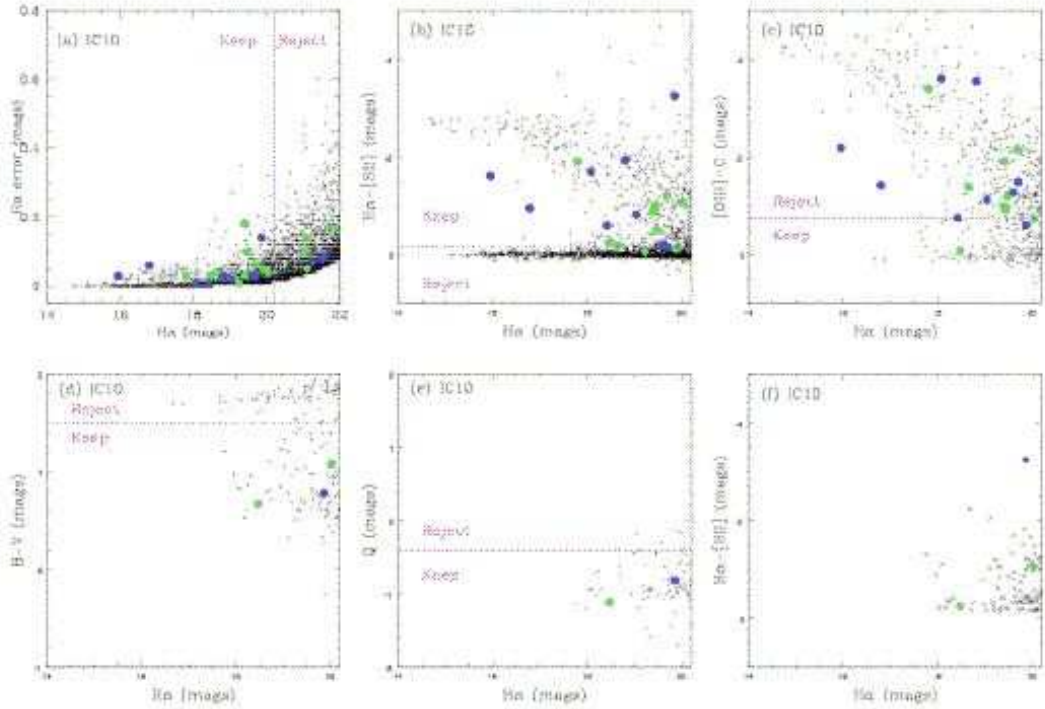}
\vskip -200pt
\caption{\label{fig:ic10cuts} Selection criteria for H$\alpha$ emission-lined stars applied to IC10.  Same as Fig.~\ref{fig:m31cuts}, except that there are no previously known LBVs, and (f) our final sample contains 81 stars. }
\end{figure}

\begin{figure}
\plotone{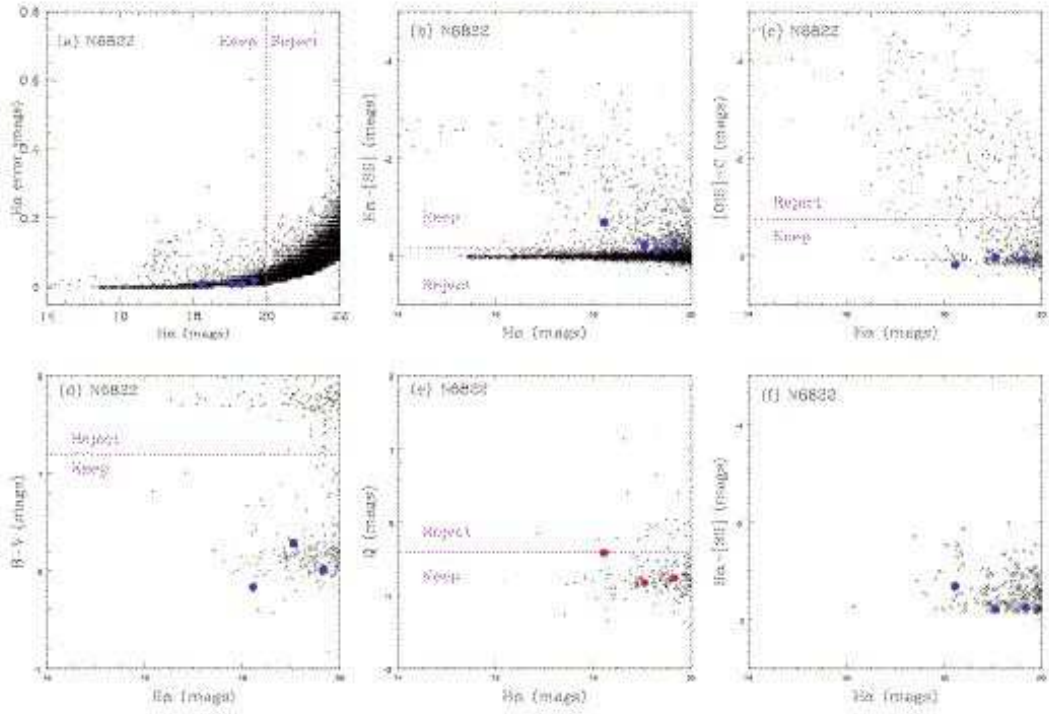}
\vskip -200pt
\caption{\label{fig:n6822cuts} Selection criteria for H$\alpha$ emission-lined stars applied to NGC 6822.  Same as Fig.~\ref{fig:m31cuts}, except that there are no previously known LBVs or WCs, and (f) our final sample contains 163 stars.}
\end{figure}

\begin{figure}
\plotone{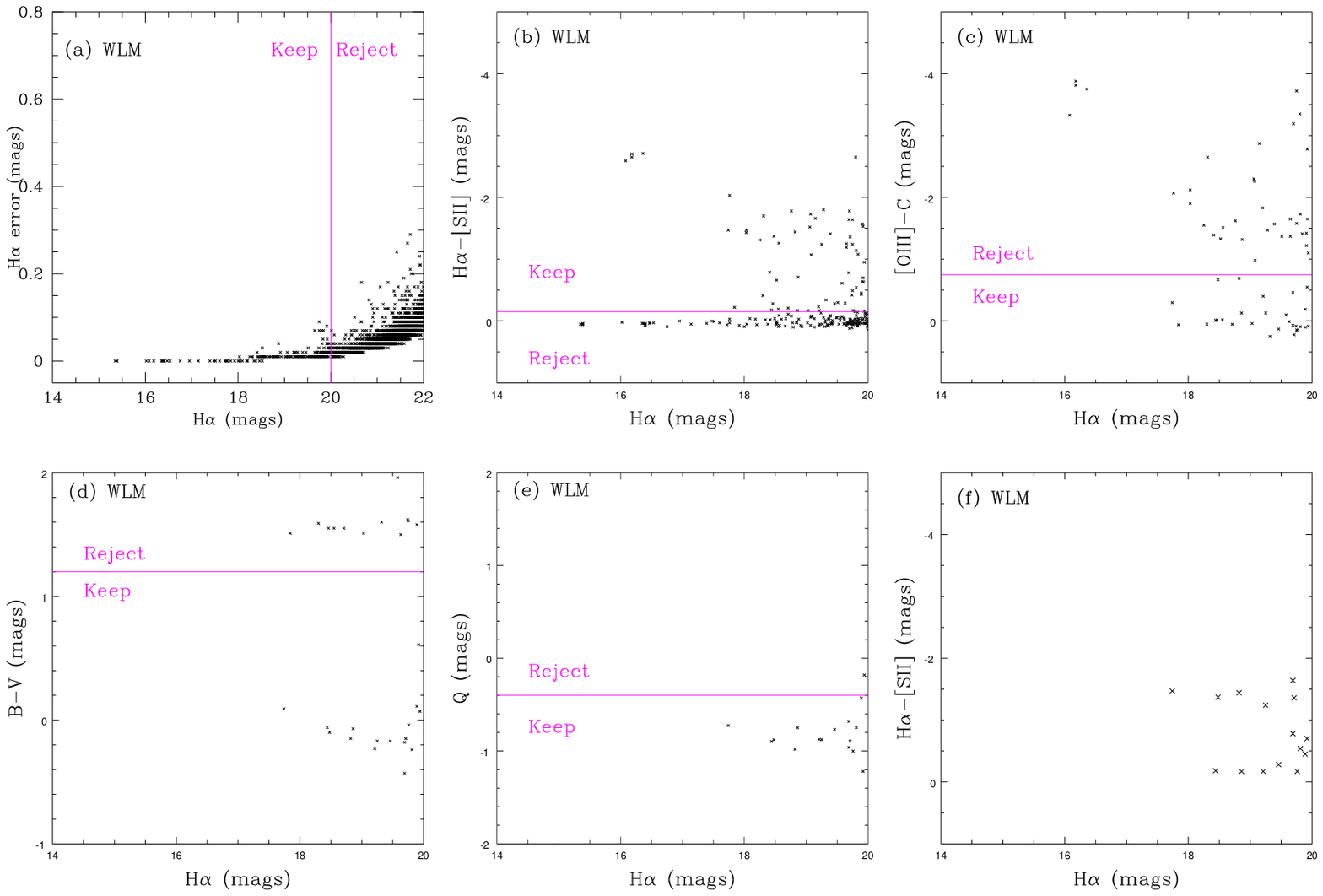}
\vskip -200pt
\caption{\label{fig:wlmcuts} Selection criteria for H$\alpha$ emission-lined stars applied to WLM.  Same as Fig.~\ref{fig:m31cuts}, except that there are no previously known
H$\alpha$ emission-lined stars, and (f) our final sample contains 15 stars.}
\end{figure}

\begin{figure}
\plotone{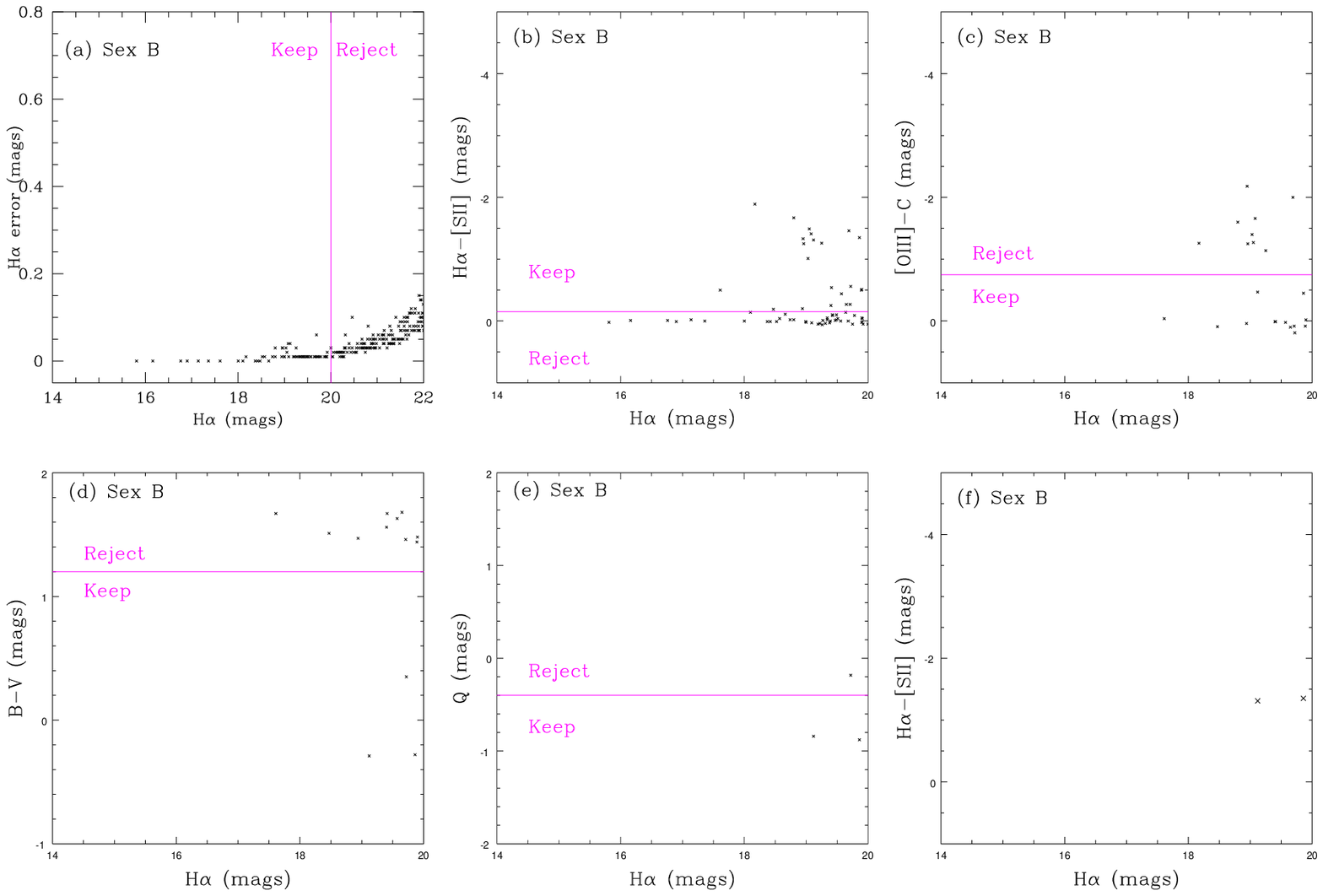}
\vskip -200pt
\caption{\label{fig:sexBcuts} Selection criteria for H$\alpha$ emission-lined stars applied to Sextans B.  Same as Fig.~\ref{fig:m31cuts}, except that there are no previously known
H$\alpha$ emission-lined stars, and (f) our final sample contains 2 stars.}
\end{figure}

\begin{figure}
\plotone{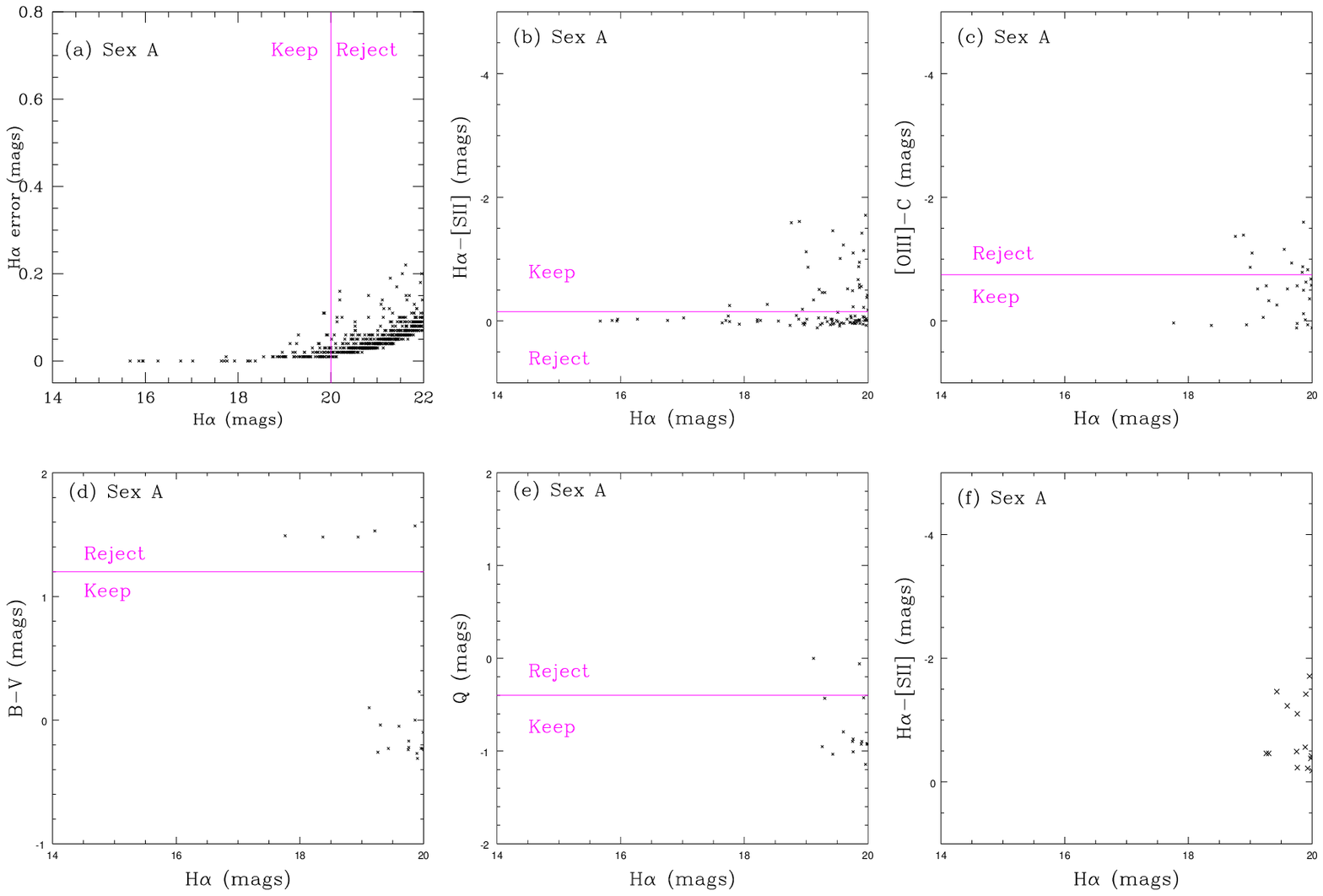}
\vskip -200pt
\caption{\label{fig:sexAcuts} Selection criteria for H$\alpha$ emission-lined stars applied to Sextans A.  Same as Fig.~\ref{fig:m31cuts}, except that there are no previously known
H$\alpha$ emission-lined stars, and (f) our final sample contains 8 stars.}
\end{figure}

\begin{figure}
\plotone{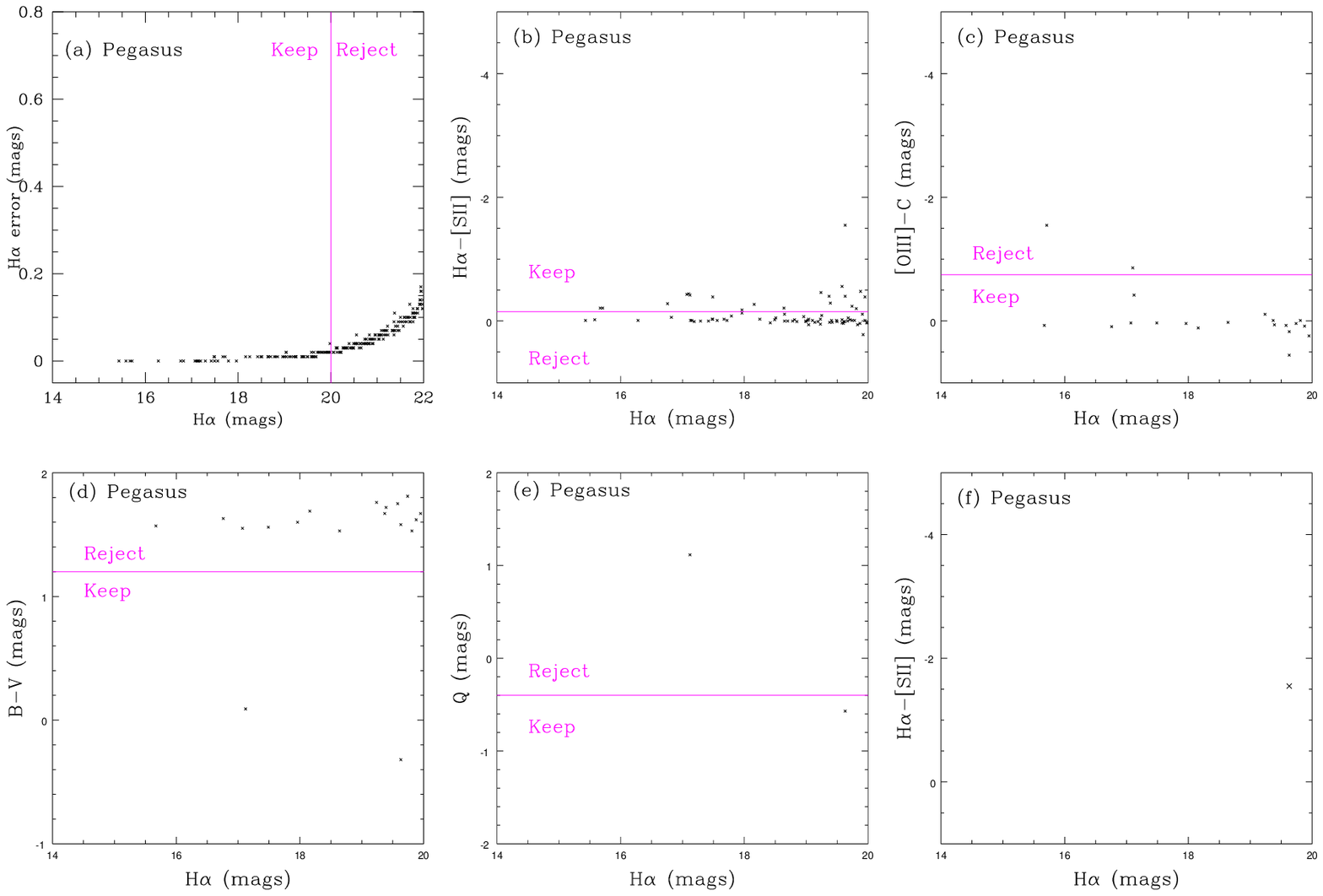}
\vskip -200pt
\caption{\label{fig:pegcuts} Selection criteria for H$\alpha$ emission-lined stars applied to Pegasus.  Same as Fig.~\ref{fig:m31cuts}, except that there are no previously known
H$\alpha$ emission-lined stars, and (f) our final sample contains 1 star.}
\end{figure}

\begin{figure}
\plotone{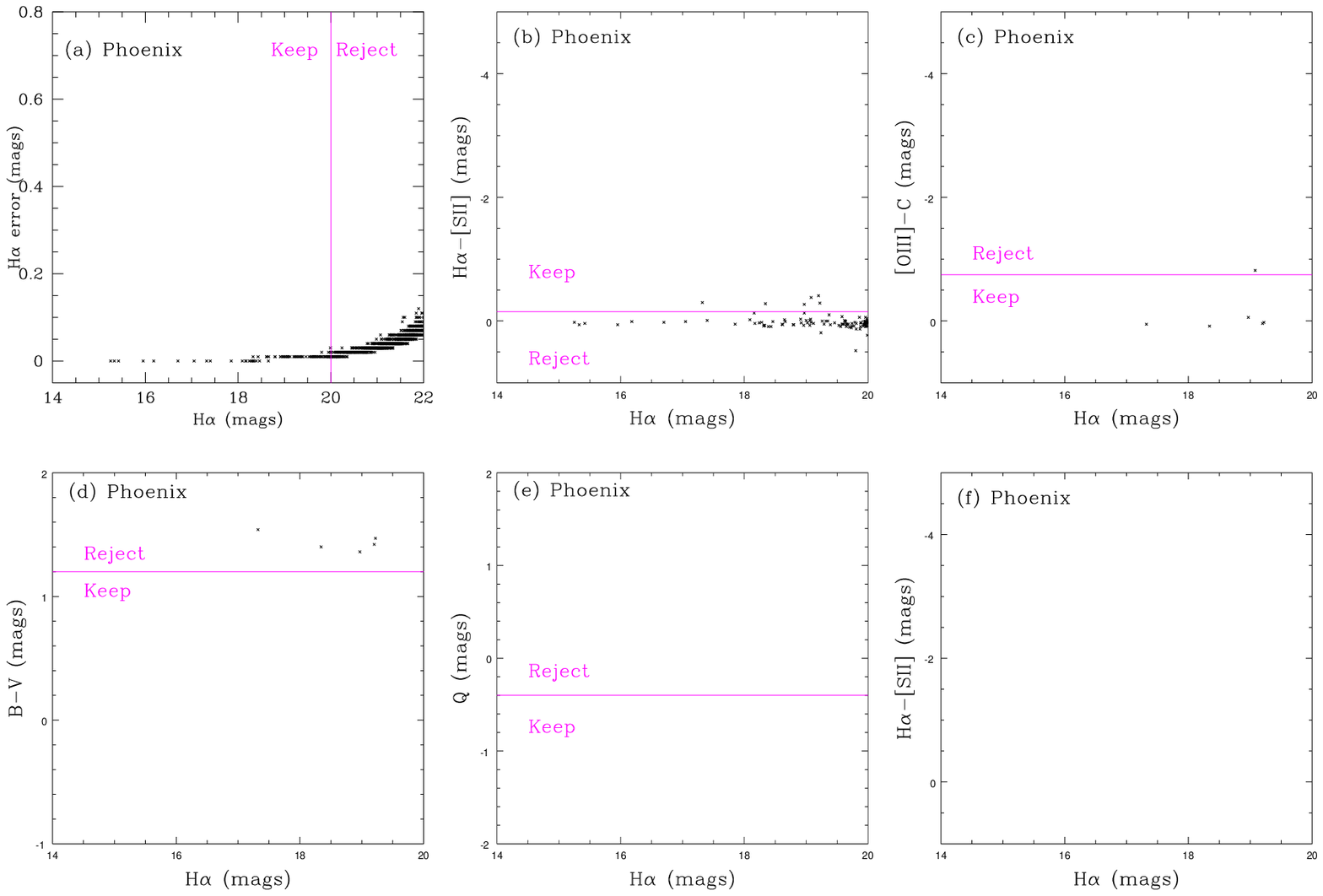}
\vskip -200pt
\caption{\label{fig:phxcuts} Selection criteria for H$\alpha$ emission-lined stars applied to Phoenix.  Same as Fig.~\ref{fig:m31cuts}, except that there are no previously known
H$\alpha$ emission-lined stars, and our final sample sadly contains no stars.}
\end{figure}

\clearpage
\begin{figure}
\epsscale{1.0}
\plotone{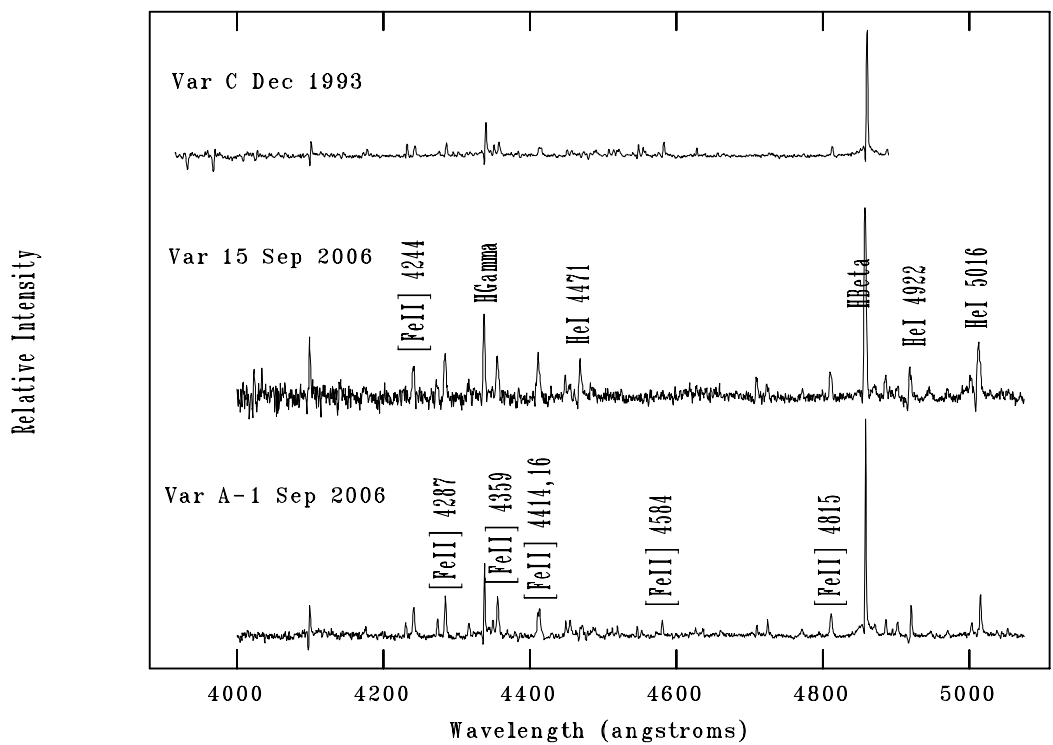}
\caption{\label{fig:lbv1} Spectra of known ``hot" LBVs with [FeII] emission.  We show spectra of Var C (M33), Var 15 (M31), and Var A-1 (M31)
in their ``quiescent" state, where the spectra are dominated by emission of the Balmer lines, He I, and forbidden Fe II.  We have identified
only the strongest lines; the wavelengths of the other emission features correspond to [Fe II]
lines listed by Kenyon \& Gallagher (1985) for AE And (their Table 3).
}
\end{figure}

\clearpage
\begin{figure}
\epsscale{0.8}
\plotone{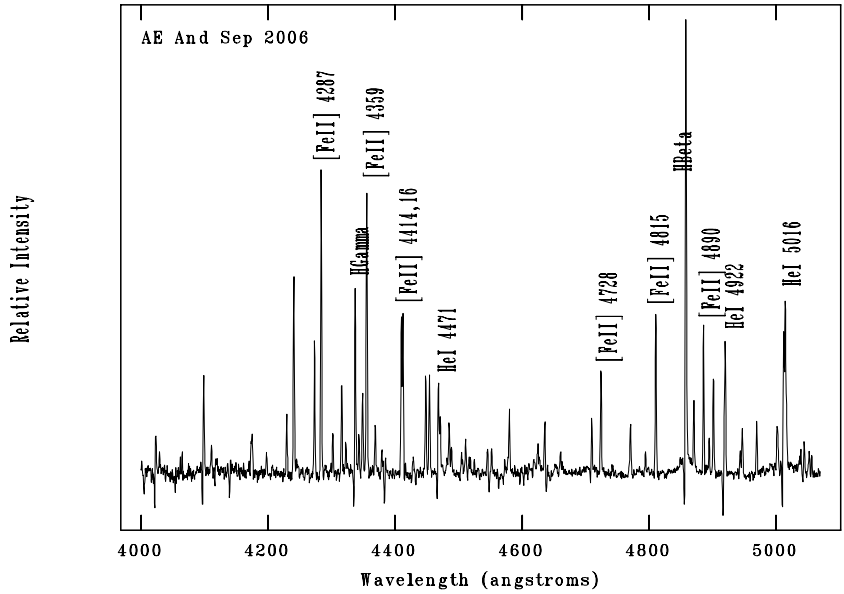}
\caption{\label{fig:AEnow} Spectrum of AE And in September 2006.  In the 1983
spectrum shown by Kenyon \& Gallagher (1985) many of the 
[FeII] lines were significantly stronger than that of H$\beta$.  In our rather noisy
spectrum we see only H$\beta$ emission, as a P Cygni line superposed on 
broad emission.
}
\end{figure}

\clearpage
\begin{figure}
\epsscale{0.8}
\plotone{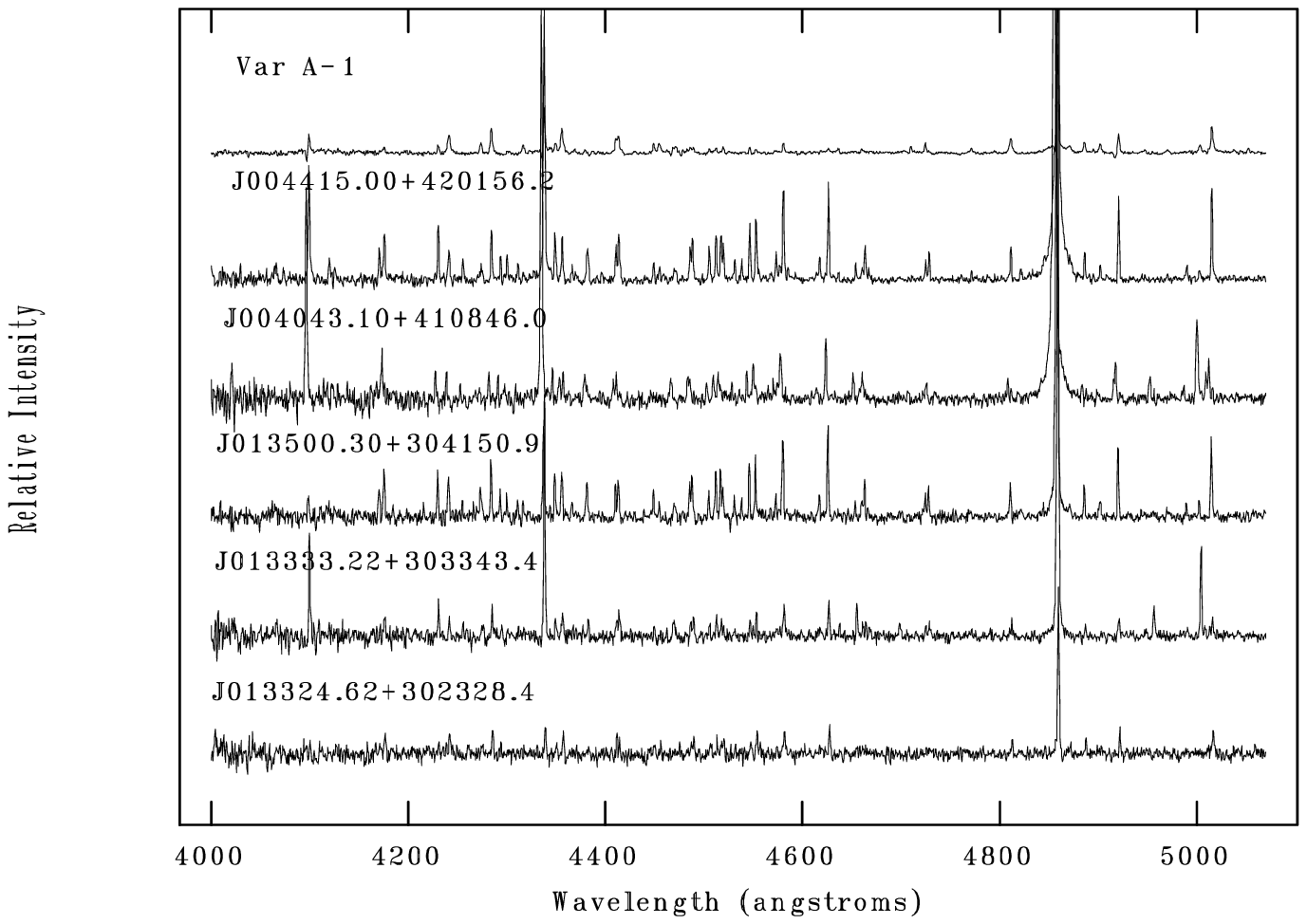}
\plotone{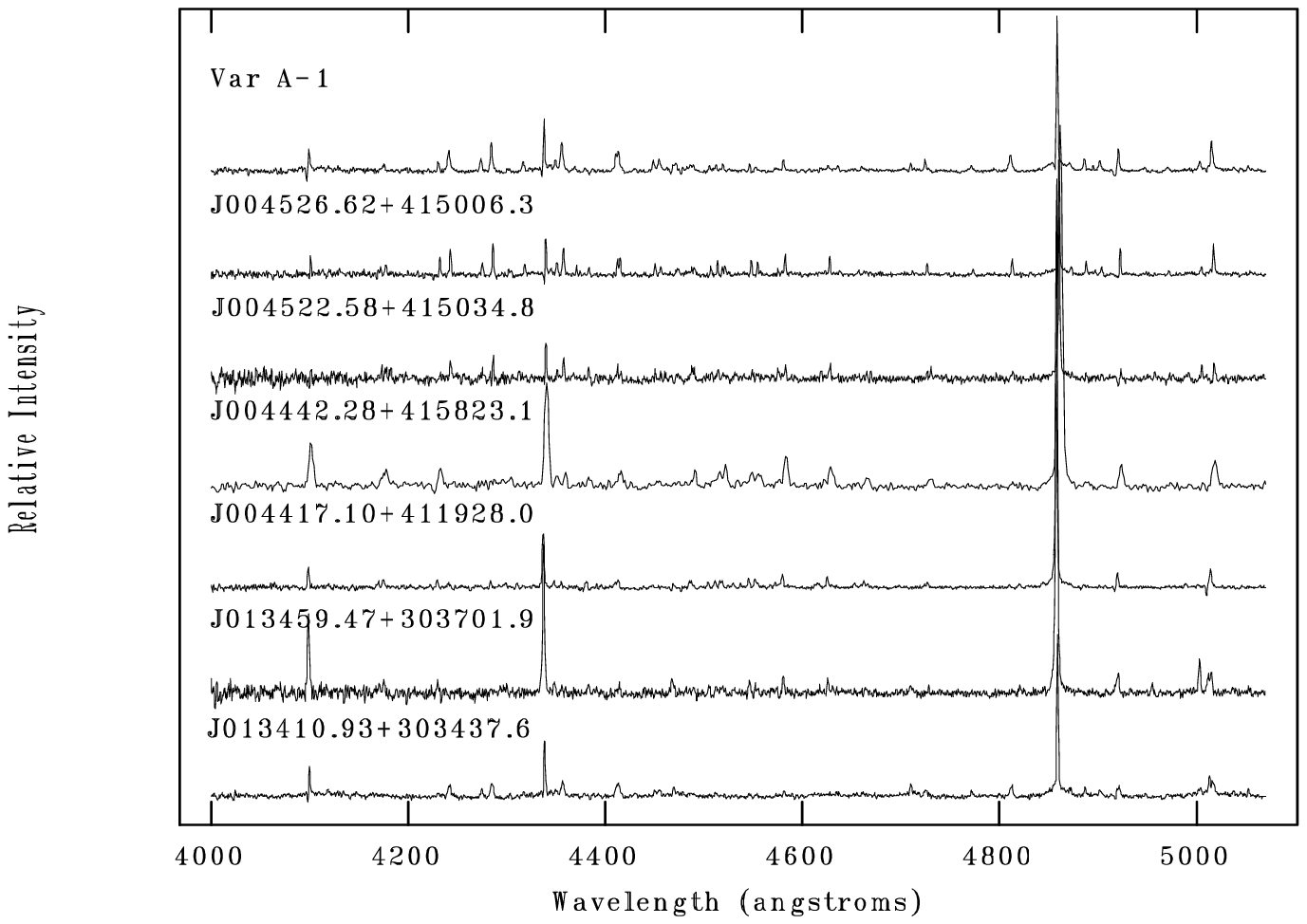}
\caption{\label{fig:lbvhot} Spectra of newly found M31 and M33
hot LBV candidates compared to Var A-1.
At top we show the stars whose spectra have strong [Fe II], while at bottom
we show stars which have relatively weak [Fe II] and Fe II lines.  For line identifications see
Fig.~\ref{fig:lbv1}.  All the spectra were obtained in September 2006, except for that of J004442.28+415823.1, which was obtained by N. Caldwell in November 2006.  The star J004417.10+411928.0
was previously identified as an LBV candidate (k350) by King et al.\ (1998).
}
\end{figure}

\clearpage
\begin{figure}
\epsscale{0.8}
\plotone{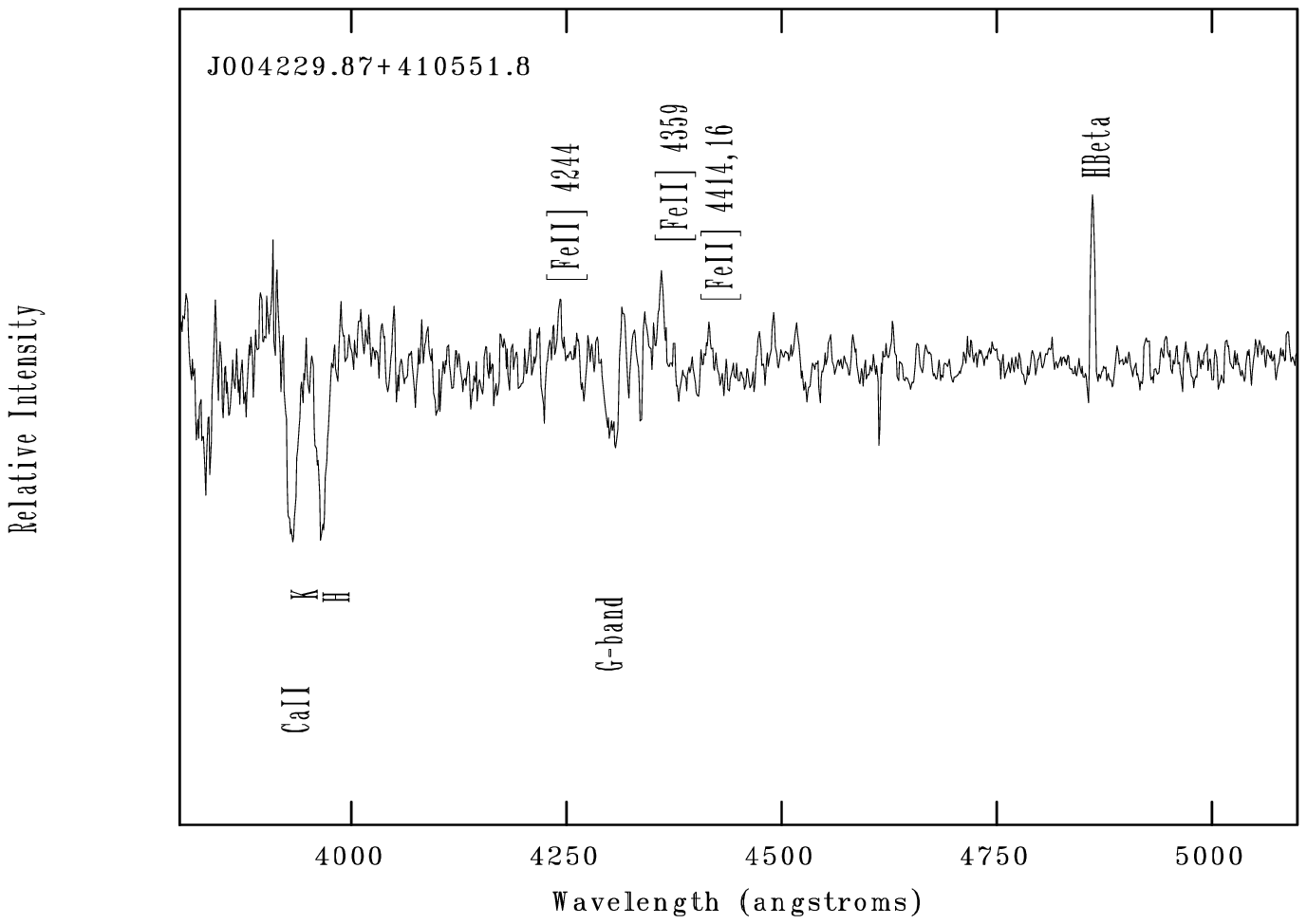}
\plotone{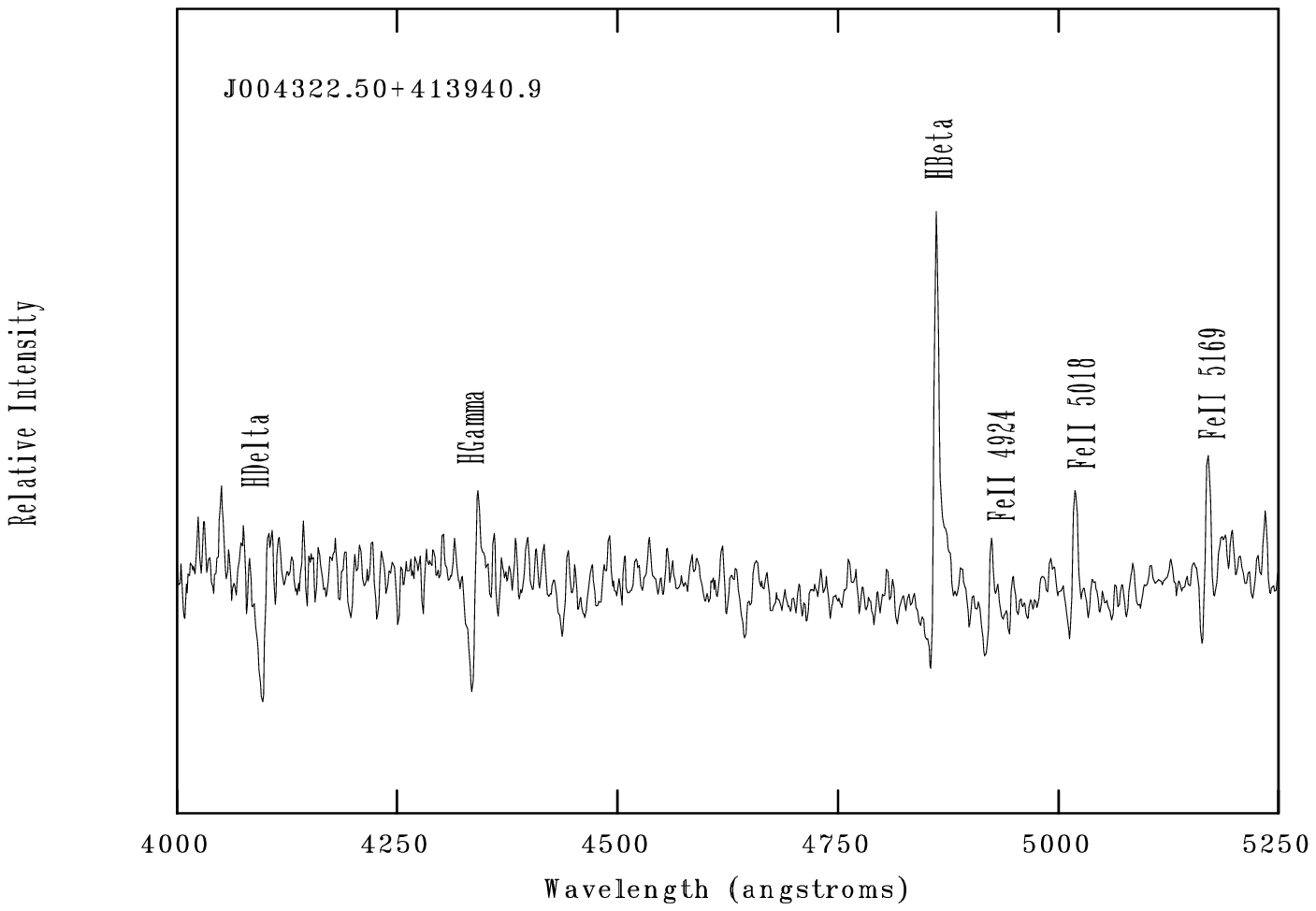}
\caption{\label{fig:nelson} Two peculiar hot LBV candidates.
The spectrum of the M31 star J004229.87+410551.8 (top)
shows [Fe II] and Balmer emission, but also an absorption spectrum
characteristic of a mid-F star or later.  The spectrum may be composite.  
The spectrum of the M31 star J004322.50+413940.9 (bottom) shows
P Cygni emission in the lower Balmer lines plus several lines of Fe II.
The spectra were
obtained by N. Caldwell in November 2006.
}
\end{figure}

\clearpage
\begin{figure}
\epsscale{0.8}
\plotone{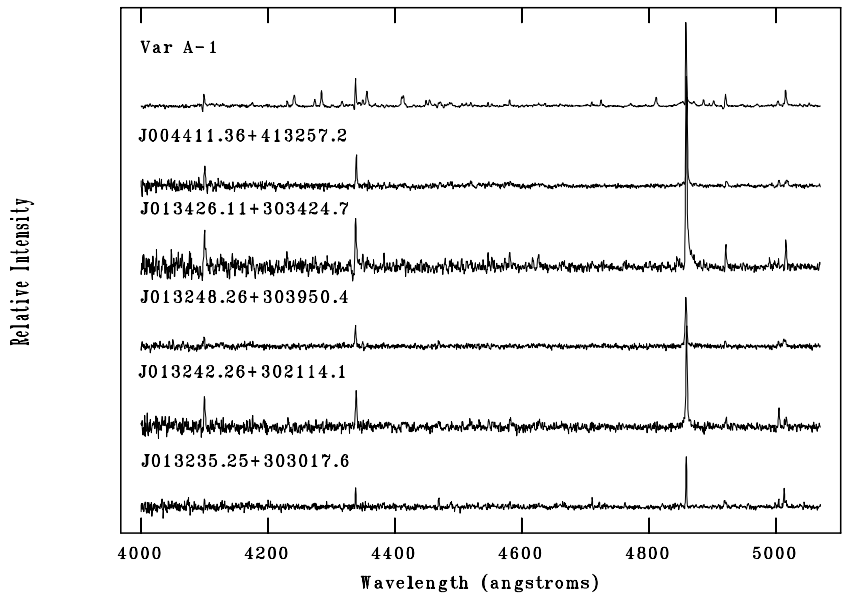}
\caption{\label{fig:hotmaybe} Spectra of newly found additional hot LBV candidates compared to Var A-1. These stars show Balmer and some He I emission,
but any [FeII] emission is incipient at best. The M31
star (J004411.36+413257.2) was  previously identified by King et al.\ (1998) as an LBV
candidate (k315a) based upon its spectrum.}
\end{figure}

\clearpage
\begin{figure}
\epsscale{0.6}
\plotone{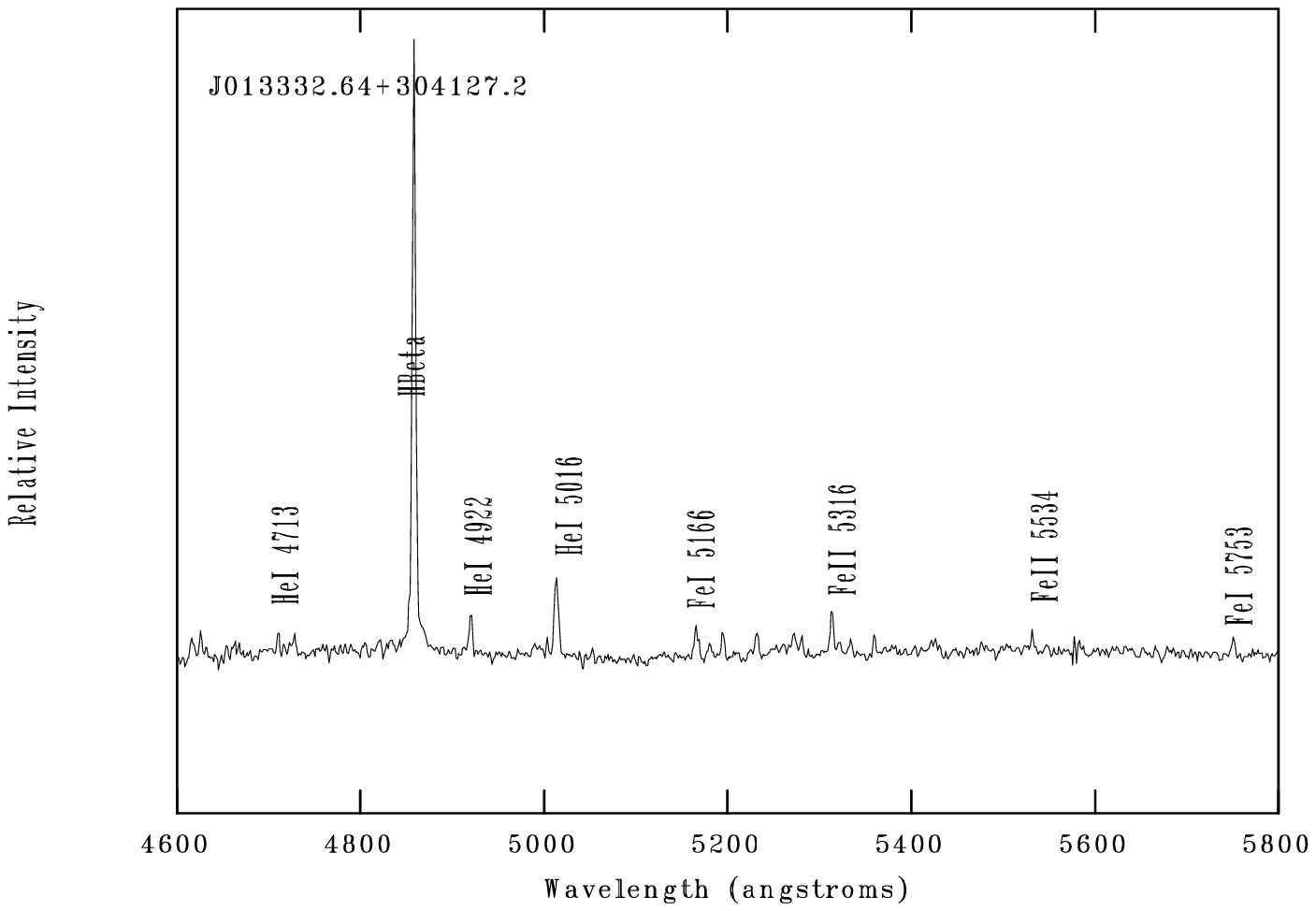}
\plotone{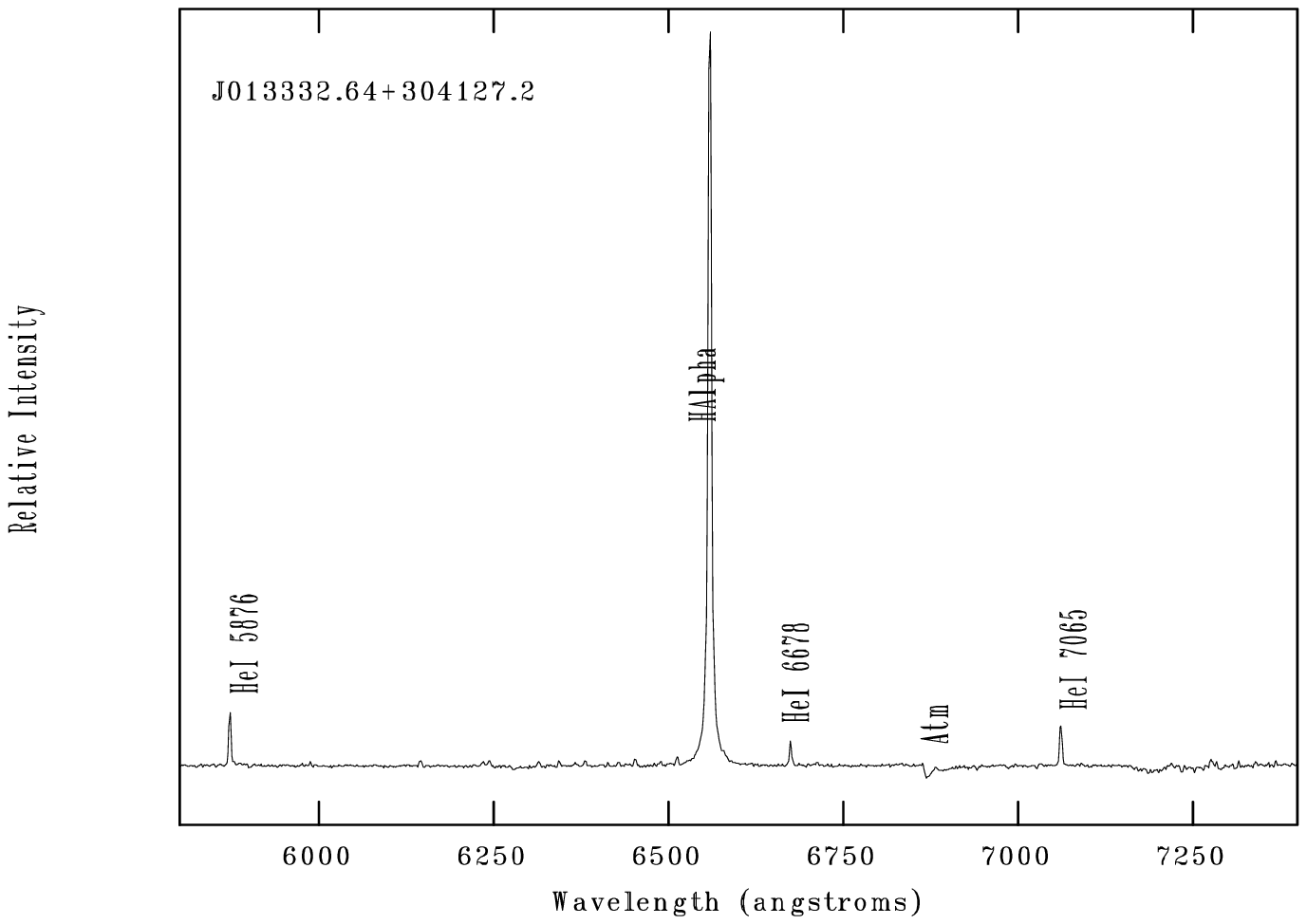}
\plotone{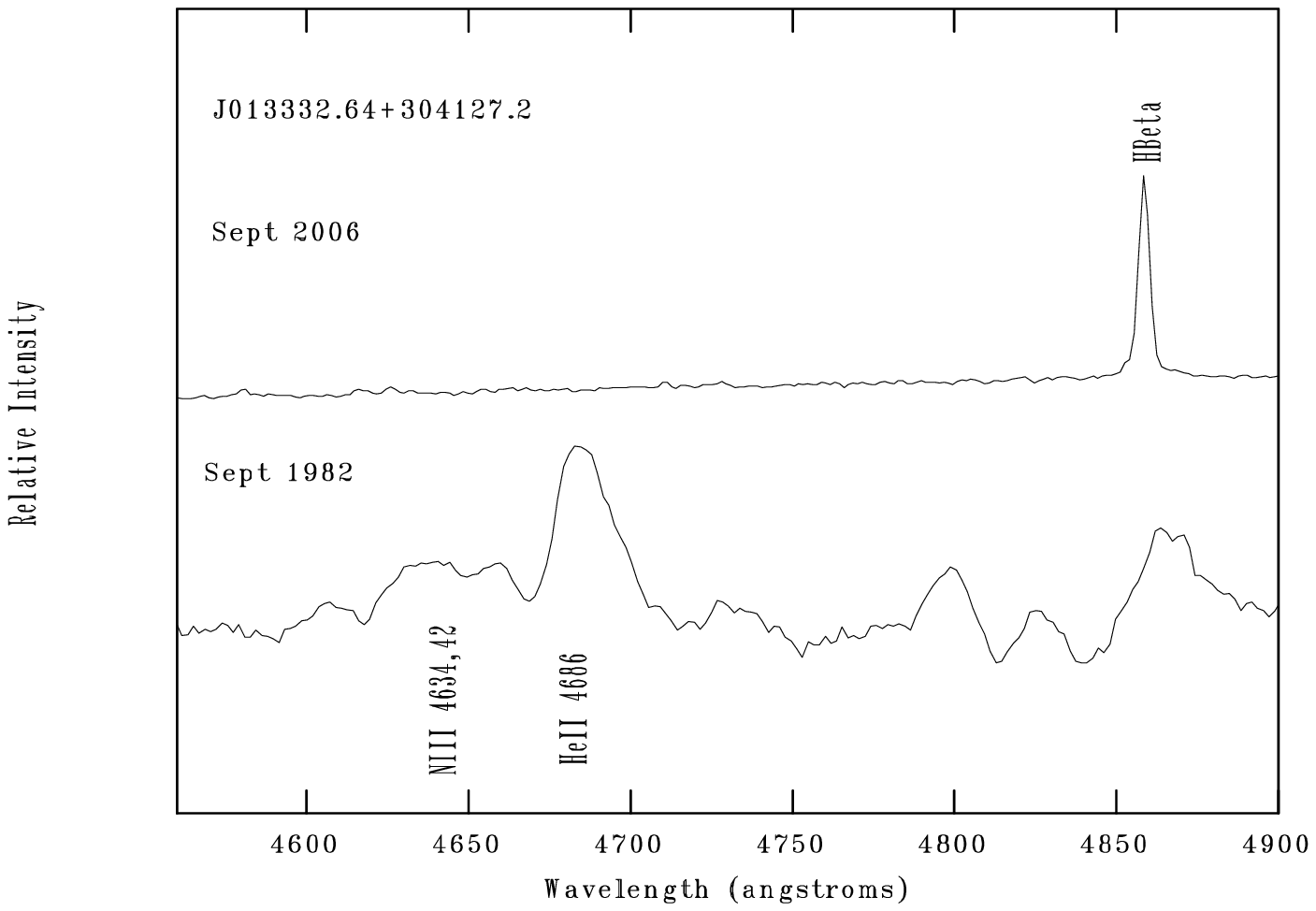}
\caption{\label{fig:am2} Spectra of the M33 star
J013332.64+304127.2 (AM2 from Armandroff \& Massey 1985, also known
to Simbad as  [MC83] 28), 
formerly
a WN star and now an LBV candidate.
{\it Top and middle:} Currently this star show Balmer, He I emission, and numerous metal lines
of Fe~I and Fe~II. {\it Bottom:} In 1982, however, its spectrum was that of a WN-type 
Wolf-Rayet star.}
\end{figure}

\clearpage
\begin{figure}
\epsscale{0.8}
\plotone{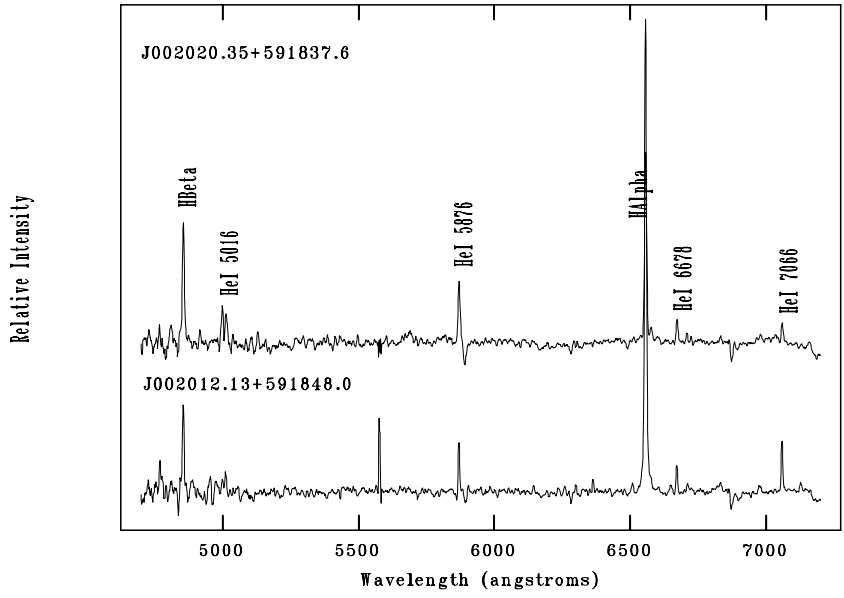}
\caption{\label{fig:ic10} Spectra of two hot LBV candidates in IC10.
The stars show He~I and Balmer emission.  Note that the absorption feature
to the red of He~I $\lambda 5876$ is due to Na D interstellar line. }
\end{figure}

\clearpage
\begin{figure}
\epsscale{0.6}
\plotone{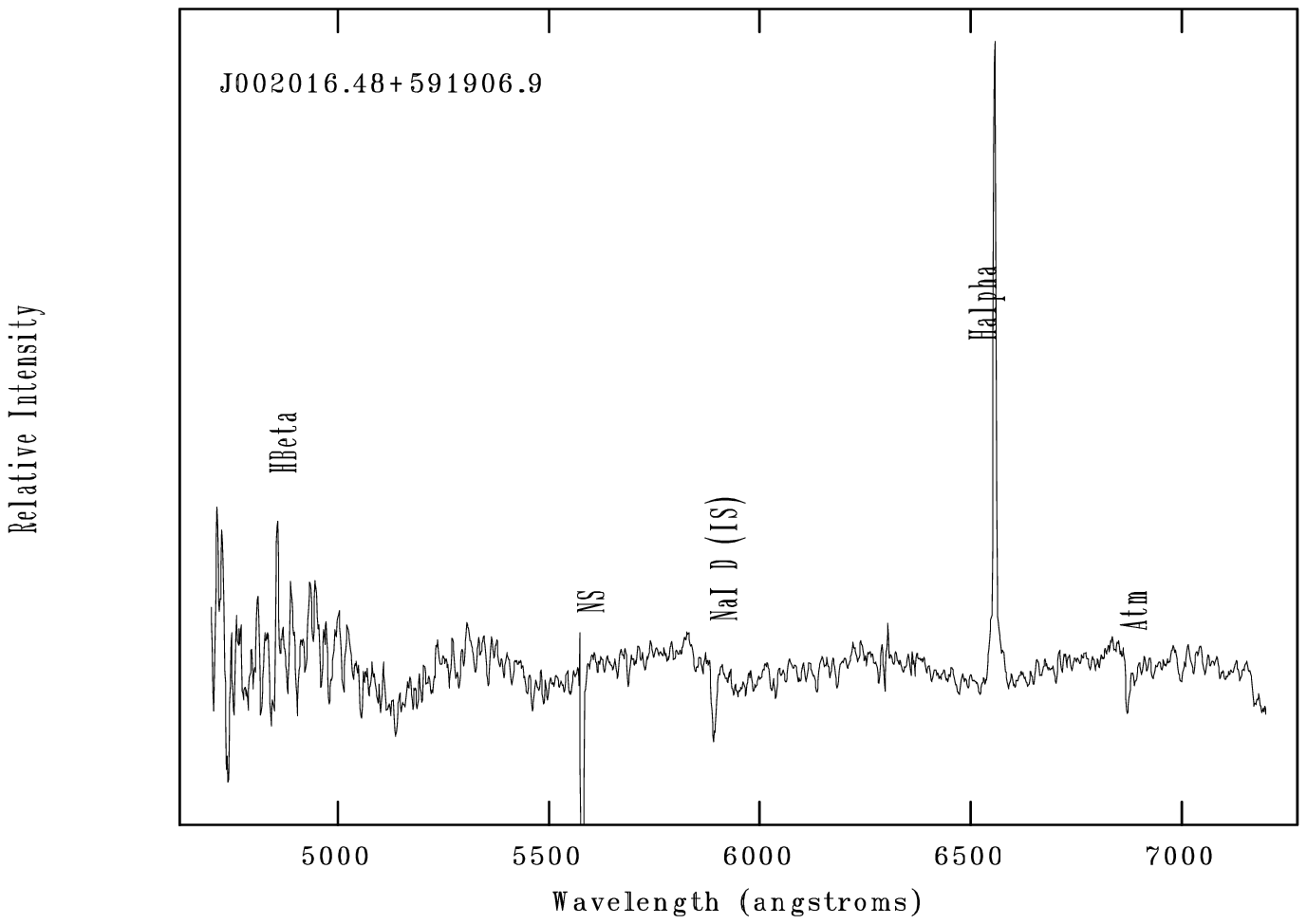}
\plotone{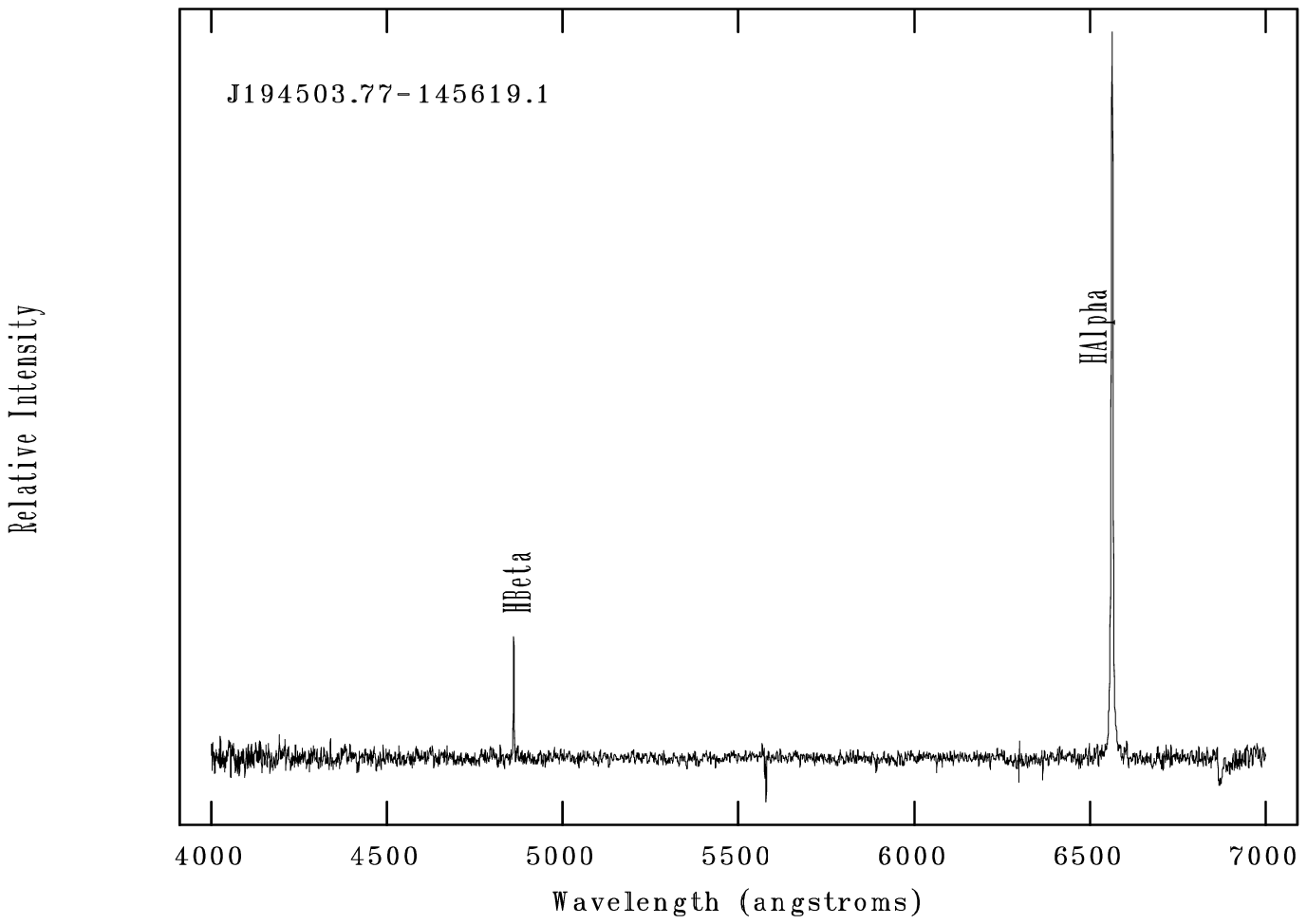}
\plotone{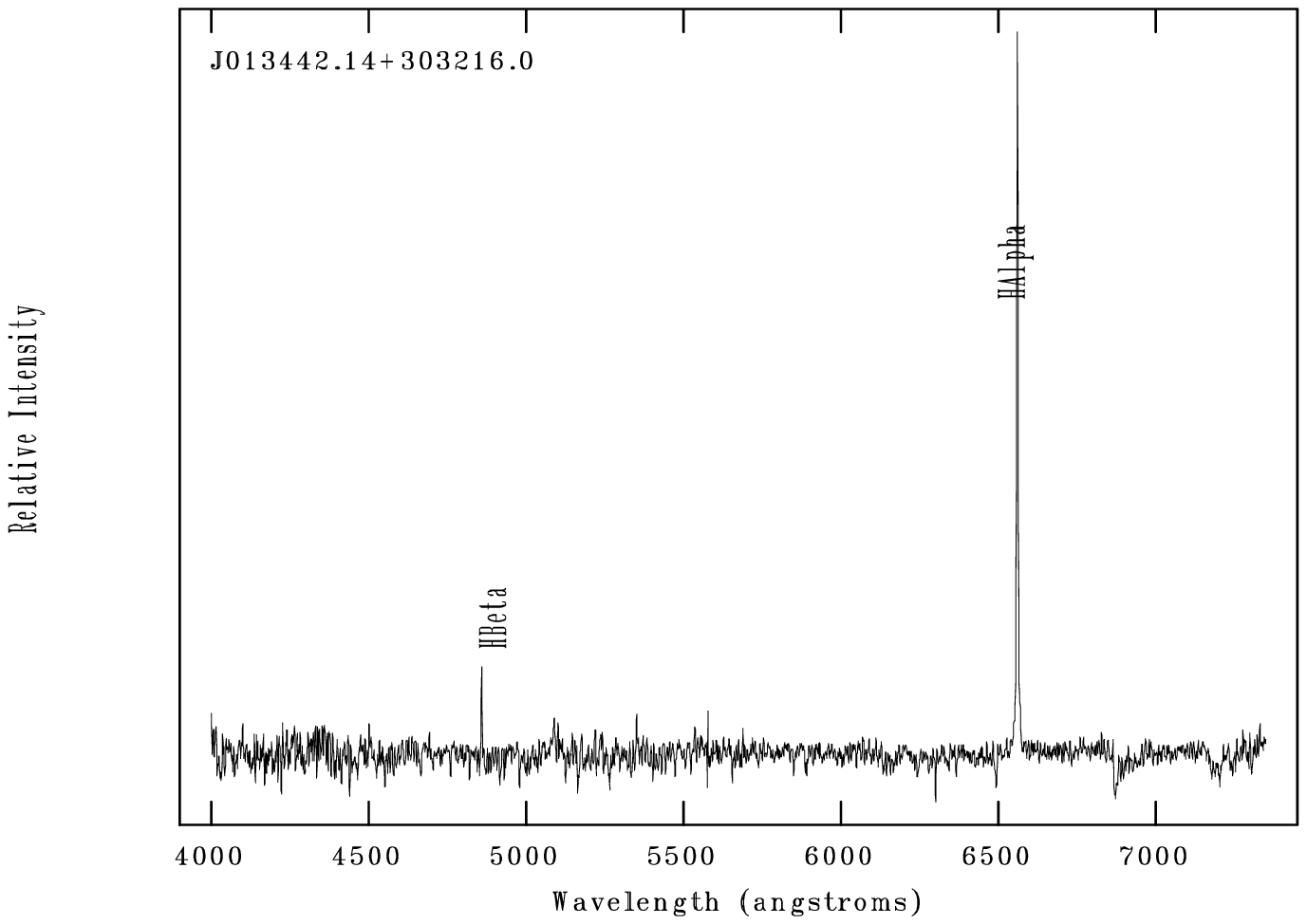}
\caption{\label{fig:n6822} Spectra of three questionable LBV candidates. 
These stars show H$\alpha$ and
H$\beta$ emission but none of the forbidden lines we would expect of an HII region.
Incomplete night-sky (NS), NaI D interstellar (IS), and the atmospheric B band (Atm)
are the only other features visible in the IC 10 spectrum at top. }
\end{figure}

\clearpage
\begin{figure}
\epsscale{0.8}
\plotone{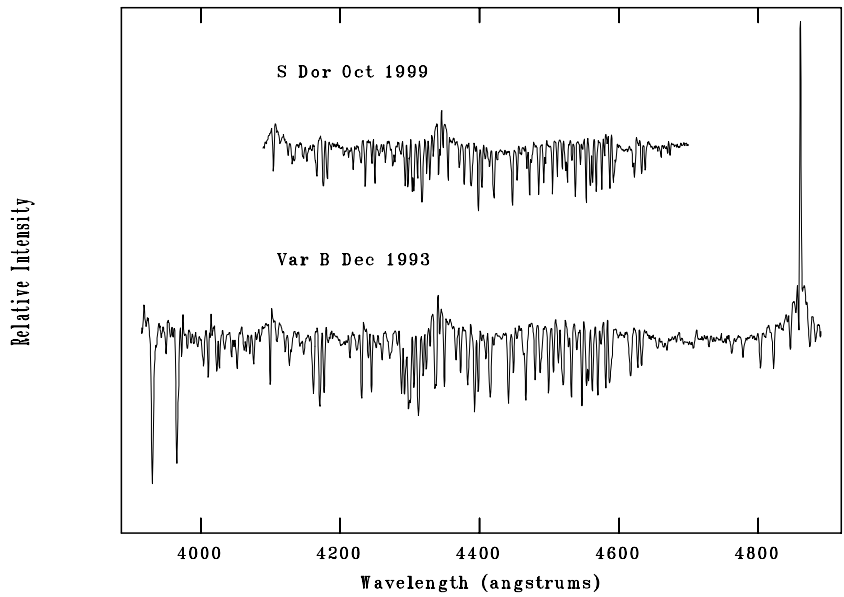}
\caption{\label{fig:lbv2} Spectra of LBVs in their cool state.   {\it Top:} We show the spectrum of the M33 LBV Var B obtained
in December 1993 (during its 1992-1993 outburst) to that of a spectrum of 
the LMC LBV S Doradus obtained in October 1996.  The absorption line spectra
resemble that of an extreme late-type F-type supergiant, with numerous metal absorption lines.
This figure is based on Fig.~1 of Massey (2000). 
}
\end{figure}

\clearpage
\begin{figure}
\epsscale{0.8}
\plotone{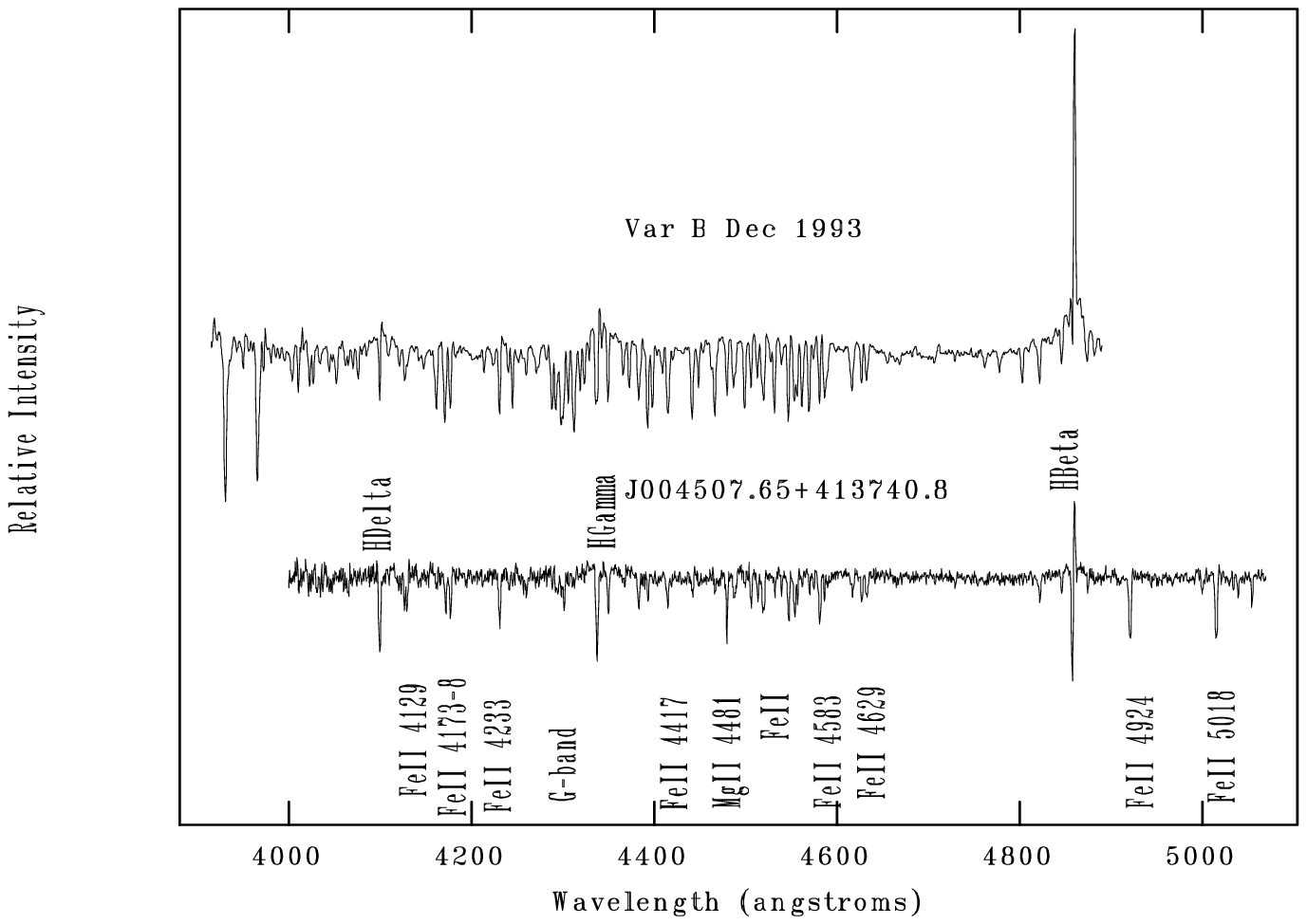}
\plotone{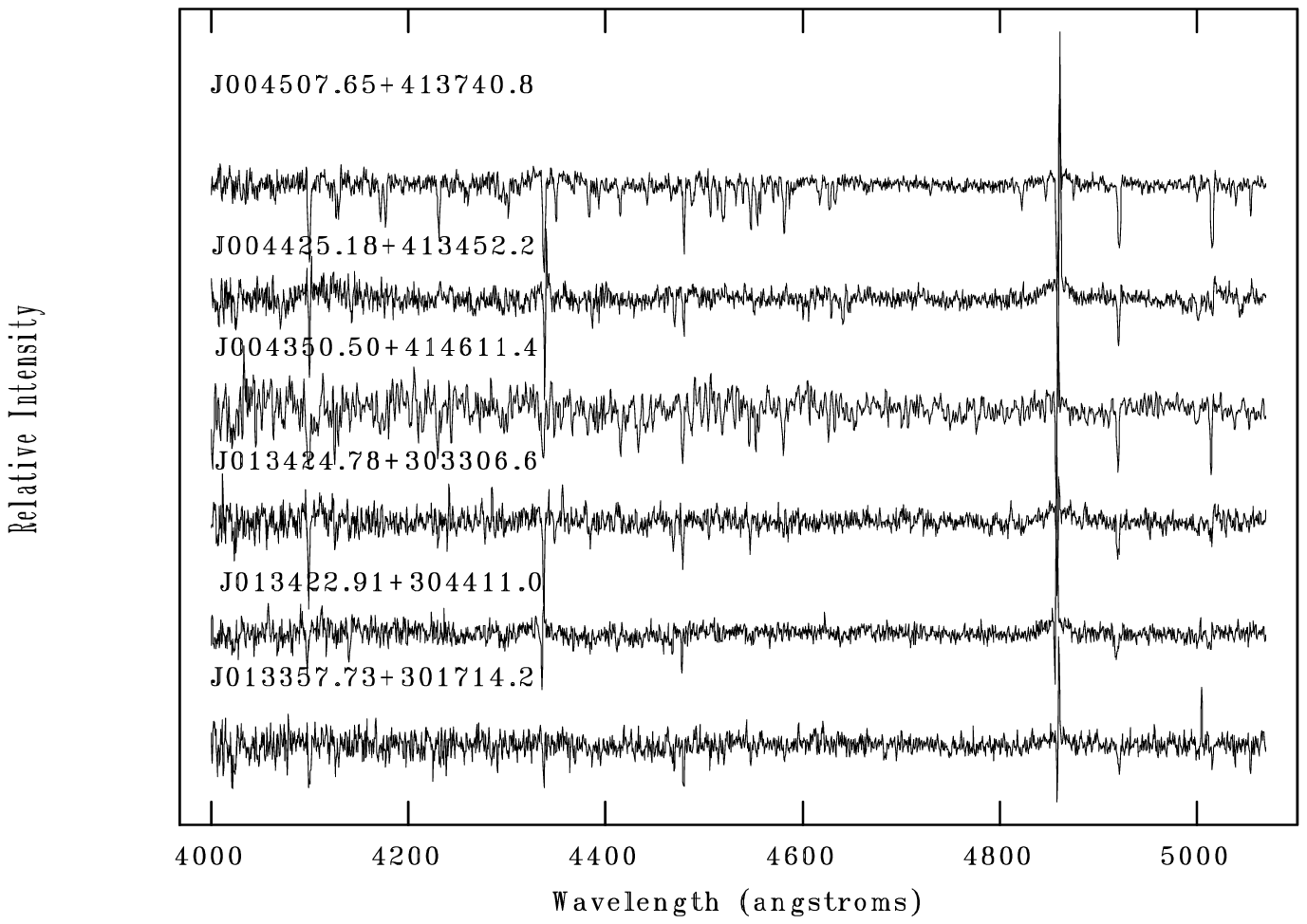}
\caption{\label{fig:coolguys}  Blue absorption 
spectra of newly found LBVs in their cool state.  In the top panel,
we compare the spectrum of the M33 LBV
Var B during outburst to that of the M31 star 
J004507.65+413740.8.  We
have identified some of the prominent lines in J004507.65+413740.8 using the very useful line
list given by Coluzzi (1993).  In the bottom panel, we show the spectra of J004507.65+413740.8
compared to those of the other 5 newly found ``cool" LBV candidates in M31 and
M33, with the spectra
scaled to emphasize the absorption components.  The star J004425.18+413452.2 was previously described as an LBV candidate (k411) by King et al.\ (1998).
See also Fig.~\ref{fig:coolprofiles}.}
\end{figure}

\clearpage
\begin{figure}
\epsscale{0.8}
\plotone{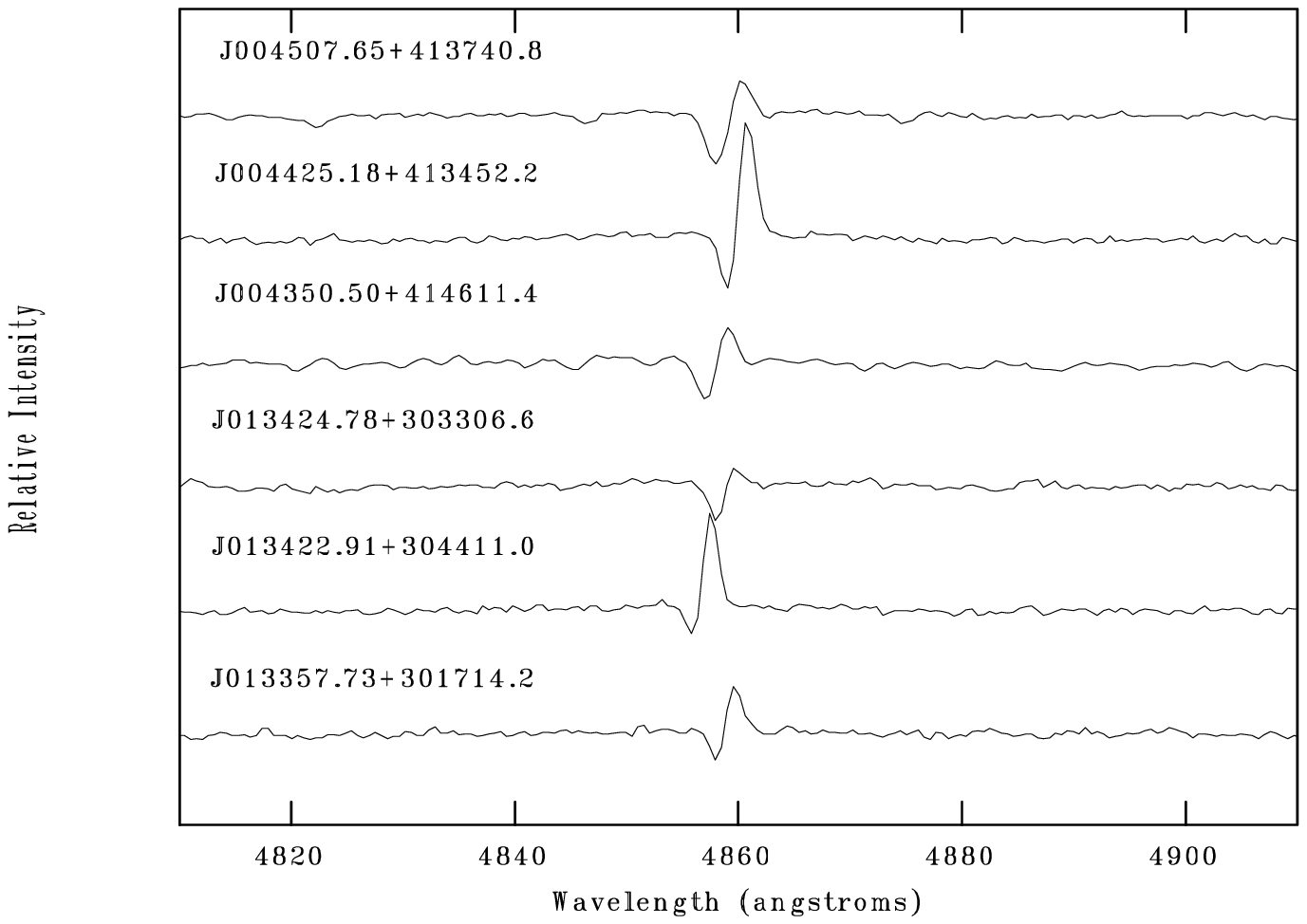}
\plotone{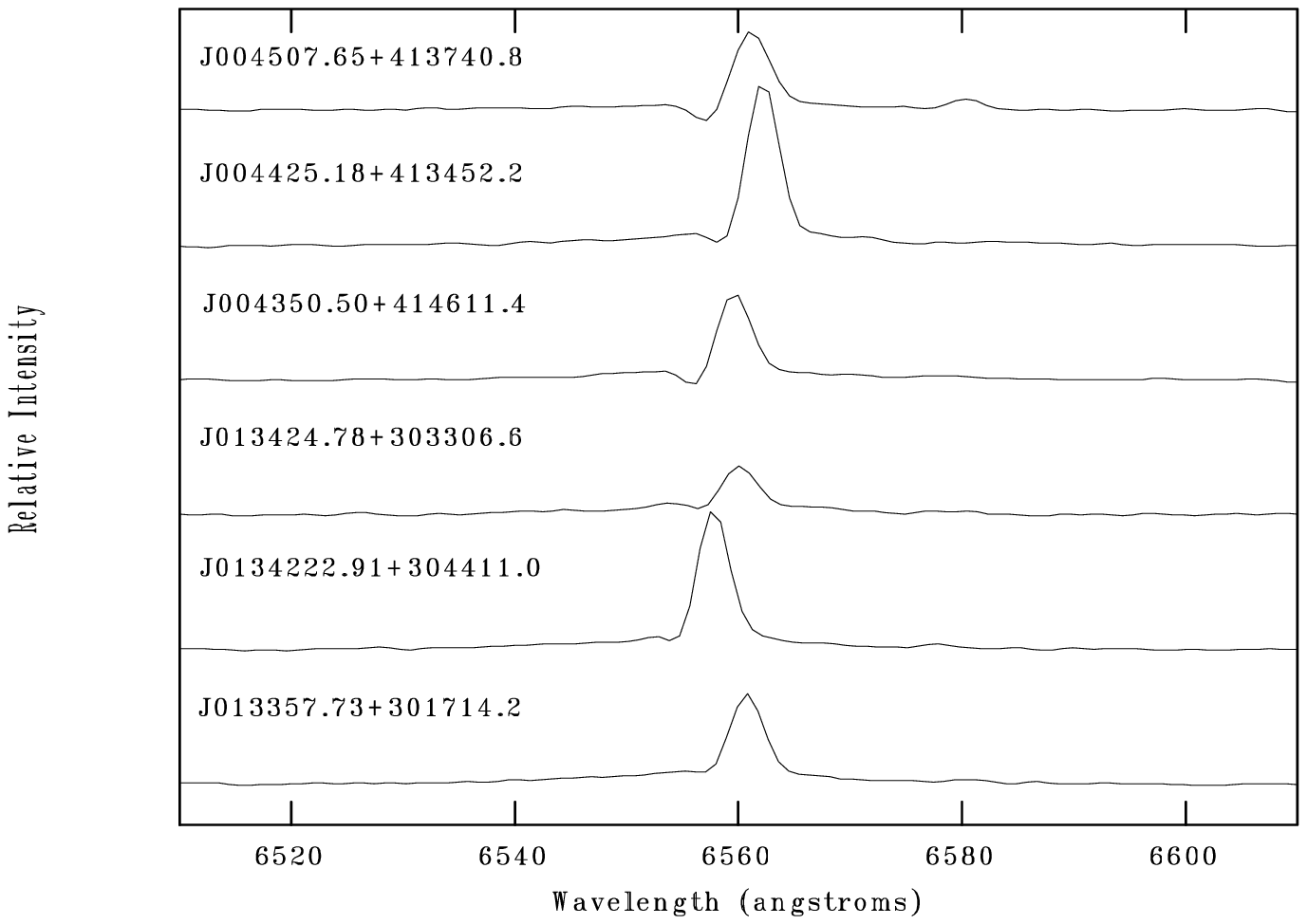}
\caption{\label{fig:coolprofiles} Emission spectra of newly found LBVs in their cool state. Here
we show the H$\beta$ profiles (top) and H$\alpha$ profiles (bottom) of our 
``cool" LBV candidates.  See also Fig.~\ref{fig:coolguys}.
}
\end{figure}

\clearpage
\begin{figure}
\epsscale{1.0}
\plotone{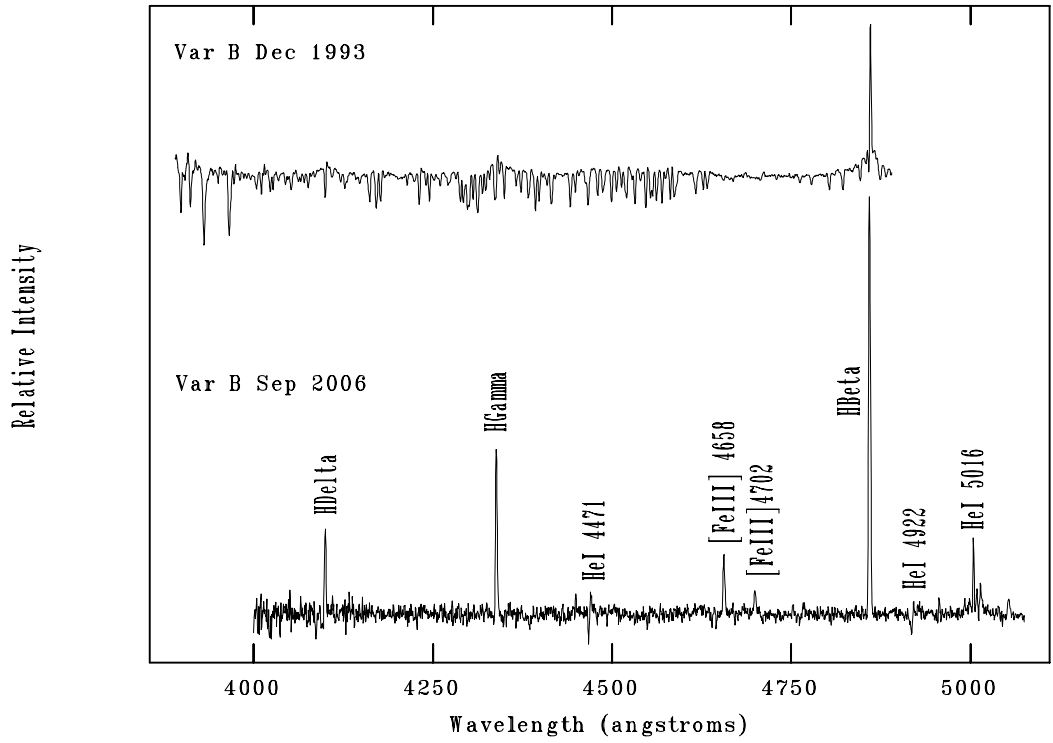}
\caption{\label{fig:varb} Changes in the spectra of the M33 LBV Var B.   We show the spectrum of Var B obtained
in December 1993 (during its 1992-1993 outburst) to that obtained in September 2006.}
\end{figure}

\clearpage
\begin{figure}
\epsscale{0.8}
\plotone{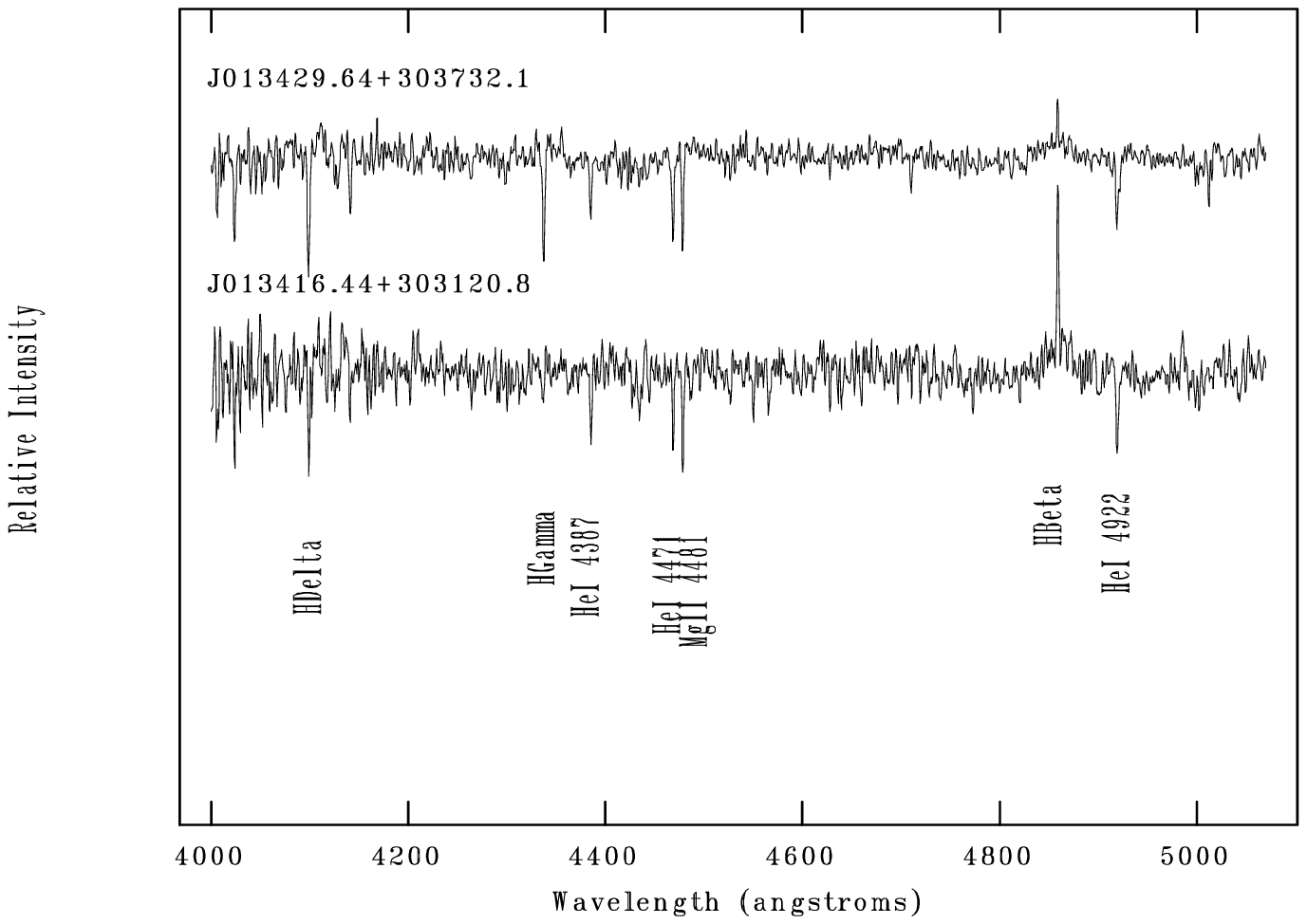}
\plotone{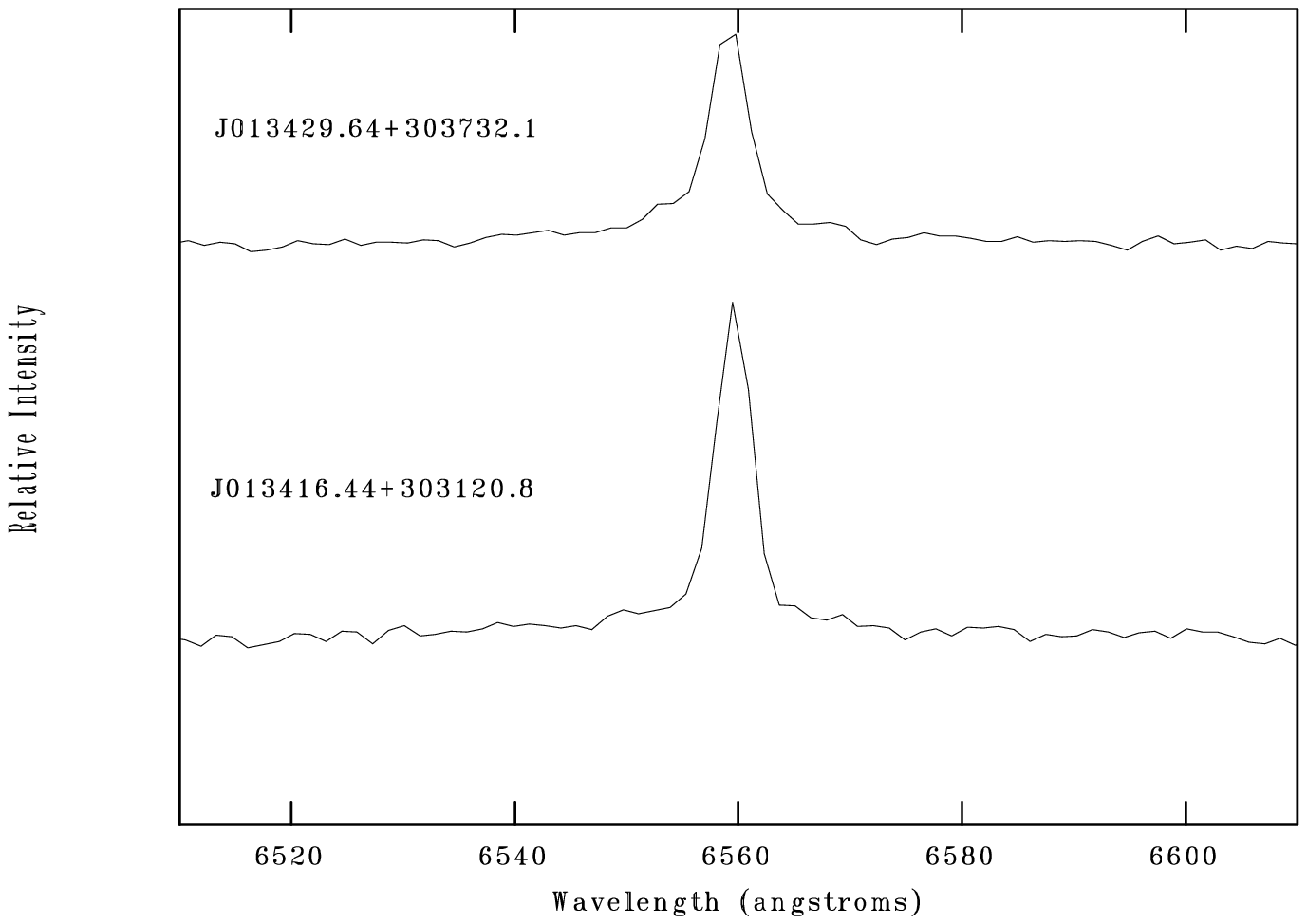}
\caption{\label{fig:weird} Two possible additional cool LBV candidates.
These two stars show B8 I absorption spectra, with H$\beta$ and H$\alpha$ emission.
Both the H$\beta$ and H$\alpha$ profiles show a very broad component with a narrow
component superposed.}
\end{figure}

\clearpage
\begin{figure}
\epsscale{1.0}
\plotone{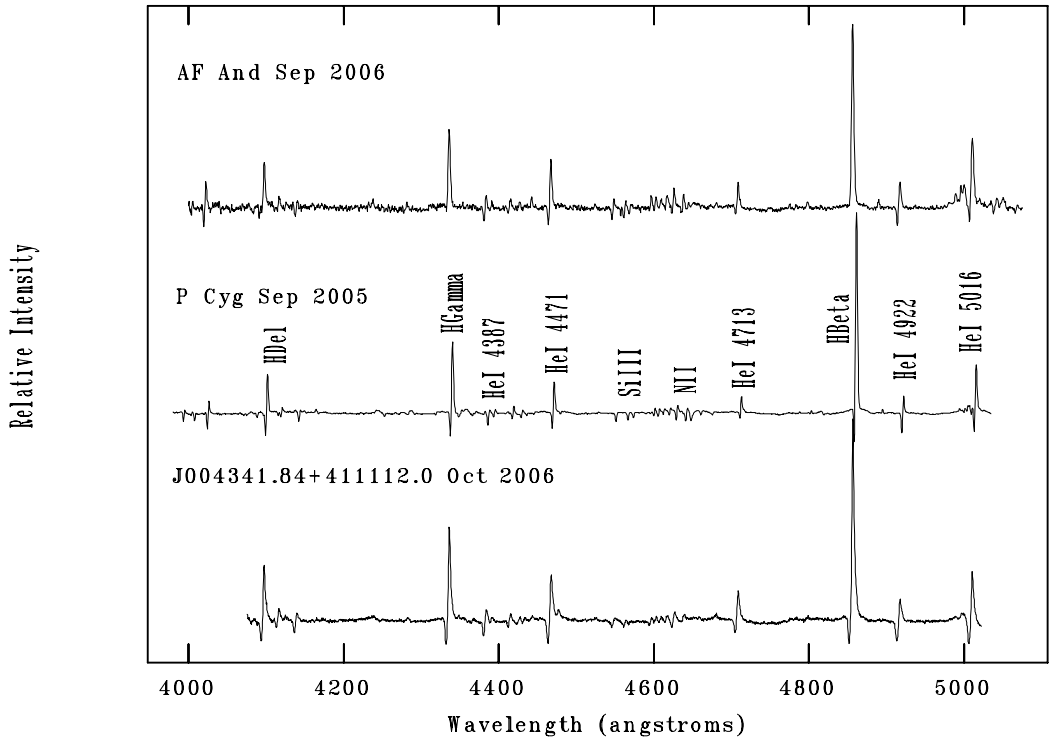}
\caption{\label{fig:pcyg} Spectra of P Cygni LBVs.  The upper spectrum is of AF And,
an LBV in M31.  The middle spectrum shows the spectrum of P Cyg itself.  The bottom
spectrum is of the star J00341.84+411112.0, an LBV candidate in M31 (Massey 2006).
Prominent lines are labeled.
}
\end{figure}

\clearpage
\begin{figure}
\epsscale{0.48}
\plotone{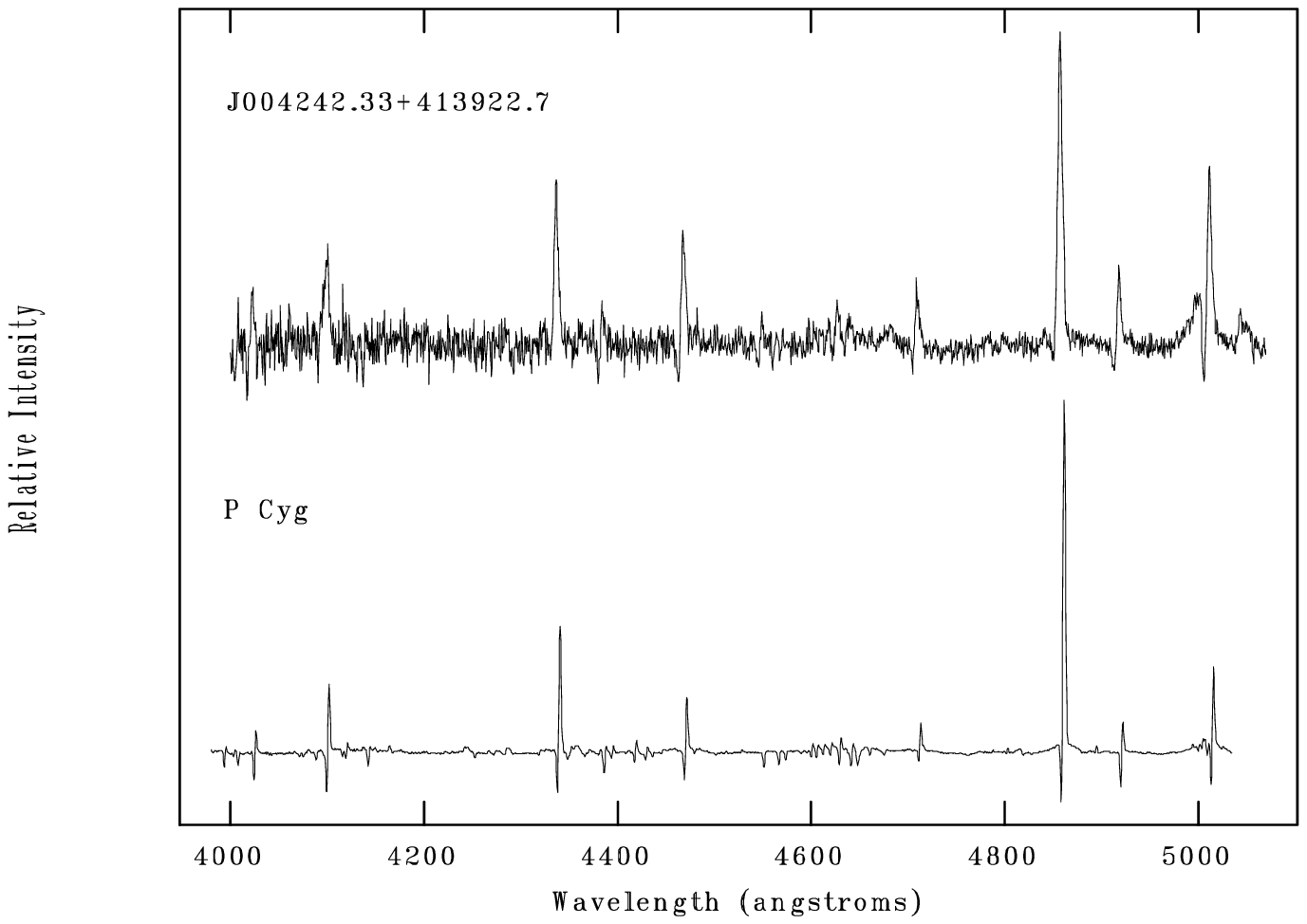}
\plotone{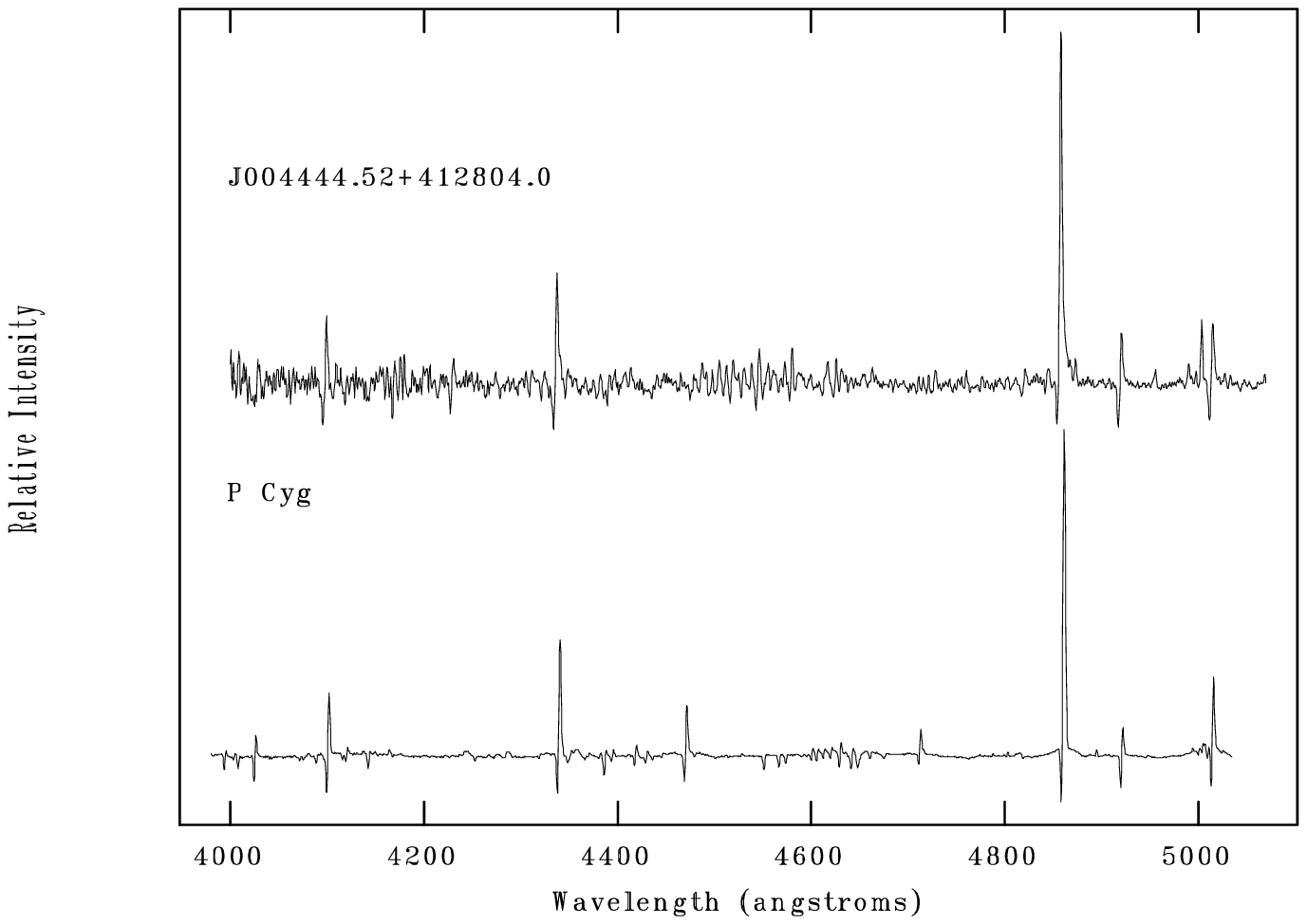}
\plotone{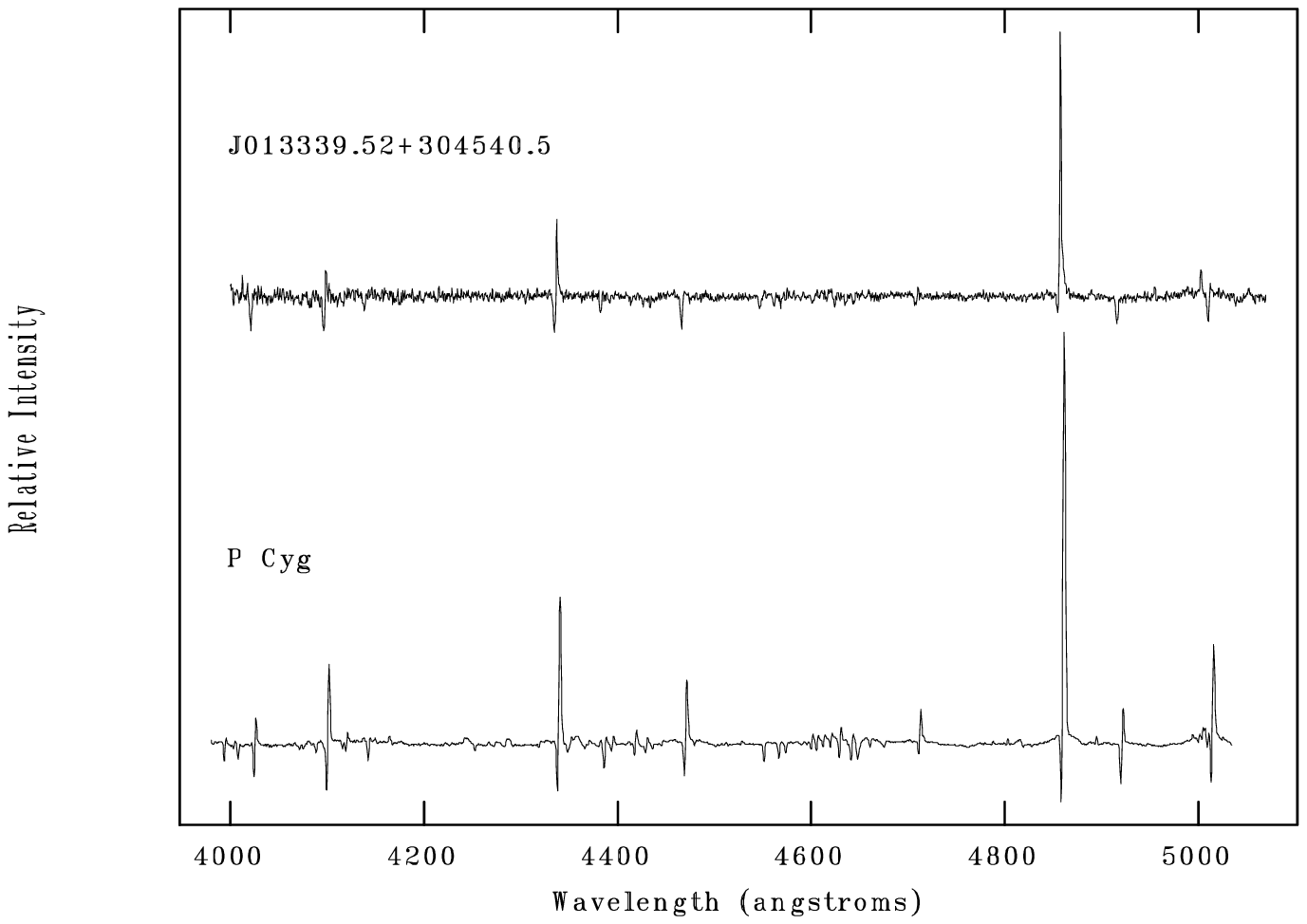}
\plotone{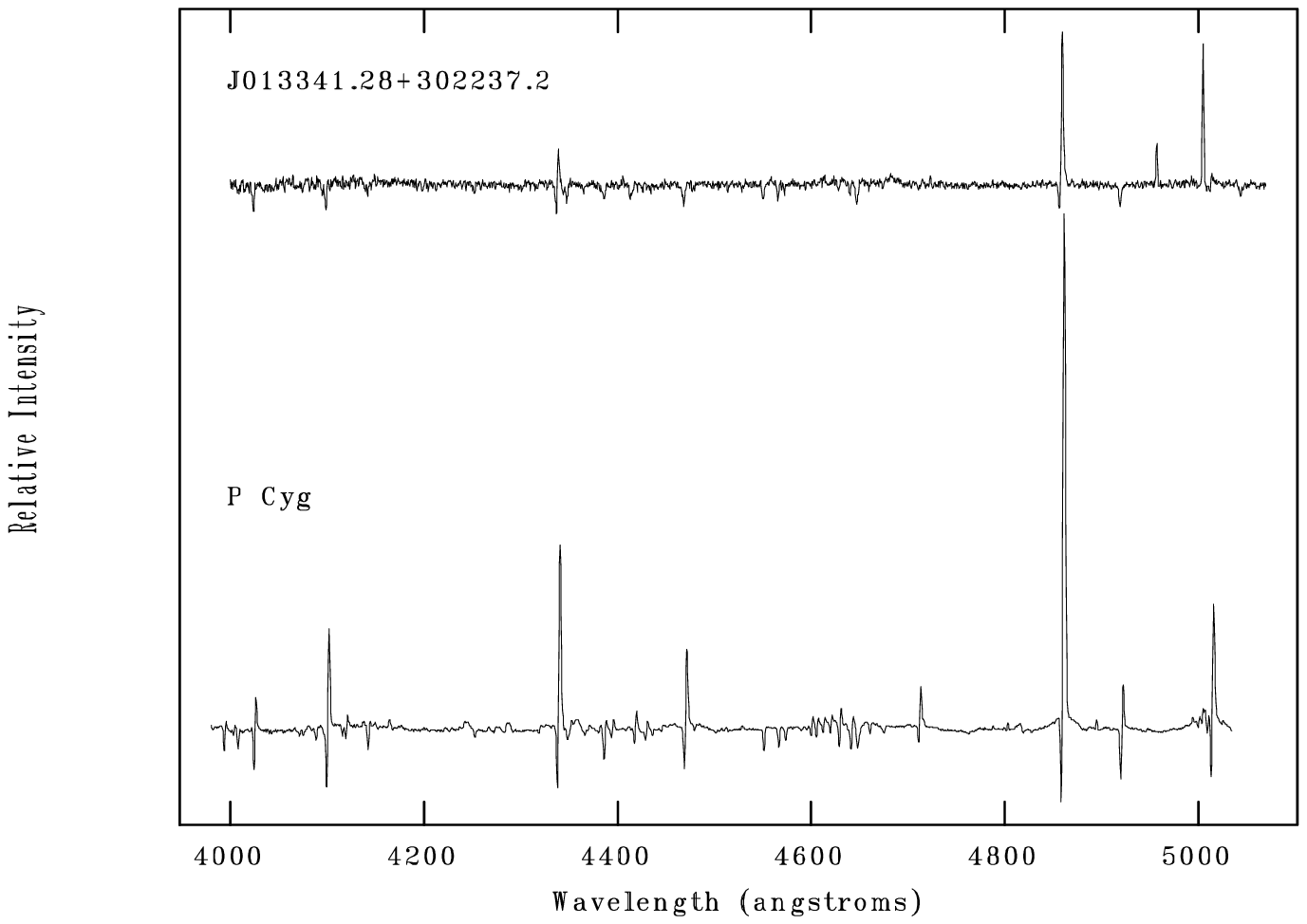}
\plotone{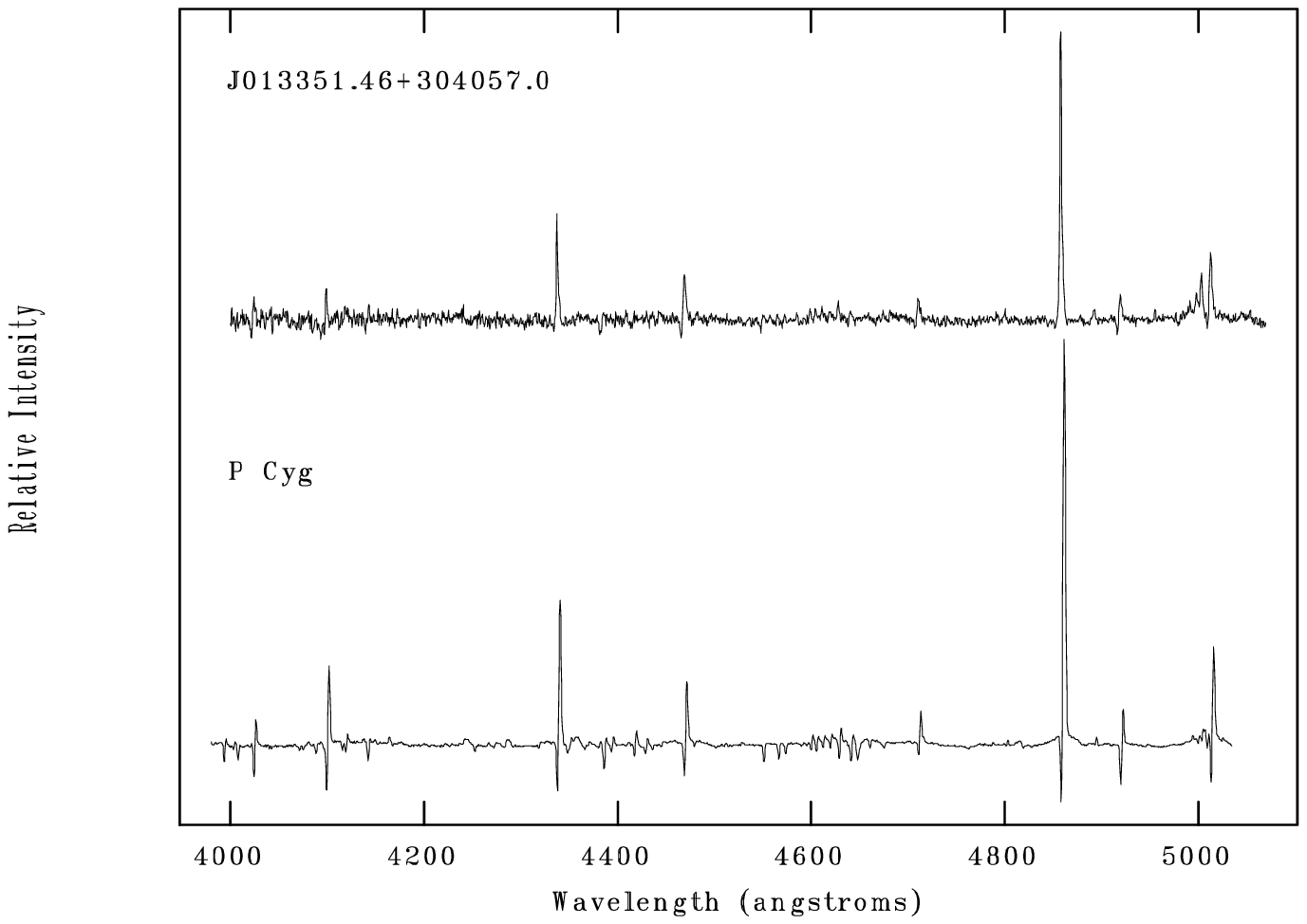}
\plotone{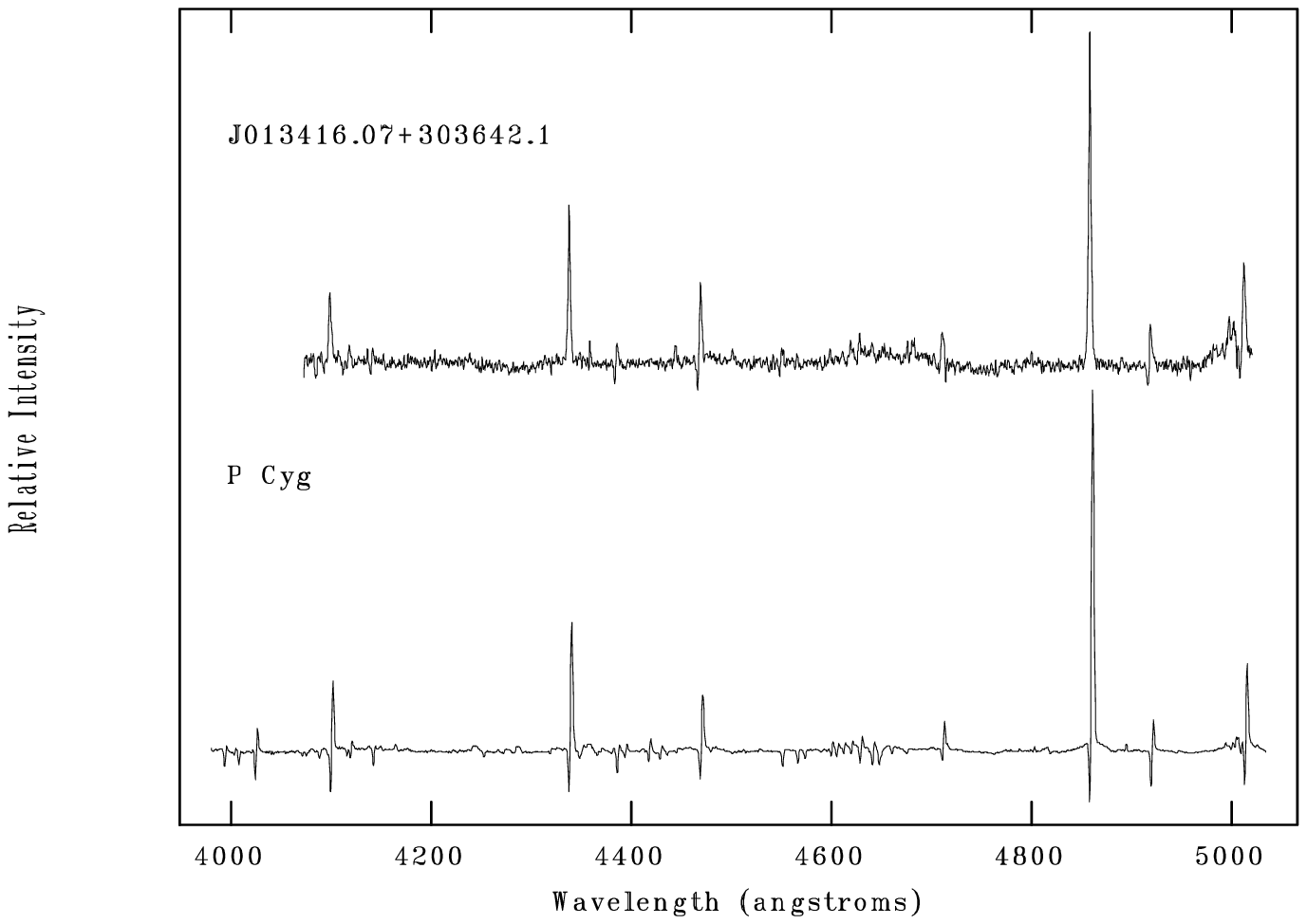}
\caption{\label{fig:pcygs} Spectra of P Cygni type LBV candidates in M31 and M33.  We compare the blue
spectra of 6 more LBV candidates to that of P Cygni.  For line identifications see Fig.~\ref{fig:pcyg}. The star J013416.07+303642.1 was also described as an LBV candidate
by Corral (1996), who called it H 108.}
\end{figure}

\clearpage
\begin{figure}
\epsscale{0.8}
\plotone{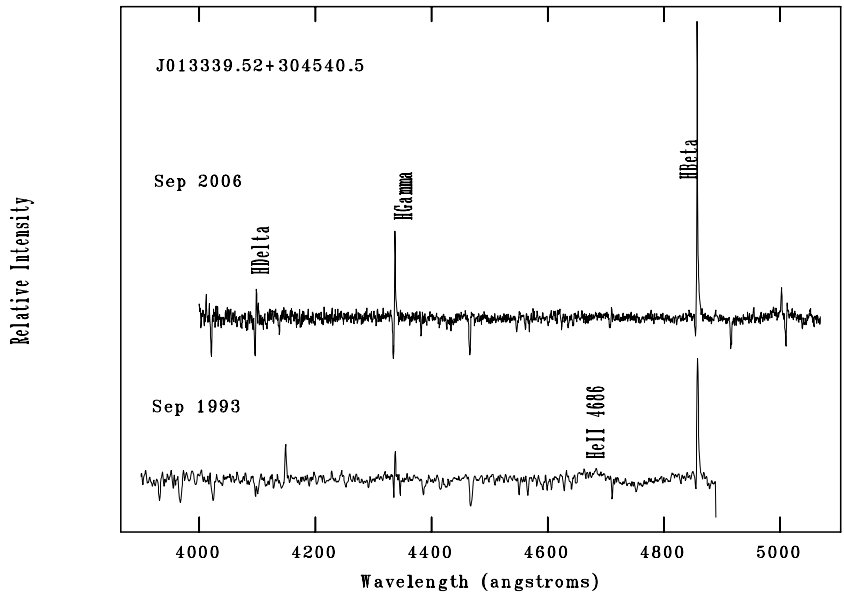}
\caption{\label{fig:pcygprob1} Spectrum of the M33 star
J013339.52+304540.5 in 1993 and 2006.
Emission in the Balmer lines has strengthened, and the broad He~II $\lambda 4686$
feature in the 1993 spectrum has disappeared.
}
\end{figure}

\clearpage
\begin{figure}
\epsscale{0.8}
\plotone{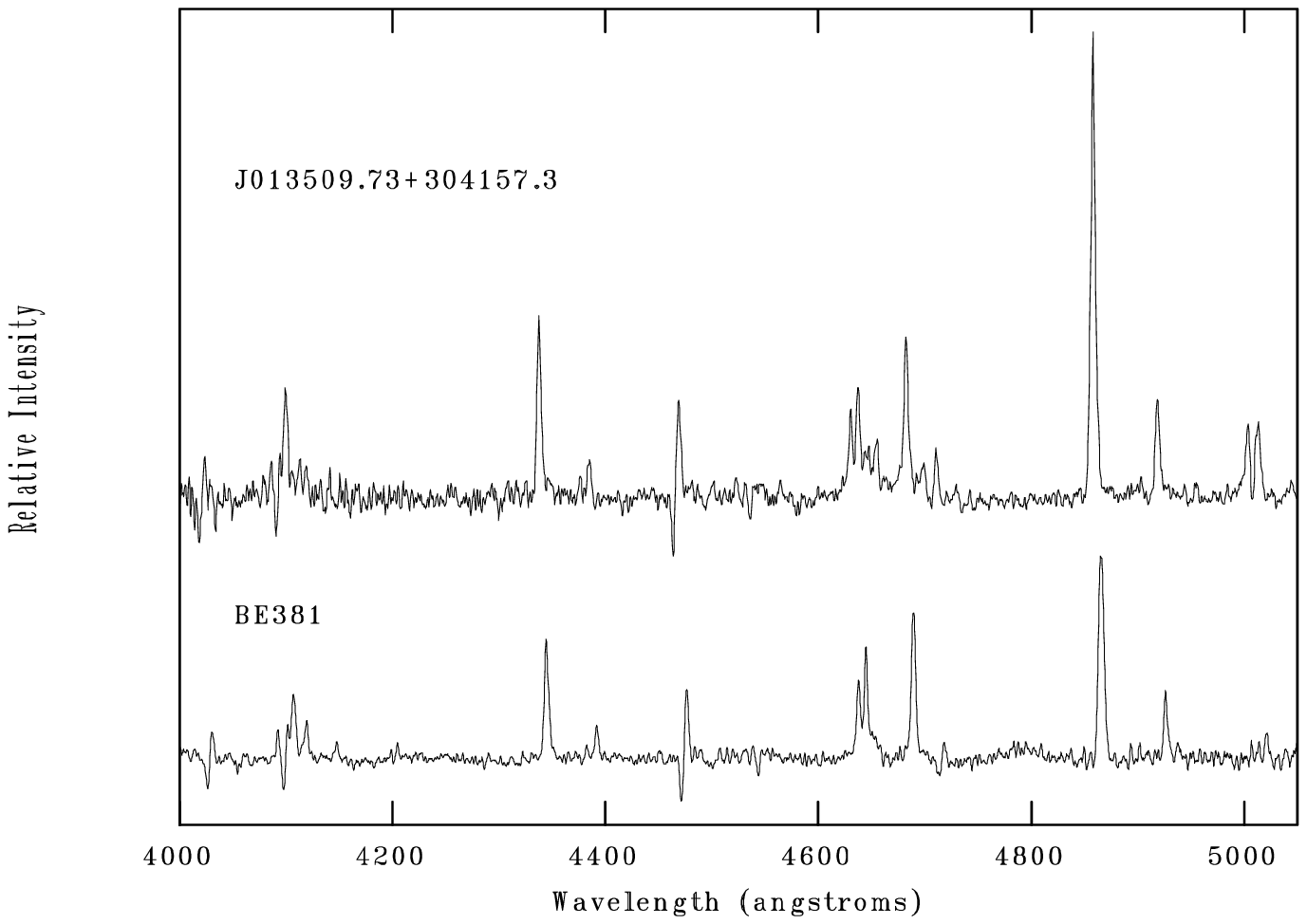}
\plotone{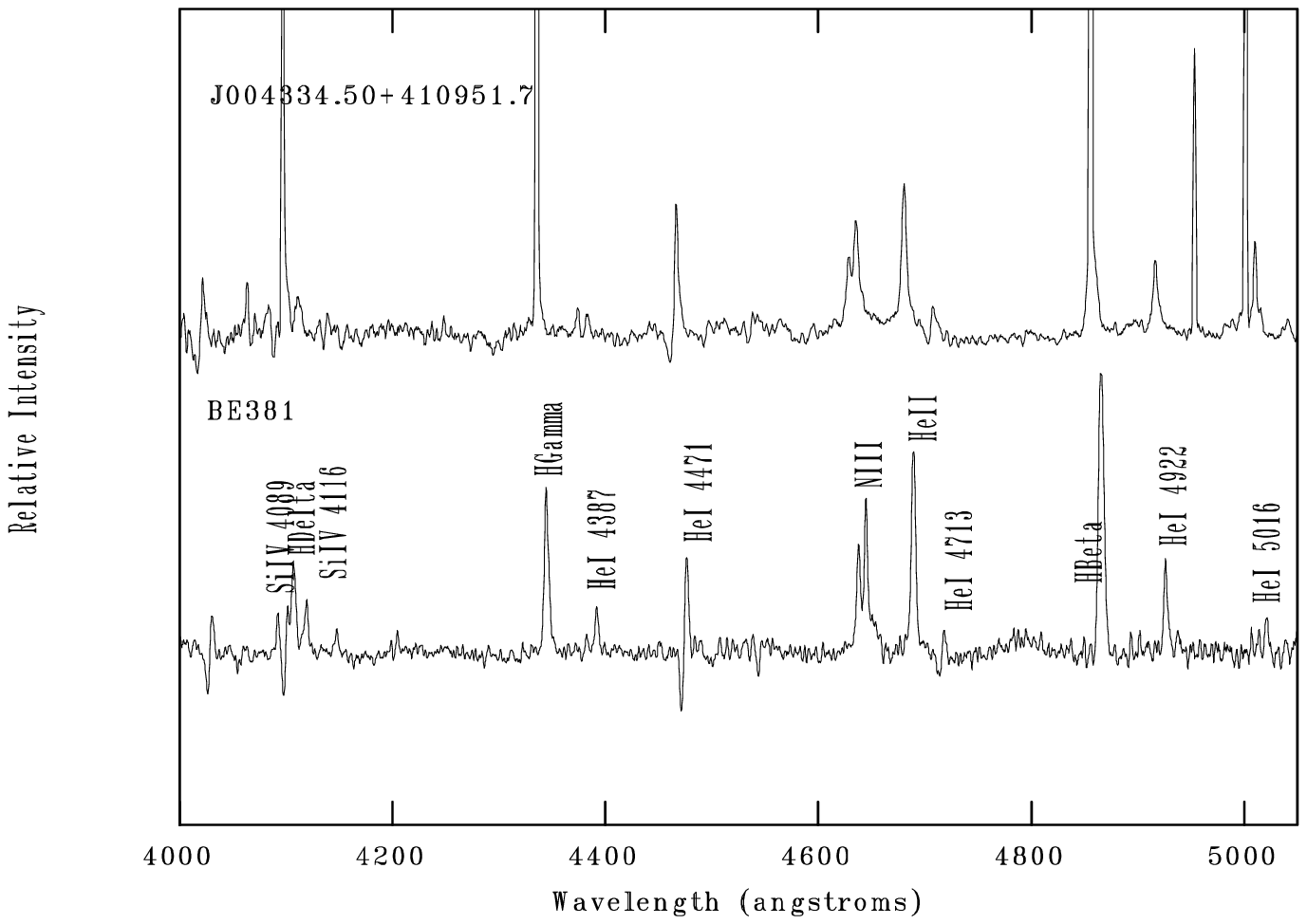}
\caption{\label{fig:ofpe} Spectrum of two newly discovered Ofpe/WN9 stars in M33
(top) and M33 (bottom).
For each star, we compare the spectra to the LMC star
BE 381, one of original Ofpe/WN9s 
(Bohannan \& Walborn 1989).  The star J013509.73+30417.3 is 
also known as Romano's Star, an LBV candidate recently found independently
to be in an Ofpe/WN9 state by Viotti et al.\ (2007).
}
\end{figure}

\clearpage
\begin{figure}
\epsscale{0.8}
\plotone{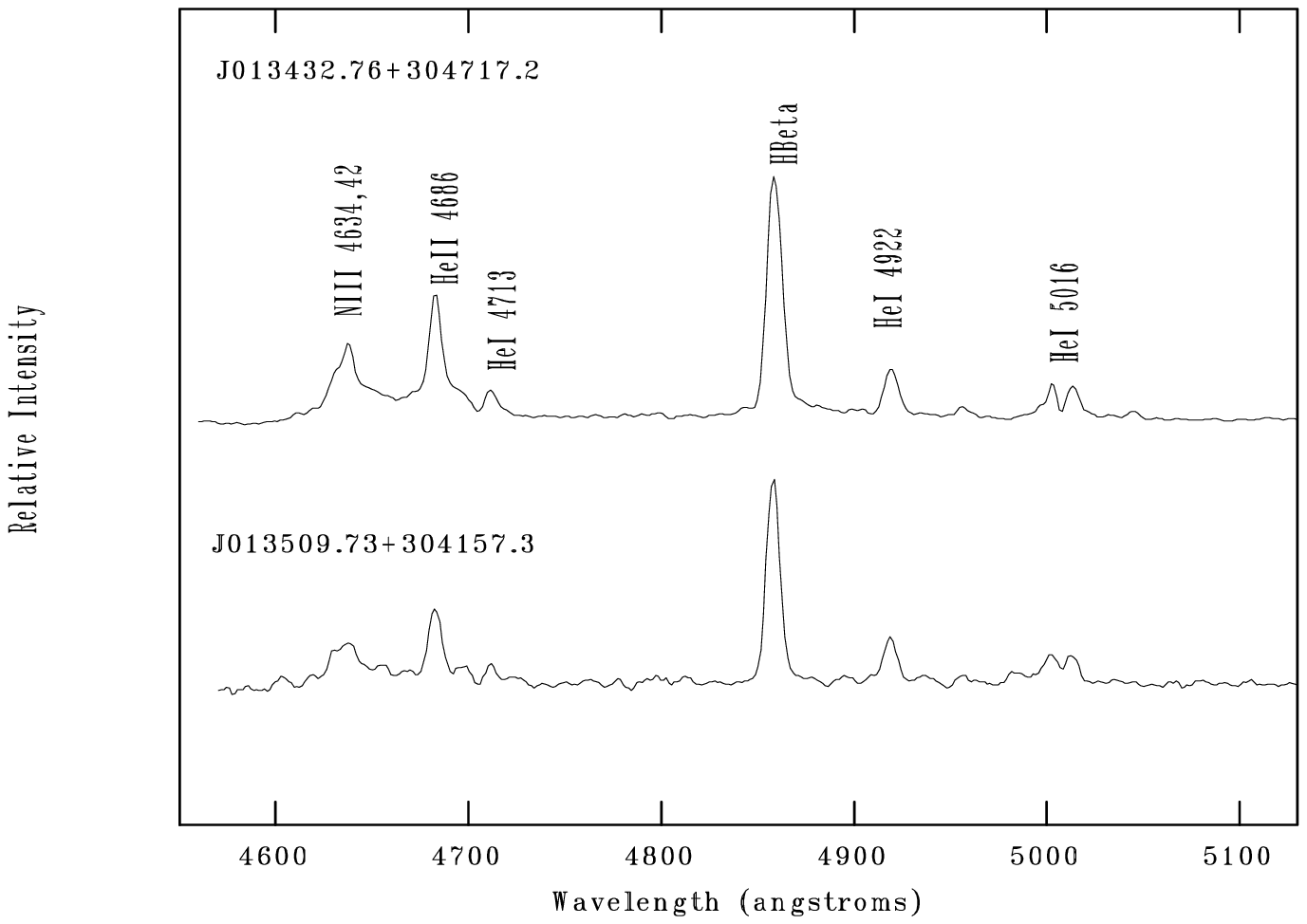}
\plotone{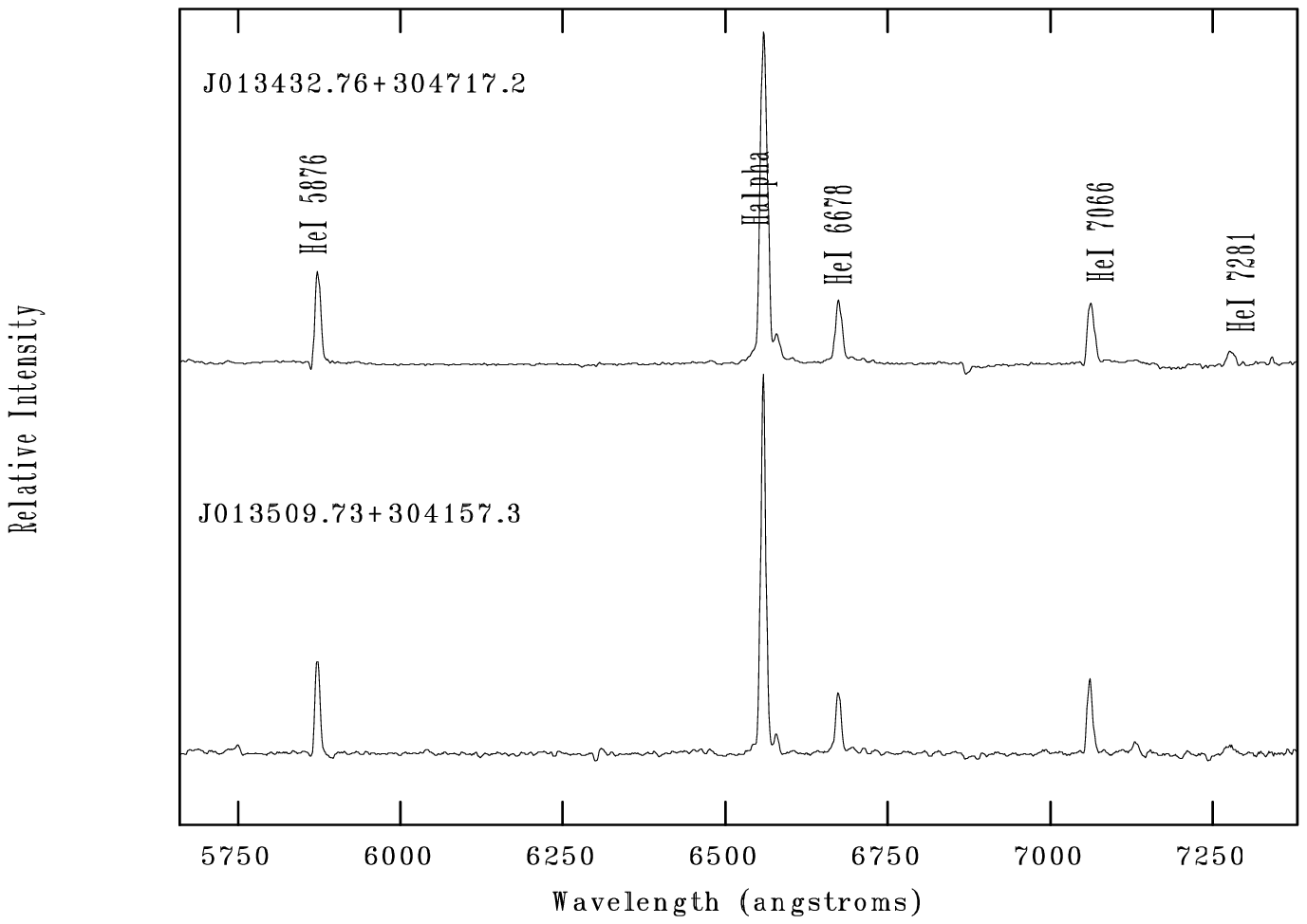}
\caption{\label{fig:ofpeRED} Red spectrum of an additional Ofpe/WN9 star in M33. 
Here we compare the spectrum of J013432.76+304717.2 to that of J013509.73+304157.3.  The blue spectrum of the latter is shown in Fig~\ref{fig:ofpe}.
}
\end{figure}

\clearpage
\begin{figure}
\epsscale{0.8}
\plotone{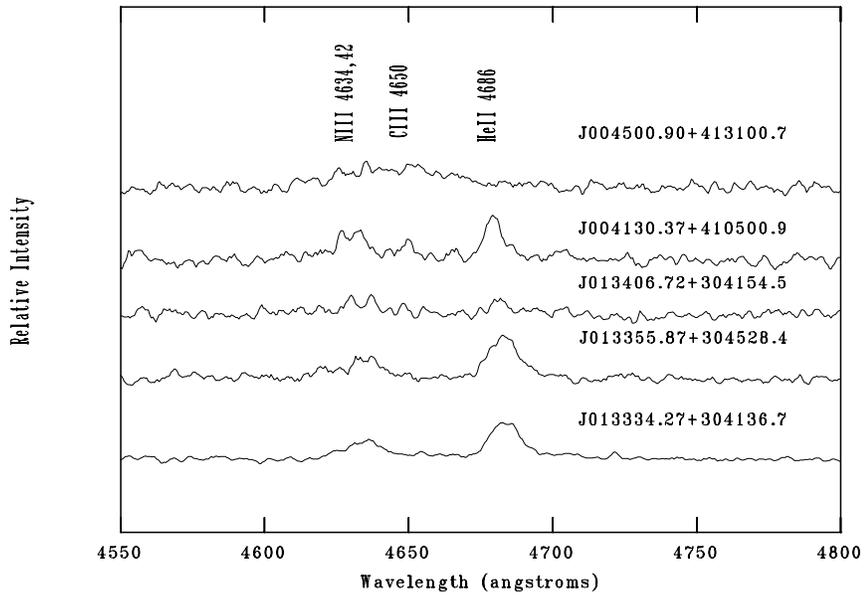}
\caption{\label{fig:wrguys} Spectra of other newly found WR stars in M31 and M33.  We classify all of
these as WNL except for J004500.90+413100.7, which is a WC.
}
\end{figure}

\clearpage
\begin{figure}
\epsscale{0.8}
\plotone{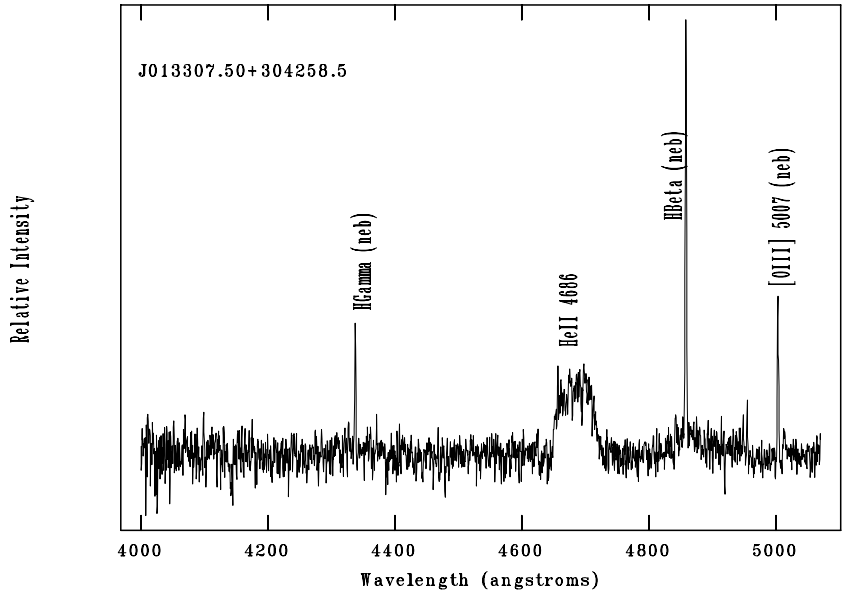}
\caption{\label{fig:UIT} Spectrum of the M33 star
J013307.50+304258.5.  This is a previously
recognized WR star, but one which was not properly classified until now. }
\end{figure}

\clearpage
\begin{figure}
\epsscale{0.6}
\plotone{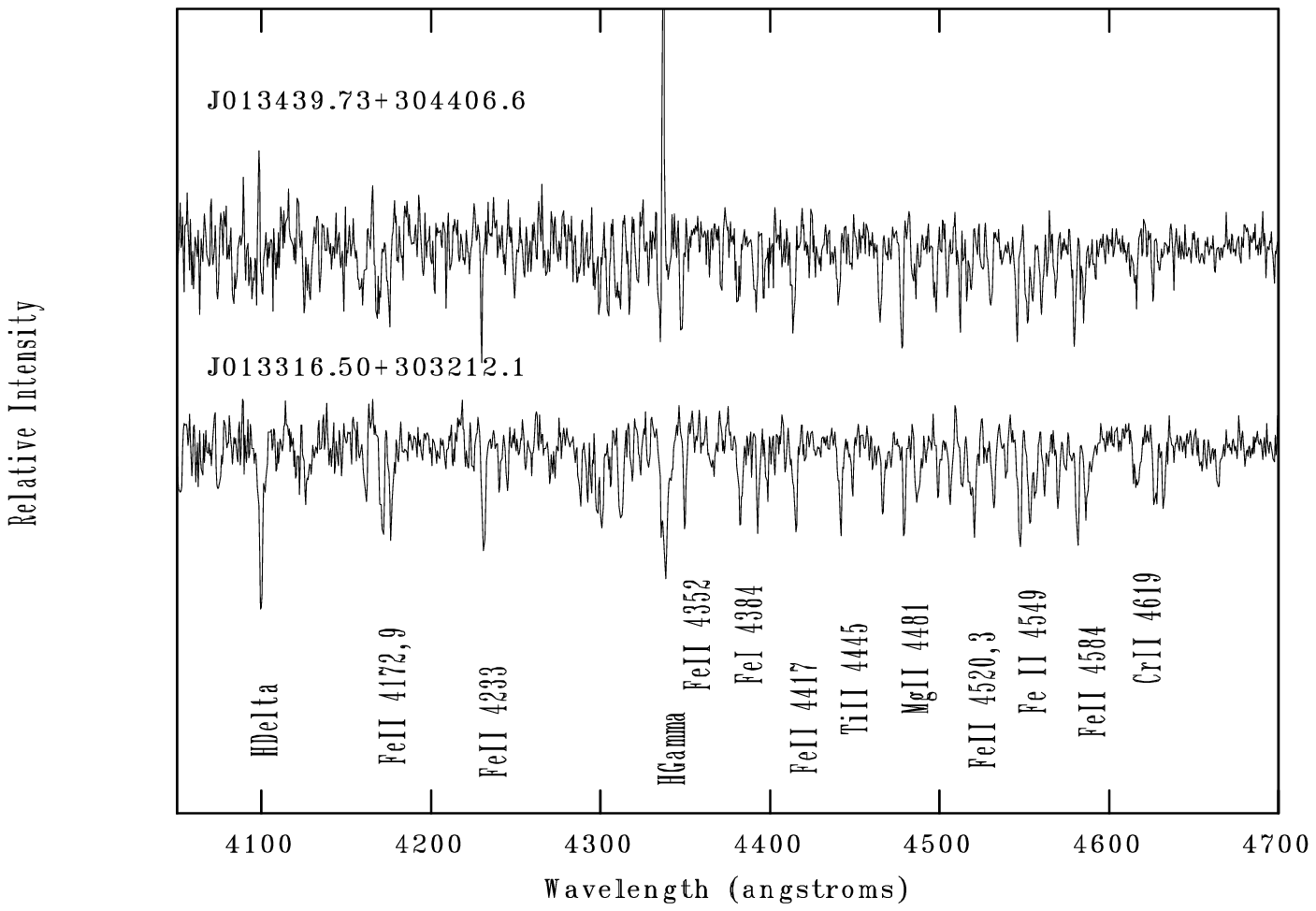}
\plotone{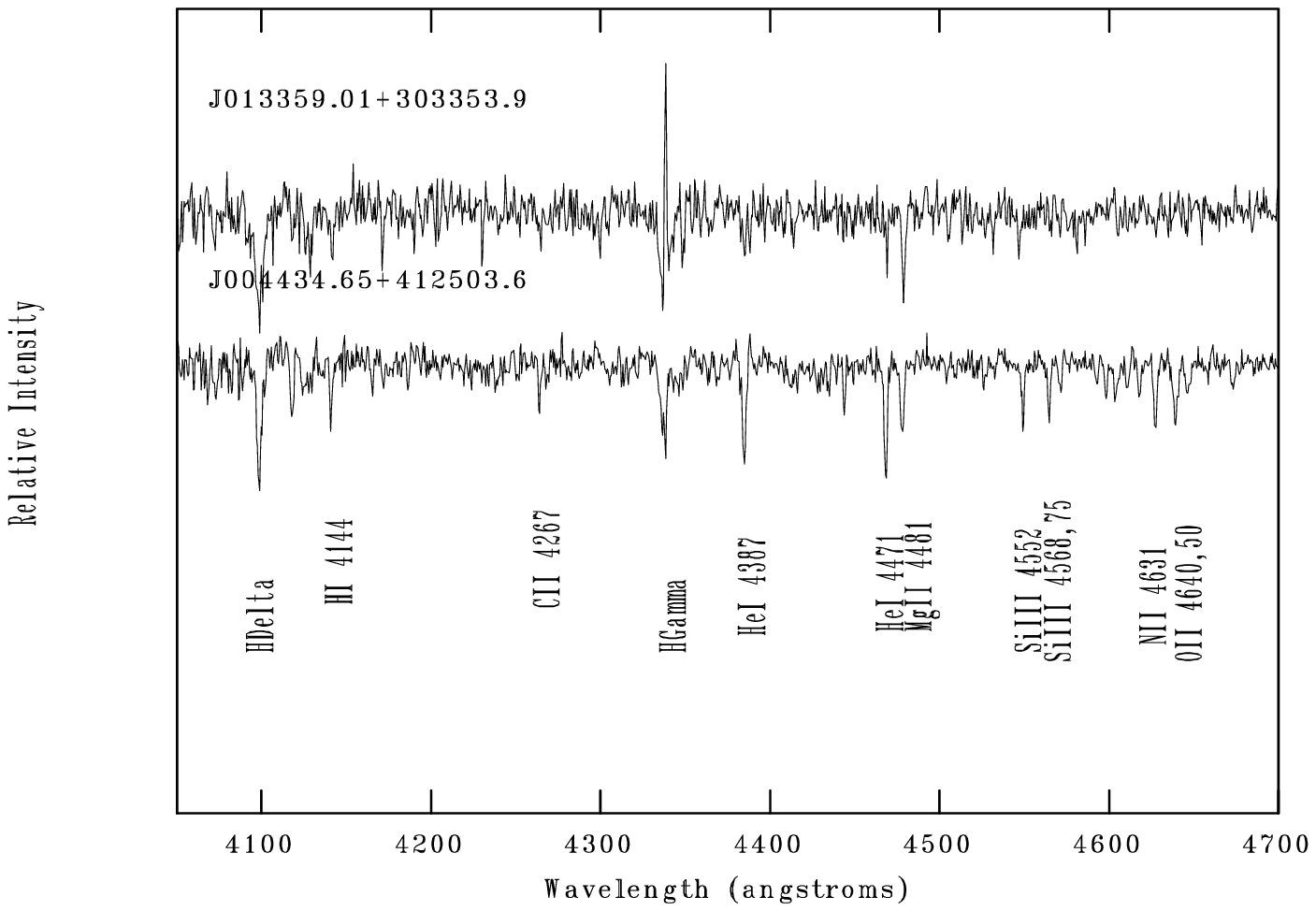}
\plotone{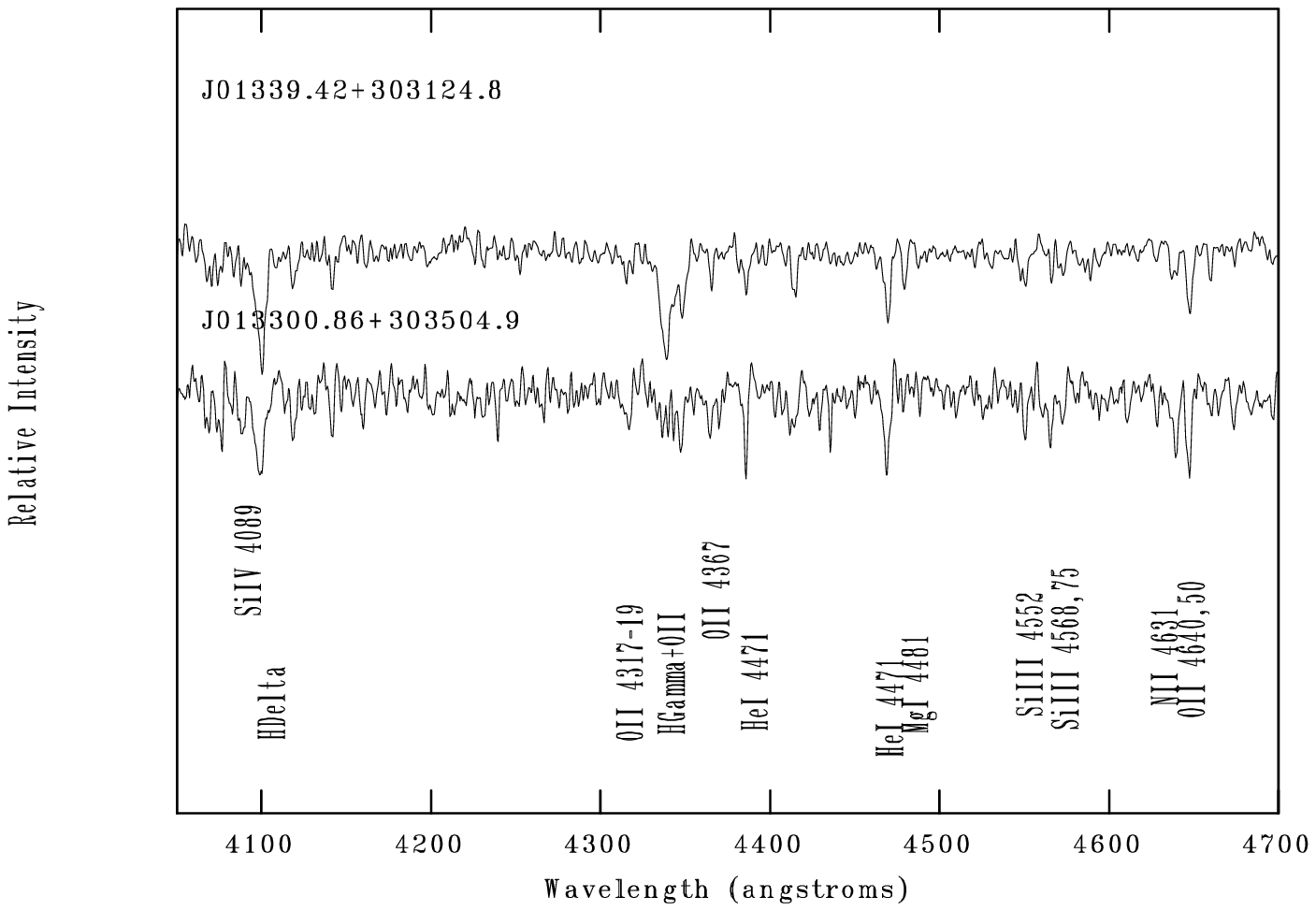}
\caption{\label{fig:nice} Spectra of  supergiants found in HII regions in M31 and M33.  In the upper
panel we show the spectra of two A/F supergiants. We estimate the spectral types
to be late A or early F.   In the middle panel we show
the spectra of two B-type supergiants.  The spectral types are B3~I (J004434.65+412503.6) and B8~I (J013359.01+303353.9). In the bottom panel we show
the spectra of two more B-type supergiants.  The spectral types are B3~I
(J013339.42+303124.8) and B1~I
(J013300.86+303504.8).}
\end{figure}

\clearpage
\begin{figure}
\epsscale{0.8}
\plotone{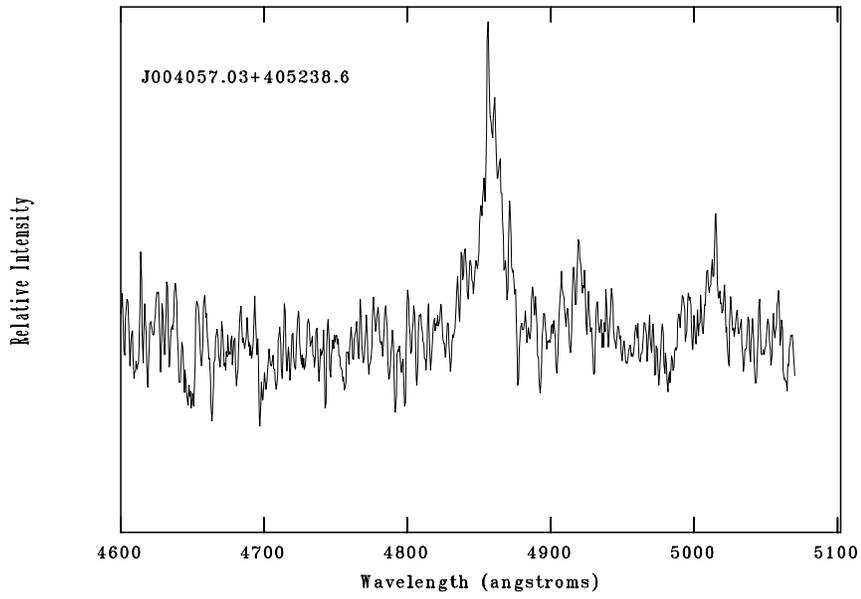}
\caption{\label{fig:bizzare} Spectra of an unusual emission-line source in M31.}
\end{figure}
\clearpage

\begin{figure}
\plotone{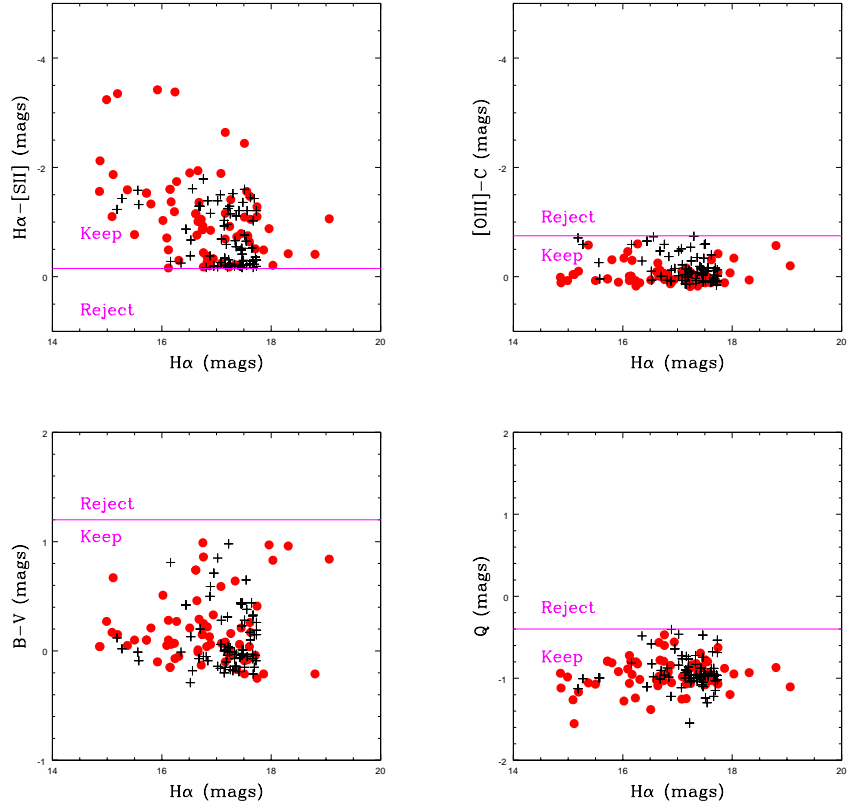}
\vskip -100pt
\caption{\label{fig:winlose} Selection criteria for H$\alpha$ emission-lined stars compared to ``winners" (red dots) and ``losers" (black +).  The ``winners" included the 
previously known and newly found LBVs, LBV candidates, and Ofpe/WN9 stars.
The ``losers" include HII regions and stars with no emission.}
\end{figure}

\begin{deluxetable}{l c c l l l l l l l l l}
\tabletypesize{\tiny}
\tablewidth{0pc}
\tablenum{1}
\tablecolumns{12}
\tablecaption{\label{tab:journal}Mosaic Observations}
\tablehead{
\colhead{Field}
&\colhead{$\alpha_{\rm 2000}$}
&\colhead{$\delta_{\rm 2000}$}
&\colhead{Obs.}
&\multicolumn{2}{c}{$H\alpha$\tablenotemark{a}}
& \colhead{}
& \multicolumn{2}{c}{[SII]\tablenotemark{a}}
& \colhead{}
& \multicolumn{2}{c}{[OIII]\tablenotemark{a}}
      \\ \cline{5-6} \cline{8-9} \cline{11-12} 
\colhead{}  & & & & \colhead{Date} & \colhead{DIQ(")} && \colhead{Date} & \colhead{DIQ(")} && \colhead{Date} & \colhead{DIQ(")} 
}
\startdata
M31-F1                    & 00 47 02.4 &+42 18 02 &KPNO& 2001 Sep 17 & 1.3 && 2001 Sep 17 & 1.1 && 2001 Sep 17 & 1.2 \\
M31-F2                    & 00 46 06.5 &+42 03 28 &KPNO& 2000 Oct 5 & 0.9 && 2000 Oct 5  & 1.0  && 2000 Oct 5 & 1.1  \\
M31-F3                    & 00 45 10.6 &+41 48 54 &KPNO& 2001 Sep 17 & 1.4 && 2001 Sep 17 & 1.5 && 2001 Sep 17 & 1.5 \\
M31-F4                    & 00 44 14.7 &+41 34 20 &KPNO& 2001 Sep 17 & 1.2 && 2001 Sep 17 & 1.2 && 2001 Sep 17 & 1.3 \\
M31-F5                    & 00 43 18.8 &+41 19 46 &KPNO& 2002 Sep 12 & 1.0 && 2002 Sep 12 & 1.3 && 2002 Sep 12 & 1.2 \\
M31-F6                    & 00 42 22.9 &+41 05 12 &KPNO& 2002 Sep 12 & 0.9 && 2002 Sep 12 & 0.9 && 2002 Sep 12 & 1.1 \\
M31-F7                    & 00 41 27.0 &+40 50 38 &KPNO& 2002 Sep 9  & 0.9 && 2002 Sep 9 & 0.8 && 2002 Sep 9 & 0.9 \\
M31-F8                    & 00 40 31.1 &+40 36 04 &KPNO& 2001 Sep 19 & 1.3 && 2001 Sep 19 & 1.1 && 2001 Sep 19 & 1.1 \\
M31-F9                    & 00 39 35.2 &+40 21 30 &KPNO& 2001 Sep 20 & 1.2 && 2001 Sep 20 & 1.0 && 2001 Sep 20 & 1.0 \\
M31-F10                   & 00 38 39.3 &+40 06 56 &KPNO& 2001 Sep 21 & 1.1 && 2001 Sep 21 & 1.3 && 2001 Sep 12 & 1.2 \\
M33-North                 & 01 34 00.1& +30 55 37 &KPNO& 2001 Sep 18 & 0.9 && 2001 Sep 18 & 0.9 && 2001 Sep 18 & 1.0 \\
M33-Center                & 01 33 50.9 &+30 39 37 &KPNO& 2000 Oct 5  & 1.0 && 2000 Oct 5 & 1.1 && 2000 Oct 5 & 1.1 \\
M33-South                 & 01 33 11.3 &+30 22 10 &KPNO& 2001 Sep 18 & 0.9 && 2001 Sep 20 & 1.1 && 2001 Sep 21 & 1.1 \\
IC10\tablenotemark{b}     & 00 20 24.5 &+59 17 30 &KPNO& 2001 Sep 22 & 0.9 &&2001 Sep 22 & 0.9 && 2001 Sep 22 &1.0 \\
NGC 6822                  & 19 44 56.1 &$-$14 48 05 &CTIO& 2000 Sep 2  & 0.9 &&2000 Sep 2 &  1.0 && 2000 Sep 2\tablenotemark{c} & 1.0   \\
WLM                       & 00 01 57.9 &$-$15 27 51 &CTIO& 2000 Sep 2  & 0.9 &&2000 Sep 2 &  0.9 && 2000 Sep 2 & 1.0 \\
Sextans B\tablenotemark{b}& 10 00 00.1 &+05 19 56 &KPNO& 2001 Feb 27& 1.2 &&2001Feb 27 & 1.5   && 2001 Feb 27 & 1.4  \\
Sextans A\tablenotemark{b}& 10 11 00.8 &$-$04 41 34 &KPNO& 2001 Feb 27& 1.3 &&2001Feb 27 & 1.1   && 2001 Feb 27 & 1.2   \\
Pegasus\tablenotemark{b}  & 23 28 36.2 &+14 44 35 &KPNO& 2000 Oct 5 & 1.4 &&2000 Oct 5 & 1.4   && 2000 Oct 5  & 1.5   \\  
Phoenix                   & 01 51 06.3 &$-$44 26 41 &CTIO&  2000 Sep 2 &0.8 &&2000 Sep 2 & 0.9   && 2000 Sep 2 & 1.0 \\
\enddata
\tablecomments{Units of right ascension are hours, minutes,
and seconds, and units of declination are degrees, arcminutes, and
arcseconds.}
\tablenotetext{a}{Series of 5 dithered 300s, unless otherwise noted.}
\tablenotetext{b}{Field was offset 540" south and 270" west in order to 
center on chip ``imt2" (Jacoby 2000).}
\tablenotetext{c}{Series of 5 dithered 360s exposures.}
\end{deluxetable}

\begin{deluxetable}{l r r r r r r l}
\tabletypesize{\scriptsize}
\tablewidth{0pc}
\tablenum{2}
\tablecolumns{8}
\tablecaption{\label{tab:calib} Adopted Calibration\tablenotemark{a} }
\tablehead{
&\multicolumn{3}{c}{Continuum Source}
&&\multicolumn{3}{c}{Emission-line Source}\\ \cline{2-4} \cline{6-8}
\colhead{Observatory}
&\colhead{$H\alpha$}
&\colhead{[SII]}
&\colhead{[OIII]}
&&\colhead{$H\alpha$}
&\colhead{[SII]}
&\colhead{[OIII]}
}
\startdata
KPNO &$2.14\times 10^{-18}$ &$2.21\times 10^{-18}$ &$6.41\times 10^{-18}$ &&$1.79\times 10^{-16}$ &$1.84\times 10^{-16}$ &$3.92\times 10^{-16}$\\
CTIO &$2.18\times 10^{-18}$ &$2.02\times 10^{-18}$ &$5.11\times 10^{-18}$ &&$1.80\times 10^{-16}$ &$1.71\times 10^{-16}$ &$2.75\times 10^{-16}$\\
\enddata
\tablenotetext{a}{These numbers are the equivalent of 1 count s$^{-1}$ in units of  
ergs s$^{-1}$ cm$^{-2}$ \AA$^{-1}$ for a point source measured with a 15-pixel radius aperture.}
\end{deluxetable}

\begin{deluxetable}{l r r r r r r r r r r r r r r r r r r r}
\rotate
\tabletypesize{\tiny}
\tablewidth{0pc}
\tablenum{3}
\tablecolumns{20}
\tablecaption{\label{tab:errors}Photometric Errors}
\tablehead{
\colhead{}
&\multicolumn{9}{c}{H$\alpha$, [SII]}
&\colhead{}
&\multicolumn{9}{c}{OIII} \\ \cline{2-10} \cline{12-20}
\colhead{Mag}
&\colhead{M31}
&\colhead{M33}
&\colhead{IC10}
&\colhead{N6822}
&\colhead{WLM}
&\colhead{Sex B}
&\colhead{Sex A}
&\colhead{Pegasus}
&\colhead{Phoenix}
&\colhead{}
&\colhead{M31}
&\colhead{M33}
&\colhead{IC10}
&\colhead{N6822}
&\colhead{WLM}
&\colhead{Sex B}
&\colhead{Sex A}
&\colhead{Pegasus}
&\colhead{Phoenix} 
}
\startdata
15  &              0.00   &   0.00  & 0.00   &   0.00      &    0.00    &    0.00  &   0.00  &  \nodata &   0.00  
&&      0.00 &  0.00&  0.00  &  0.00  &  0.00  &      0.00  &          0.00 &     \nodata & \nodata \\
16  &              0.00   &   0.00  & 0.00   &   0.00     &     0.00      &  0.00   &  0.00  &  0.00  &  0.00  
&&      0.00  & 0.00&  0.00 &   0.00 &   0.00   &     0.00  &          0.00   &   0.00  &      0.00\\
17  &              0.00   &    0.00 &  0.00  &    0.00     &     0.00     &   0.00  &   0.00  &  0.00 &   0.00  
&&     0.00 &  0.00&  0.00  &  0.00 &   0.00   &     0.00  &          0.00   &   0.00  &      0.00\\
18 &               0.00     &  0.01 &  0.00   &   0.01     &     0.00    &    0.00  &   0.00  &  0.01 &   0.00  
&&     0.00 &  0.01&  0.00  &  0.01&    0.01   &     0.00   &          0.00  &   0.01  &      0.01\\
19  &              0.01     &  0.01 &  0.01   &   0.01     &     0.01    &    0.01   &  0.01  &  0.01  &  0.01 
&&      0.01 &  0.01&  0.01  &  0.01 &  0.01    &     0.01   &         0.01   &   0.01 &       0.01\\
20   &             0.04     &  0.03 &  0.02   &   0.03     &     0.02    &    0.02 &    0.02 &   0.02  &  0.02 
&&      0.02 &  0.02&  0.02  &  0.02   & 0.02   &     0.02     &       0.02   &   0.03  &      0.02\\
21  &              0.08     &  0.06 &  0.05   &   0.06     &     0.05    &    0.04  &   0.04  &  0.06 &   0.03  
&&     0.05 &  0.05&  0.04  &  0.06  &  0.06   &     0.04    &        0.04    &  0.07  &      0.05\\
22 &               0.12     &  0.11 &  0.12    &  0.12     &     0.10     &   0.10  &   0.09  &  0.14 &   0.07  
&&    0.10 &  0.10 & 0.10  &  0.12  &  0.12    &    0.08   &         0.07    &  0.13  &      0.10  \\
\enddata
\tablecomments{Median error (mags) for stars within 0.5~mag of the stated value; i.e., 14.5-15.5
for the first row.}
\end{deluxetable}

\begin{deluxetable}{l r}
\tabletypesize{\scriptsize}
\tablewidth{0pc}
\tablenum{4}
\tablecolumns{2}
\tablecaption{\label{tab:cuts} Selection Criteria}
\tablehead{
\colhead{Index}
&\colhead{Used}
}
\startdata
H$\alpha$          & $\leq 20.0$\tablenotemark{a} \\
H$\alpha-$[SII]   & $\leq -0.15$ \\
\[[OIII]-C               & $\geq -0.75$ \\
$B-V$              & $\leq 1.2$\tablenotemark{b}\\
$Q$                   & $\leq 0.4$\tablenotemark{c} \\
\enddata
\tablenotetext{a}{For IC10, $\leq 20.2$.}
\tablenotetext{b}{For IC10, $\leq 1.5$.}
\tablenotetext{c}{For IC10, stars without $U-B$ were also retained.}
\end{deluxetable}

\begin{deluxetable}{l c c c c}
\tabletypesize{\scriptsize}
\tablewidth{0pc}
\tablenum{5}
\tablecolumns{5}
\tablecaption{\label{tab:galaxies}Adopted Distances and Reddenings}
\tablehead{
\colhead{Galaxy}
&\colhead{$(m-M)_0$\tablenotemark{a}}
&\colhead{$E(B-V)$\tablenotemark{b}}
&\colhead{$(m-M)_{H\alpha}$\tablenotemark{c}}
&\colhead{$M_{H\alpha}=-6$}
}
\startdata
M31 &24.4 & 0.13 & 24.7 & 18.7 \\
M33 &24.5 & 0.12 & 24.8 & 18.8 \\
IC10 &24.1 & 0.81 & 26.2 & 20.2 \\
NGC 6822 &23.5 &0.25 & 24.1 & 18.1\\
WLM & 24.8 &0.07 & 25.0  & 19.0 \\
Sextans B &25.6 &0.09 & 25.8  & 19.8 \\
Sextans A &25.8 &0.05 & 25.9  & 19.9 \\
Pegasus  &24.4 &0.15 & 24.8  & 18.8\\
Phoenix  &23.0 & 0.15 & 23.4  & 17.4  \\
\enddata
\tablenotetext{a}{True distance moduli are from van den Bergh 2000.}
\tablenotetext{b}{Typical color excess of a blue star, from Paper II.}
\tablenotetext{c}{Apparent distance moduli at H$\alpha$ follow from
$(m-M)_0 + 2.54 E(B-V).$}
\end{deluxetable}

\begin{deluxetable}{l r r}
\tabletypesize{\scriptsize}
\tablewidth{0pc}
\tablenum{6}
\tablecolumns{3}
\tablecaption{\label{tab:shifts} Photometric Shifts applied}
\tablehead{
\colhead{Galaxy}
&\colhead{$H\alpha-[SII]$}
&\colhead{\[[OIII]-C}
}
\startdata
M31       &-0.10 &-0.32\\
M33       &-0.06 &-0.29\\
IC10      &-0.17 &-0.10\\
NGC 6822  &-0.20 &+0.07\\
WLM       &-0.16 &+0.00\\
Sextans B &+0.34 &-0.51\\
Sextans A &-0.17 &-0.37\\
Pegasus   &-0.02 &-0.33\\
Phoenix   &-0.09 &-0.05\\     
\enddata
\end{deluxetable}

\begin{deluxetable}{l r r r r}
\tablewidth{0pc}
\tablenum{7}
\tablecaption{\label{tab:nums} Number of Potential H$\alpha$ Emission-Lined Objects with $M_{H\alpha}\le -6$}
\tablehead{
\colhead{Galaxy}
&\colhead{$M_V$\tablenotemark{a}}
&\colhead{$\log \dot{M}$\tablenotemark{b}}
&\colhead{Num.}
&\colhead{Num.\ with spectra}
}
\startdata 
M31 & -21.2& -1.3 & 498 & 59 \\
M33 & -18.9 & -1.0 & 1068 & 136\\
IC10 & -16.3 & -1.3 &  96 & 7\\
NGC 6822 & -16.0 & -2.0 & 9 & 2\\
WLM & -14.4 & -2.8 & 6 & 0\\
Sextans B & -14.3 & -3.0 & 1 & 0 \\
Sextans A & -14.2 & -2.2 & 9 & 0 \\
Pegasus & -12.3 & -4.4 & 0 & 0 \\
Phoenix & -9.8 & \nodata & 0 & 0\\
\enddata
\tablenotetext{a}{Absolute visual magnitude from van den Bergh 2000 and references
therein.}
\tablenotetext{b}{Star formation rate in terms of $M_\odot$ yr$^{-1}$ integrated over the entire
galaxy, from Table 1 of Paper II and references therein.}
\end{deluxetable}

\begin{deluxetable}{l r r r r r r r r r l l l}
\tabletypesize{\tiny}
\rotate
\tablewidth{0pc}
\tablenum{8}
\tablecolumns{13}
\tablecaption{\label{tab:m31}H$\alpha$ Emission-lined Stars in M31}
\tablehead{
\colhead{LGGS}
&\colhead{$\alpha_{\rm 2000}$}
&\colhead{$\delta_{\rm 2000}$}
&\colhead{$H\alpha$}
&\colhead{$H\alpha$-[SII]}
&\colhead{\[[OIII]$-$ C}
&\colhead{$V$}
&\colhead{$B-V$}
&\colhead{$U-B$}
&\colhead{$Q$}
&\colhead{Spectral Type}
&\colhead{Cross ID}
&\colhead{Ref.}
}
\startdata
J004140.32+411730.9&00 41 40.32&+41 17 30.9&  18.02&  -0.87&  -0.06&  19.10&   0.33&  -0.80&  -1.04& \nodata & \nodata & \nodata  \\
J004140.46+411714.1&00 41 40.46&+41 17 14.1&  19.88&  -0.23&   0.04&  19.92&   0.28&  -0.77&  -0.97& \nodata & \nodata  & \nodata \\
J004142.91+411822.1&00 41 42.91&+41 18 22.1&  19.90&  -0.43&  -0.20&  20.27&   0.13&  -0.75&  -0.84& \nodata & \nodata & \nodata  \\
J004142.96+412043.9&00 41 42.96&+41 20 43.9&  19.81&  -0.35&  -0.56&  20.20&  -0.03&  -0.97&  -0.95& \nodata & \nodata & \nodata  \\
J004143.00+412042.7&00 41 43.00&+41 20 42.7&  19.56&  -0.25&  -0.33&  19.70&  -0.04&  -0.99&  -0.96& \nodata & \nodata & \nodata  \\
J004143.13+411551.9&00 41 43.13&+41 15 51.9&  19.03&  -0.35&   0.10&  18.95&   0.01&  -0.97&  -0.98& \nodata & \nodata  & \nodata \\
J004143.44+411555.3&00 41 43.44&+41 15 55.3&  19.66&  -1.81&  -0.16&  22.50&   0.24&  -0.87&  -1.04& \nodata & \nodata & \nodata  \\
J004143.54+411820.1&00 41 43.54&+41 18 20.1&  19.53&  -0.35&  -0.36&  19.98&   0.08&  -0.89&  -0.95& \nodata & \nodata & \nodata  \\
J004143.56+411815.8&00 41 43.56&+41 18 15.8&  18.91&  -0.78&  -0.27&  20.16&   0.86&  -0.33&  -0.95& \nodata & \nodata & \nodata  \\
J004143.71+411826.3&00 41 43.71&+41 18 26.3&  20.00&  -1.07&   0.08&  21.20&   0.08&  -0.78&  -0.84& \nodata & \nodata & \nodata  \\
\enddata
\tablecomments{Units of right ascension are hours, minutes, and seconds, and units
of declination are degrees, arcminutes, and arcseconds. 
Table~\ref{tab:m31} is published in its entirety in the electronic edition of the 
{\it Astronomical Journal}.  A portion is shown here for guidance regarding form and
content.}
\tablerefs{For spectral types and cross-IDs:
(1) Paper I; (2) This paper; (3) Massey et al.\ 1995; (4) Bianchi et al.\ 1994; (5) Hubble \& Sandage 1953; (6) Massey \& Johnson 1998
and references therein.}

\end{deluxetable}

\begin{deluxetable}{l r r r r r r r r r l l l}
\tabletypesize{\tiny}
\rotate
\tablewidth{0pc}
\tablenum{9}
\tablecolumns{13}
\tablecaption{\label{tab:m33}H$\alpha$ Emission-lined Stars in M33}
\tablehead{
\colhead{LGGS}
&\colhead{$\alpha_{\rm 2000}$}
&\colhead{$\delta_{\rm 2000}$}
&\colhead{$H\alpha$}
&\colhead{$H\alpha$-[SII]}
&\colhead{\[[OIII]$-$ C}
&\colhead{$V$}
&\colhead{$B-V$}
&\colhead{$U-B$}
&\colhead{$Q$}
&\colhead{Spectral Type}
&\colhead{Cross ID}
&\colhead{Ref.}
}
\startdata
J013225.51+302652.3&01 32 25.51&+30 26 52.3&  18.94&  -1.10&  -0.66&  20.60&  -0.01&  -0.94&  -0.93& \nodata & \nodata & \nodata \\
J013226.94+302538.1&01 32 26.94&+30 25 38.1&  18.90&  -1.22&  -0.46&  20.68&   0.04&  -0.55&  -0.58& \nodata & \nodata & \nodata  \\
J013226.99+302413.2&01 32 26.99&+30 24 13.2&  18.56&  -1.21&  -0.15&  20.41&  -0.02&  -1.14&  -1.13& \nodata & \nodata & \nodata  \\
J013227.89+302542.8&01 32 27.89&+30 25 42.8&  20.00&  -1.05&  -0.05&  21.23&   0.02&  -0.90&  -0.91& \nodata & \nodata & \nodata  \\
J013228.78+303044.8&01 32 28.78&+30 30 44.8&  19.80&  -0.21&  -0.12&  19.73&  -0.16&  -1.09&  -0.97& \nodata & \nodata & \nodata  \\
J013229.01+303453.7&01 32 29.01&+30 34 53.7&  19.47&  -0.23&   0.02&  19.83&  -0.05&  -1.14&  -1.10& \nodata & \nodata  & \nodata \\
J013229.03+302819.6&01 32 29.03&+30 28 19.6&  18.48&  -0.76&   0.10&  19.00&   0.04&  -0.77&  -0.80& \nodata & \nodata & \nodata  \\
J013229.24+303445.3&01 32 29.24&+30 34 45.3&  19.18&  -0.41&  -0.14&  19.56&  -0.03&  -1.03&  -1.01& \nodata & \nodata & \nodata  \\
J013229.35+303445.4&01 32 29.35&+30 34 45.4&  19.90&  -0.29&  -0.44&  20.42&  -0.14&  -0.83&  -0.73& \nodata & \nodata & \nodata  \\
J013229.57+303412.8&01 32 29.57&+30 34 12.8&  18.81&  -0.16&  -0.08&  18.87&  -0.09&  -0.81&  -0.75& \nodata & \nodata & \nodata  \\
\enddata
\tablecomments{Units of right ascension are hours, minutes, and seconds, and units
of declination are degrees, arcminutes, and arcseconds. 
Table~\ref{tab:m33} is published in its entirety in the electronic edition of the 
{\it Astronomical Journal}.  A portion is shown here for guidance regarding form and
content.}
\tablerefs{For spectral types and cross-IDs:
(1) Paper I; (2) This paper; (3) Massey et al.\ 1996; (4) Massey et al.\ 1995; (5) Monteverde et al.\ 1996; (6) Hubble \& Sandage 1953; 
(7) Massey \& Johnson 1998 and references therein; (8) Corral 1996; (9) Viotti et al.\ 2007 and references therein.}
\end{deluxetable}

\begin{deluxetable}{l r r r r r r r r r l  l l}
\tabletypesize{\tiny}
\rotate
\tablewidth{0pc}
\tablenum{10}
\tablecolumns{12}
\tablecaption{\label{tab:ic10}H$\alpha$ Emission-lined Stars in IC 10}
\tablehead{
\colhead{LGGS}
&\colhead{$\alpha_{\rm 2000}$}
&\colhead{$\delta_{\rm 2000}$}
&\colhead{$H\alpha$}
&\colhead{$H\alpha$-[SII]}
&\colhead{\[[OIII]$-$ C}
&\colhead{$V$}
&\colhead{$B-V$}
&\colhead{$U-B$}
&\colhead{$Q$}
&\colhead{Spectral Type}
&\colhead{Cross ID}
&\colhead{Refs.}
}
\startdata
J002003.24+591343.7&00 20 03.24&+59 13 43.7&  19.84&  -0.33&   0.00&  20.69&   0.73&  -0.51&  -1.04&  \nodata & \nodata &\nodata                             \\
J002024.68+591648.3&00 20 24.68&+59 16 48.3&  20.00&  -0.16&  -0.04&  20.75&   1.11&   0.38&  -0.42&    \nodata & \nodata &\nodata                            \\
J002027.96+591659.4&00 20 27.96&+59 16 59.4&  19.61&  -0.80&  -0.74&  21.80&   1.32& 100.00&  99.99&   \nodata & \nodata &\nodata                             \\
J002030.95+591702.3&00 20 30.95&+59 17 02.3&  18.95&  -0.78&  -0.13&  20.21&   0.85&  -0.40&  -1.01&     \nodata & \nodata &\nodata                           \\
J002026.65+591714.4&00 20 26.65&+59 17 14.4&  19.93&  -0.35&  -0.01&  22.06&   0.99& 100.00&  99.99&  \nodata & \nodata &\nodata                              \\
J002028.07+591714.3&00 20 28.07&+59 17 14.3&  19.86&  -3.24&  -0.60&  21.54&   0.78&  -0.26&  -0.82& WN7-8 & RSMV2 & 1          \\
J002020.20+591724.1&00 20 20.20&+59 17 24.1&  19.99&  -0.35&  -0.50&  21.72&   0.94& 100.00&  99.99&    \nodata & \nodata &\nodata                            \\
J002030.85+591728.4&00 20 30.85&+59 17 28.4&  20.06&  -1.48&   0.58&  21.76&   0.88& 100.00&  99.99&  \nodata & \nodata &\nodata                              \\
J002011.89+591737.6&00 20 11.89&+59 17 37.6&  19.49&  -0.85&   0.05&  20.88&   1.04&  -0.26&  -1.01&    \nodata & \nodata &\nodata                            \\
J002024.66+591744.6&00 20 24.66&+59 17 44.6&  19.81&  -0.87&  -0.50&  22.06&   0.70&  -0.55&  -1.05&    \nodata & \nodata &\nodata                            \\
\enddata
\tablecomments{Units of right ascension are hours, minutes, and seconds, and units
of declination are degrees, arcminutes, and arcseconds. 
An entry of ``99.99" denotes no measurement.
Table~\ref{tab:ic10} is published in its entirety in the electronic edition of the 
{\it Astronomical Journal}.  A portion is shown here for guidance regarding form and
content.}
\tablerefs{For spectral types and cross-IDs:
(1) Crowther et al.\ 2003; (2) This paper; (3) Massey \& Armandroff 1995}
\end{deluxetable}

\begin{deluxetable}{l r r r r r r r r r l l l}
\tabletypesize{\tiny}
\rotate
\tablewidth{0pc}
\tablenum{11}
\tablecolumns{13}
\tablecaption{\label{tab:n6822}H$\alpha$ Emission-lined Stars in NGC 6822}
\tablehead{
\colhead{LGGS}
&\colhead{$\alpha_{\rm 2000}$}
&\colhead{$\delta_{\rm 2000}$}
&\colhead{$H\alpha$}
&\colhead{$H\alpha$-[SII]}
&\colhead{\[[OIII]$-$ C}
&\colhead{$V$}
&\colhead{$B-V$}
&\colhead{$U-B$}
&\colhead{$Q$}
&\colhead{Spectral Type}
&\colhead{Cross ID}
&\colhead{Ref.}
}
\startdata
J194434.02-144229.0&19 44 34.02&-14 42 29.0&  19.93&  -0.50&  -0.47&  20.04&   0.09&  -0.74&  -0.80& \nodata & \nodata & \nodata                        \\
J194434.10-144224.5&19 44 34.10&-14 42 24.5&  18.91&  -0.33&  -0.10&  19.37&   0.04&  -0.82&  -0.85& \nodata & \nodata & \nodata                        \\
J194434.17-144229.8&19 44 34.17&-14 42 29.8&  18.55&  -0.63&  -0.69&  19.13&   0.01&  -0.78&  -0.79& \nodata & \nodata & \nodata                        \\
J194434.24-144149.0&19 44 34.24&-14 41 49.0&  19.93&  -0.57&   0.17&  20.54&   0.20&  -0.72&  -0.86&  \nodata & \nodata & \nodata                       \\
J194434.39-144227.5&19 44 34.39&-14 42 27.5&  18.38&  -0.16&   0.02&  18.50&   0.18&  -0.85&  -0.98&  \nodata & \nodata & \nodata                       \\
J194435.78-144620.0&19 44 35.78&-14 46 20.0&  19.97&  -0.48&   0.14&  20.37&   0.18&  -0.68&  -0.81& \nodata & \nodata & \nodata                        \\
J194436.37-144820.8&19 44 36.37&-14 48 20.8&  19.87&  -0.44&   0.11&  20.38&   0.24&  -0.70&  -0.87&  \nodata & \nodata & \nodata                       \\
J194437.40-145044.0&19 44 37.40&-14 50 44.0&  19.70&  -0.31&  -0.21&  19.98&   0.16&  -0.43&  -0.55&   \nodata & \nodata & \nodata                      \\
J194437.97-145106.2&19 44 37.97&-14 51 06.2&  19.66&  -0.26&   0.08&  19.83&   0.01&  -0.76&  -0.77& WN&N6822-WR4 & 1                  \\
J194438.39-145147.0&19 44 38.39&-14 51 47.0&  19.70&  -0.68&  -0.31&  20.36&   0.04&  -0.75&  -0.78&   \nodata & \nodata & \nodata                      \\
\enddata
\tablecomments{Units of right ascension are hours, minutes, and seconds, and units
of declination are degrees, arcminutes, and arcseconds. 
Table~\ref{tab:n6822} is published in its entirety in the electronic edition of the 
{\it Astronomical Journal}.  A portion is shown here for guidance regarding form and
content.}
\tablerefs{For spectral types and cross-IDs:
(1) Massey \& Johnson 1998.  Note that for the WR stars, the CDS lists these
as ``[AM85] N".  (2) This paper.  (3) Westerlund et al.\ 1983.}
\end{deluxetable}

\begin{deluxetable}{l r r r r r r r r r l l l}
\tabletypesize{\tiny}
\rotate
\tablewidth{0pc}
\tablenum{12}
\tablecolumns{13}
\tablecaption{\label{tab:wlm}H$\alpha$ Emission-lined Stars in WLM}
\tablehead{
\colhead{LGGS}
&\colhead{$\alpha_{\rm 2000}$}
&\colhead{$\delta_{\rm 2000}$}
&\colhead{$H\alpha$}
&\colhead{$H\alpha$-[SII]}
&\colhead{\[[OIII]$-$ C}
&\colhead{$V$}
&\colhead{$B-V$}
&\colhead{$U-B$}
&\colhead{$Q$}
&\colhead{Spectral Type}
&\colhead{Cross ID}
&\colhead{Ref.}
}
\startdata
J000153.57-152732.6&00 01 53.57&-15 27 32.6&  19.81&  -0.54&   0.09&  20.29&  -0.24&  -0.92&  -0.75&  \nodata & \nodata & \nodata                       \\
J000154.96-152831.4&00 01 54.96&-15 28 31.4&  19.69&  -1.64&   0.07&  22.24&  -0.43&  -1.27&  -0.96&   \nodata & \nodata & \nodata                      \\
J000155.24-152718.7&00 01 55.24&-15 27 18.7&  19.76&  -0.17&   0.15&  19.78&  -0.04&  -1.03&  -1.00&  \nodata & \nodata & \nodata                       \\
J000155.90-152839.3&00 01 55.90&-15 28 39.3&  19.21&  -0.17&  -0.40&  20.09&  -0.23&  -1.04&  -0.87&   \nodata & \nodata & \nodata                      \\
J000156.08-152837.5&00 01 56.08&-15 28 37.5&  18.48&  -1.37&  -0.67&  20.77&  -0.10&  -0.95&  -0.88&   \nodata & \nodata & \nodata                      \\
J000156.16-152841.9&00 01 56.16&-15 28 41.9&  18.86&  -0.17&  -0.13&  18.92&  -0.07&  -0.80&  -0.75&   \nodata & \nodata & \nodata                      \\
J000156.16-152841.9&00 01 56.16&-15 28 41.9&  18.86&  -0.17&  -0.13&  18.92&  -0.07&  -0.80&  -0.75&   \nodata & \nodata & \nodata                      \\
J000156.53-152703.2&00 01 56.53&-15 27 03.2&  18.44&  -0.18&  -0.01&  18.50&  -0.06&  -0.94&  -0.90&   \nodata & \nodata & \nodata                      \\
J000156.75-152636.6&00 01 56.75&-15 26 36.6&  19.89&  -0.45&   0.10&  20.27&   0.11&  -0.35&  -0.43& A3II  & B8 & 1              \\
J000157.14-152700.9&00 01 57.14&-15 27 00.9&  19.92&  -0.70&  -0.55&  21.52&   0.61&  -0.78&  -1.22&    \nodata & \nodata & \nodata                     \\
J000157.20-152648.2&00 01 57.20&-15 26 48.2&  18.82&  -1.44&  -0.69&  21.25&  -0.15&  -1.09&  -0.98&    \nodata & \nodata & \nodata                     \\
J000159.58-152728.9&00 01 59.58&-15 27 28.9&  19.25&  -1.24&  -0.13&  21.19&  -0.17&  -1.00&  -0.88&   \nodata & \nodata & \nodata                      \\
J000159.62-153016.0&00 01 59.62&-15 30 16.0&  19.71&  -1.36&   0.22&  21.43&  -0.15&  -1.00&  -0.89&    \nodata & \nodata & \nodata                     \\
J000200.42-152935.7&00 02 00.42&-15 29 35.7&  19.69&  -0.78&  -0.46&  21.07&  -0.18&  -0.81&  -0.68&   \nodata & \nodata & \nodata                      \\
J000201.95-152744.8&00 02 01.95&-15 27 44.8&  19.46&  -0.28&   0.13&  19.47&  -0.17&  -0.89&  -0.77&   \nodata & \nodata & \nodata                      \\
J000202.33-152743.2&00 02 02.33&-15 27 43.2&  17.74&  -1.47&  -0.30&  19.46&   0.09&  -0.66&  -0.72&    \nodata & \nodata & \nodata                     \\
\enddata
\tablecomments{Units of right ascension are hours, minutes, and seconds, and units
of declination are degrees, arcminutes, and arcseconds.}
\tablerefs{For spectral types and Cross-IDs:
(1) Bresolin et al.\ 2006.}
\end{deluxetable}

\begin{deluxetable}{l r r r r r r r r r l l l}
\tabletypesize{\tiny}
\rotate
\tablewidth{0pc}
\tablenum{13}
\tablecolumns{10}
\tablecaption{\label{tab:sexB}H$\alpha$ Emission-lined Stars in Sextans B}
\tablehead{
\colhead{LGGS}
&\colhead{$\alpha_{\rm 2000}$}
&\colhead{$\delta_{\rm 2000}$}
&\colhead{$H\alpha$}
&\colhead{$H\alpha$-[SII]}
&\colhead{\[[OIII]$-$ C}
&\colhead{$V$}
&\colhead{$B-V$}
&\colhead{$U-B$}
&\colhead{$Q$}
}
\startdata
J100002.93+052022.8&10 00 02.93&+05 20 22.8&  19.12&  -1.31&  -0.47&  19.86&  -0.29&  -1.05&  -0.84\\
J100005.54+051802.5&10 00 05.54&+05 18 02.5&  19.86&  -1.35&  -0.45&  21.68&  -0.28&  -1.08&  -0.88\\
\enddata
\tablecomments{Units of right ascension are hours, minutes, and seconds, and units
of declination are degrees, arcminutes, and arcseconds. }
\end{deluxetable}
\begin{deluxetable}{l r r r r r r r r r l l l}
\tabletypesize{\tiny}
\rotate
\tablewidth{0pc}
\tablenum{14}
\tablecolumns{10}
\tablecaption{\label{tab:sexA}H$\alpha$ Emission-lined Stars in Sextans A}
\tablehead{
\colhead{LGGS}
&\colhead{$\alpha_{\rm 2000}$}
&\colhead{$\delta_{\rm 2000}$}
&\colhead{$H\alpha$}
&\colhead{$H\alpha$-[SII]}
&\colhead{\[[OIII]$-$ C}
&\colhead{$V$}
&\colhead{$B-V$}
&\colhead{$U-B$}
&\colhead{$Q$}
}
\startdata
J101053.60-044117.8&10 10 53.60&-04 41 17.8&  19.60&  -1.23&  -0.52&  20.97&  -0.05&  -0.83&  -0.79  \\
J101053.67-044118.4&10 10 53.67&-04 41 18.4&  19.76&  -1.10&  -0.57&  21.07&  -0.17&  -1.13&  -1.01   \\
J101053.90-044111.0&10 10 53.90&-04 41 11.0&  19.99&  -0.42&  -0.58&  20.32&  -0.10&  -0.99&  -0.92   \\
J101053.94-044110.1&10 10 53.94&-04 41 10.1&  19.98&  -0.38&  -0.68&  20.53&  -0.23&  -1.09&  -0.92 \\
J101054.08-044111.5&10 10 54.08&-04 41 11.5&  19.76&  -0.23&   0.04&  19.49&  -0.22&  -1.03&  -0.87   \\
J101100.56-043930.9&10 11 00.56&-04 39 30.9&  19.93&  -0.22&   0.06&  19.87&   0.23&  -0.26&  -0.43   \\
J101105.07-044214.6&10 11 05.07&-04 42 14.6&  20.00&  -0.18&   0.11&  19.74&  -0.24&  -1.10&  -0.93   \\
J101105.17-044236.0&10 11 05.17&-04 42 36.0&  19.30&  -0.46&  -0.33&  19.36&  -0.04&  -0.46&  -0.43  \\
J101105.30-044210.1&10 11 05.30&-04 42 10.1&  19.89&  -0.56&  -0.02&  19.98&  -0.27&  -1.12&  -0.93   \\
J101105.38-044240.1&10 11 05.38&-04 42 40.1&  19.26&  -0.46&  -0.57&  19.46&  -0.26&  -1.14&  -0.95  \\
J101105.69-044213.6&10 11 05.69&-04 42 13.6&  19.75&  -0.49&   0.11&  19.70&  -0.24&  -1.07&  -0.90   \\
J101106.56-044217.1&10 11 06.56&-04 42 17.1&  19.90&  -1.42&  -0.63&  21.29&  -0.31&  -1.12&  -0.90\\
J101107.34-044231.7&10 11 07.34&-04 42 31.7&  19.43&  -1.46&  -0.26&  20.57&  -0.23&  -1.20&  -1.03   \\
J101109.27-044053.3&10 11 09.27&-04 40 53.3&  19.96&  -1.71&  -0.36&  21.80&  -0.23&  -1.31&  -1.14  \\

\enddata
\tablecomments{Units of right ascension are hours, minutes, and seconds, and units
of declination are degrees, arcminutes, and arcseconds. }
\end{deluxetable}
\begin{deluxetable}{l r r r r r r r r r l l l}
\tabletypesize{\tiny}
\rotate
\tablewidth{0pc}
\tablenum{15}
\tablecolumns{10}
\tablecaption{\label{tab:peg}H$\alpha$ Emission-lined Stars in Pegasus}
\tablehead{
\colhead{LGGS}
&\colhead{$\alpha_{\rm 2000}$}
&\colhead{$\delta_{\rm 2000}$}
&\colhead{$H\alpha$}
&\colhead{$H\alpha$-[SII]}
&\colhead{\[[OIII]$-$ C}
&\colhead{$V$}
&\colhead{$B-V$}
&\colhead{$U-B$}
&\colhead{$Q$}
}
\startdata
J232834.97+144356.9&23 28 34.97&+14 43 56.9&  19.63&  -1.55&   0.55&  22.44&  -0.32&  -0.80&  -0.57  \\
\enddata
\tablecomments{Units of right ascension are hours, minutes, and seconds, and units
of declination are degrees, arcminutes, and arcseconds. }
\end{deluxetable}

\begin{deluxetable}{l c c c c}
\tablewidth{0pc}
\tablenum{15}
\tablecolumns{5}
\tablecaption{\label{tab:spectexp}Spectroscopic Exposures}
\tablehead{
\colhead{Field}
&\colhead{$\alpha_{\rm 2000}$}
&\colhead{$\delta_{\rm 2000}$}
&\multicolumn{2}{c}{Exposures} \\ \cline{4-5}
\multicolumn{3}{c}{}
&\colhead{Blue}
&\colhead{Red} 
}
\startdata
M31-NE & 00 43.8 &  +41 33 & 6x1800s & 3x1800s \\
M31-SW & 00 41.2 & +40 44 & 4x1800s,1000s & \nodata \\
M33-N    & 01 33.7 & +30 44 &  3x1800s,3x1600s & 4x1800s \\
M33-S    & 01 33.7 & +30 34 &  5x1800, 1600s & 4x1800s \\
IC10       & 00 20.2 & +59 18 &  \nodata & 3x1200s \\
NGC6822 & 19 44.7 & $-$14 53 & 4x1800s, 1200s & 3x1600s \\
\enddata
\end{deluxetable}
\begin{deluxetable}{l l l}
\tabletypesize{\tiny}
\tablewidth{0pc}
\tablenum{16}
\tablecolumns{3}
\tablecaption{\label{tab:newsp}New Spectroscopic Identifications}
\tablehead{
\colhead{LGGS}
&\colhead{Spectra}
&\colhead{Figure} 
}
\startdata
\cutinhead{IC10}
J002012.13+591848.0&hot LBV candidate           &\ref{fig:ic10} \\
J002016.48+591906.9&hot LBV candidate?          &\ref{fig:n6822}\\
J002020.35+591837.6&hot LBV candidate           &\ref{fig:ic10}\\
\cutinhead{M31}
J003910.85+403622.4&HII?                        &\nodata\\
J003944.71+402056.2&HII&\nodata\\
J004030.28+404233.1&HII/B1.5\tablenotemark{a} &\nodata\\
J004032.37+403859.8&HII/B\tablenotemark{a} &\nodata\\
J004033.80+405717.2&HII/B0.2\tablenotemark{a}&\nodata\\
J004043.10+410846.0&hot LBV candidate           &\ref{fig:lbvhot}\\
J004052.19+403116.6&no emission--star/B8\tablenotemark{a}&\nodata\\
J004057.03+405238.6&broad-lined peculiar        &\ref{fig:bizzare}\\
J004058.04+410327.9&HII/B8\tablenotemark{a} &\nodata\\
J004109.26+404906.0&HII&\nodata\\
J004129.74+405100.8&HII&\nodata\\
J004130.37+410500.9&WNL                         &\ref{fig:wrguys}\\
J004220.31+405123.2&HII&\nodata\\
J004229.87+410551.8&hot LBV candidate           &\ref{fig:nelson}\\
J004242.33+413922.7&P Cyg LBV candidate         &\ref{fig:pcygs}\\
J004253.42+412700.5&star in HII region&\nodata\\
J004259.31+410629.1&HII&\nodata \\
J004303.21+410433.8&HII&\nodata\\
J004313.27+410257.4&no emission---star\\
J004322.50+413940.9&hot LBV candidate        &\ref{fig:nelson}\\
J004334.50+410951.7&Ofpe/WN9                    &\ref{fig:ofpe}\\
J004339.28+411019.4&HII? &\nodata\\
J004350.50+414611.4&cool LBV candidate          &\ref{fig:coolguys}\\
J004410.90+413203.2&star in HII region &\nodata\\
J004411.36+413257.2&hot LBV candidate (k315a)   &\ref{fig:hotmaybe} \\
J004415.00+420156.2&hot LBV candidate           &\ref{fig:lbvhot}\\
J004416.28+412106.6&HII&\nodata\\
J004417.10+411928.0&hot LBV candidate (k350)    &\ref{fig:lbvhot}\\
J004425.18+413452.2&cool LBV candidate (k411)   &\ref{fig:coolguys}\\
J004433.58+415248.0&HII&\nodata \\
J004434.65+412503.6&B3 I/B1:\tablenotemark{b} in HII                 &\ref{fig:nice}\\
J004438.55+412511.1&HII &\nodata\\
J004442.07+412732.3&HII&\nodata\\
J004442.28+415823.1&hot LBV candidate           &\ref{fig:lbvhot}\\
J004443.57+412616.5&star in HII&\nodata \\
J004444.52+412804.0&P Cyg LBV candidate         &\ref{fig:pcygs} \\
J004500.90+413100.7&WC in HII                   &\ref{fig:wrguys}\\
J004507.65+413740.8&cool LBV candidate          &\ref{fig:coolguys}\\
J004511.60+413716.8&HII&\nodata \\
J004522.58+415034.8&hot LBV candidate           &\ref{fig:lbvhot}\\
J004526.62+415006.3&hot LBV candidate           &\ref{fig:lbvhot}\\
J004545.94+415030.5&AVe (foreground)&\nodata  \\
\cutinhead{M33}
J013235.25+303017.6&hot LBV candidate           &\ref{fig:hotmaybe}\\
J013241.30+302231.2&HII&\nodata\\
J013242.26+302114.1&hot LBV candidate           &\ref{fig:hotmaybe}\\
J013245.00+303456.7&HII&\nodata \\
J013248.26+303950.4&hot LBV candidate           &\ref{fig:hotmaybe}\\
J013259.74+303854.8&HII?&\nodata\\
J013300.86+303504.9&B1~I in HII/B1.5Ia\tablenotemark{b }                 &\ref{fig:nice}\\
J013301.24+303051.3&HII&\nodata\\
J013303.09+303101.8&HII&\nodata\\
J013307.50+304258.5&WN (UIT041=M33-WR19)        &\ref{fig:UIT}\\
J013311.26+304515.3&HII&\nodata\\
J013311.45+302951.3&HII&\nodata \\
J013315.21+305318.5&HII&\nodata\\
J013316.50+303212.1&late A~I/early F~I          &\ref{fig:nice}\\
J013317.22+303201.6&HII&\nodata\\
J013324.62+302328.4&hot LBV candidate           &\ref{fig:lbvhot}\\ 
J013327.03+303841.6&HII&\nodata\\
J013329.88+303147.3&HII/O+neb\tablenotemark{a}&\nodata \\
J013332.64+304127.2&hot LBV candidate\tablenotemark{c}           &\ref{fig:am2}\\
J013333.22+303343.4&hot LBV candidate          &\ref{fig:lbvhot}\\
J013334.06+304744.3&HII plus HeII $\lambda$ 4686?&\nodata\\
J013334.27+304136.7&WNL in HII                  &\ref{fig:wrguys}\\
J013334.29+303400.1&HII&\nodata\\
J013334.39+303208.4&HII&\nodata\\
J013335.32+303931.0&HII&\nodata\\
J013337.56+303202.3&HII&\nodata\\
J013339.08+302010.7&star in HII&\nodata\\
J013339.42+303124.8&B3 I/B1~Ia\tablenotemark{b}                 &\ref{fig:nice}\\
J013339.42+303810.8&HII&\nodata \\
J013339.52+304540.5&P Cyg LBV candidate\tablenotemark{d}          &\ref{fig:pcygs}\\
J013341.28+302237.2&P Cyg LBV candidate\tablenotemark{e} (101-A)        &\ref{fig:pcygs}\\
J013342.03+304733.6&star in HII&\nodata\\
J013342.52+303258.6&HII&\nodata\\
J013343.19+303906.4&HII&\nodata\\
J013343.50+303911.5&HII&\nodata\\
J013344.52+304432.3&HII/OB+neb\tablenotemark{a}&\nodata \\
J013344.56+303201.3&HII&\nodata\\
J013344.79+304432.4&HII/OB+neb\tablenotemark{a}&\nodata\\
J013344.85+303600.4&HII&\nodata \\
J013345.25+303626.6&HII/B\tablenotemark{a}&\nodata\\
J013347.33+303306.8&HII&\nodata\\
J013349.28+305250.2&HII&\nodata\\
J013349.72+303730.6&HII&\nodata\\
J013349.94+302928.8&HII&\nodata\\
J013350.21+303347.6&HII&\nodata\\
J013351.46+304057.0&P Cyg LBV candidate         &\ref{fig:pcygs}\\
J013352.19+303636.6&HII&\nodata \\
J013352.39+303920.9&HII/OB+neb&\nodata\\
J013355.51+304526.8&HII&\nodata\\
J013355.87+304528.4&WNL in HII\\
J013357.73+301714.2&cool LBV candidate          &\ref{fig:coolguys}\\
J013359.01+303353.9&B8I in HII                  &\ref{fig:nice}\\
J013359.11+303437.2&star in HII&\nodata\\
J013359.40+302311.0&HII/A0Ia\tablenotemark{a}&\nodata\\
J013401.44+303630.8&HII&\nodata\\
J013401.68+303720.0&HII&\nodata\\
J013406.72+304154.5&Of/early O\tablenotemark{b}                         &\ref{fig:wrguys}\\
J013407.32+304732.4&HII&\nodata\\
J013408.21+303405.2&HII&\nodata\\
J013410.93+303437.6&hot LBV candidate           &\ref{fig:lbvhot}\\
J013414.21+303343.3&HII&\nodata\\
J013415.43+303707.4&HII&\nodata\\
J013416.07+303642.1&P Cyg LBV candidate (H108)        &\ref{fig:pcygs}\\
J013416.35+303712.3&HII/WN7\tablenotemark{a}&\nodata\\
J013416.44+303120.8&cool LBV candidate          &\ref{fig:weird}\\
J013422.91+304411.0&cool LBV candidate          &\ref{fig:coolguys}\\
J013424.78+303306.6&cool LBV candidate          &\ref{fig:coolguys}\\
J013426.11+303424.7&hot LBV candidate           &\ref{fig:hotmaybe}\\
J013429.64+303732.1&cool LBV candidate          &\ref{fig:weird}\\
J013430.29+304039.8&star in HII&\nodata\\
J013432.76+304717.2&Ofpe/WN9                    &\ref{fig:ofpeRED}\\
J013433.10+304659.0&HII&\nodata\\
J013435.15+304705.1&HII&\nodata\\
J013438.76+304358.8&HII/Late O\tablenotemark{a}&\nodata\\
J013439.73+304406.6&late A~I/early F~I          &\ref{fig:nice}\\
J013442.14+303216.0&hot LBV candidate?          &\ref{fig:n6822}\\
J013459.47+303701.9&hot LBV candidate           &\ref{fig:lbvhot}\\
J013500.30+304150.9&hot LBV candidate           &\ref{fig:lbvhot}\\
J013509.73+304157.3&Ofpe/WN9  (Romano's Star)                  &\ref{fig:ofpe},\ref{fig:ofpeRED}\\
\cutinhead{NGC 6822}
J194452.97-144305.1&HII&\nodata\\
J194503.77-145619.1&hot LBV candidate?          &\ref{fig:n6822}\\
\enddata
\tablenotetext{a}{In a few cases where our fiber spectroscopy revealed only the
spectra of an HII region, previous long-slit spectra had permitted a spectral type
to be determined of the underlying star.  We include these here, and retain the
original in Tables~\ref{tab:m31} through \ref{tab:n6822}.}
\tablenotetext{b}{Previous spectral type is shown, but we adopt the new one (given first) here.}
\tablenotetext{c}{Previously called ``WN" by Massey \& Conti 1983; see text.}
\tablenotetext{d}{Previously called ``B0.5Ia+WNE by Massey et al.\  1996; see text.}
\tablenotetext{e}{Previously called B1~Ia (101-A) by Monteverde et al.\ 1996; see text.}
\end{deluxetable}

\begin{deluxetable}{l l l l l l l l}
\tabletypesize{\tiny}
\rotate
\tablewidth{0pc}
\tablenum{17}
\tablecolumns{8}
\tablecaption{\label{tab:LBVs}Spectroscopic Luminous Blue Variables in IC10, M31, M33, and NGC 6822}
\tablehead{
\colhead{LGGS}
&\colhead{Type}
&\colhead{Cross-ID\tablenotemark{a}} 
&\colhead{Ref}
&\multicolumn{2}{c}{$V$}
&\colhead{$\Delta V$}
&\colhead{Other Var.} \\ \cline{5-6}
&&&&\colhead{LGGS}
&\colhead{Older} \\
}
\startdata
\cutinhead{IC10}
J002012.13+591848.0&hot LBVcand     & \nodata &                         1     & 19.37  & 19.48 & -0.11 & \nodata \\
J002016.48+591906.9&hot LBVcand?    & \nodata &                         1     &19.19 & 19.84 & -0.65  & \nodata \\
J002020.35+591837.6&hot LBVcand     &  \nodata &                        1      &19.10 &
\nodata & \nodata & \nodata \\
\cutinhead{M31}
J004043.10+410846.0&hot LBVCand     &  \nodata &                        1      & 18.62 & \nodata & \nodata   & \nodata \\
J004051.59+403303.0&P Cyg LBVCand   &  \nodata &                        2      & 16.99 & \nodata & \nodata   & \nodata  \\
J004056.49+410308.7&Ofpe/WN9        & OB69WR2                 &3      & 18.09   & 18.13 &  -0.04    & \nodata  \\
J004229.87+410551.8&hot LBVCand &  \nodata &                        1      & 18.78   & 17.51 &   1.27    & \nodata \\
J004242.33+413922.7&P Cyg LBVCand   &  \nodata &                        1      & 18.56   & \nodata & \nodata  & \nodata \\                                                               
J004302.52+414912.4&LBV             &AE And                   &4      & 17.43   & \nodata & \nodata & Spect var, DIRECT IX D31J04302.5+414912.3 \\
J004320.97+414039.6&LBVCand         &k114a                    &5      & 19.22   & 19.12 &   0.10    & \nodata  \\
J004322.50+413940.9&hot LBV Cand & \nodata &                    1    & 20.35 & \nodata &\nodata & \nodata \\
J004333.09+411210.4&hot LBV         &AF And                   &4      & 17.33   & 16.44 &   0.88    & \nodata \\
J004334.50+410951.7&Ofpe/WN9        &  \nodata &                        1      & 18.14   & 18.17 &  -0.04    & \nodata \\
J004341.84+411112.0&P Cyg LBVCand   &  \nodata &                        6      & 17.55   & 17.53 &   0.02    & \nodata \\
J004350.50+414611.4&cool LBVCand    &  \nodata &                        1      & 17.70   & 17.54 &   0.16    & \nodata \\
J004411.36+413257.2&hotLBVCand      &k315a                    &1,5    & 18.07   & 18.02 &   0.05    & DIRECT IX D31J04411.4+413257.2\tablenotemark{d}\\
J004415.00+420156.2&hot LBVCand     &  \nodata &                        1      & 18.29   & \nodata & \nodata & \nodata \\                                                                
J004417.10+411928.0&hot LBVCand     &k350                     &1,5    & 17.11   & 17.20 &  -0.09    & \nodata \\
J004419.43+412247.0&LBV             &Var15                    &4      & 18.45   & 16.97 &   1.48    & \nodata \\
J004425.18+413452.2&cool LBVCand    &k411                     &1,5    & 17.48   & 17.50 &  -0.03    & DIRECT IX D31J04425.2+413452.1 \\
J004442.28+415823.1&hot LBVCand     & \nodata &                         1      & 19.68   & \nodata & \nodata & \nodata \\                                                                
J004444.52+412804.0&P Cyg LBVCand   &  \nodata &                        1      & 18.07   & \nodata & \nodata & DIRECT IV V13833 D31C \\                                                              
J004450.54+413037.7&LBV             &VarA-1                   &7,8    & 17.14   & 16.50 &   0.64    & DIRECT IX D31J04450.6+413037.7\\
J004507.65+413740.8&cool LBVCand    &  \nodata &                        1      & 16.15   & 16.34 &  -0.19  & \nodata \\
J004522.58+415034.8&hot LBVCand     &   \nodata &                       1      & 18.47   & 18.52 &  -0.05    & \nodata \\
J004526.62+415006.3&hot LBVCand     &  \nodata &                        1      & 17.16   & 16.31 &   0.84   & \nodata \\
J004621.08+421308.2&LBVCand         &k895                     &5      & 18.16   & 17.84 &   0.32   & \nodata \\
\cutinhead{M33}
J013235.25+303017.6&hot LBVCand     &  \nodata &                        1      & 18.01   & \nodata & \nodata & Hartman 250024 \\
J013237.72+304005.6&Ofpe/WN9        &M33WR2,MCA 1B,UIT003,H235&9,10,11& 17.63   & 17.50 &   0.13 & \nodata \\
J013242.26+302114.1&hot LBVCand     &  \nodata &                        1      & 17.44   & 18.30 &  -0.86 & \nodata \\
J013245.41+303858.3&Ofpe/WN9        &M33WR5,UIT 008          & 10     & 17.61   & 17.40 &   0.21 & Hartman  250427\\
J013248.26+303950.4&hot LBVcand     & \nodata &                         1      & 17.25 \\
J013300.02+303332.4&hot LBVcand     &UIT026                  & 10     & 18.32   & 18.00 &   0.32 & Hartman 251233 \\
J013309.14+304954.5&Ofpe/WN9        &M33WR22,UIT 045         & 10     & 17.91   & 17.81 &   0.10 & Hartman 150200 \\
J013324.62+302328.4&hot LBVCand     & \nodata &                         1      & 19.58   & \nodata & \nodata & \nodata \\                                                                
J013327.26+303909.1&Ofpe/WN9        &M33WR39,MJ C7           & 12     & 17.95   & \nodata & \nodata & Hartman 242552, Wise 31347 \\
J013332.64+304127.2&hot LBVCand     &M33WR41,AM2             & 1      & 18.99   & 19.20 &  -0.21 & Spect var \\
J013333.22+303343.4&hot LBVCand     &  \nodata &                        1      & 19.40   & \nodata & \nodata  & DIRECT VIII D33J013333.2+303344.5\tablenotemark{e} \\                                       
J013335.14+303600.4&LBV             &VarC                    & 4      & 16.43   & 15.20 &   1.23 & Wise 31284 \\
J013339.52+304540.5&P Cyg LBVCand   &B517,S193               & 1,11,13& 17.50   & 17.68 &  -0.18 & Spec var \\
J013340.60+304137.1&hot LBV Cand    &S204                    & 11     & 18.31   & 18.20 &   0.11 & \nodata \\
J013341.28+302237.2&P Cyg LBVCand   &  \nodata &                        1      & 16.28   & 16.10 &   0.18 & \nodata \\
J013349.23+303809.1&LBV             &VarB                    & 1,4    & 16.21   & 16.40 &  -0.19 & Spec var \\
J013350.12+304126.6&hot LBVCand     &UIT212,S95              & 10,11  & 16.82   & 16.60 &   0.22 & \nodata \\
J013350.92+303936.9&hot LBVCan      &UIT218                  & 10     & 14.17   & \nodata & \nodata & \nodata \\
J013351.46+304057.0&P Cyg LBVCand   & \nodata &                         1      & 17.73   & 17.80 &  -0.07 & \nodata \\
J013353.60+303851.6&Ofpe/WN9        &M33WR103,MJ X15         & 12     & 18.09   & 18.50 &  -0.41 & Hartman 23654 \\
J013355.96+304530.6&cool LBVCand    &UIT247,B324             & 10     & 14.86   & 15.20 &  -0.34 & Wise 10327 \\
J013357.73+301714.2&cool LBVCand    & \nodata &                         1      & 17.39   & \nodata & \nodata & \nodata \\                                                                
J013406.63+304147.8&hot LBVcand     &UIT301                  & 10     & 16.08   & 16.30 &  -0.22 & \nodata \\
J013410.93+303437.6&hot LBVCand     &  \nodata &                        1      & 16.03   & 16.48 &  -0.45& Wise 21206 \\
J013416.07+303642.1&P Cyg LBVCand   &H108                    & 1,11   & 17.95   & 18.10 &  -0.15& DIRECT VIII D33J013416.1+303641.8, Hartman 221349\\
J013416.10+303344.9&LBVCand         &UIT341,B526             & 10     & 17.12   & 16.70 &   0.42& \nodata \\
J013416.44+303120.8&cool LBVCand    &  \nodata &                        1      & 17.10   & 17.00 &   0.10& Wise 22199\\
J013418.74+303411.8&Ofpe/WN9        &M33WR132,UIT 349        & 10     & 19.58   & 19.20 &   0.38& \nodata \\
J013422.91+304411.0&cool LBVCand    &  \nodata &                        1      & 17.22   & 17.08 &   0.14& \nodata \\
J013424.78+303306.6&cool LBVCand    &  \nodata &                        1      & 16.84   & \nodata & \nodata & \nodata \\
J013426.11+303424.7&hot LBVCand     & \nodata &                         1      & 18.97   & 18.85 &   0.12 & Hartman 222305 \\
J013429.64+303732.1&cool LBVCand    &  \nodata &                        1      & 17.10   & 17.42 &  -0.32 & \nodata \\
J013432.76+304717.2&Ofpe/WN9        &  \nodata &                        1      & 19.09   & 18.90 &   0.19 & \nodata \\
J013442.14+303216.0&hot LBVcand?    &  \nodata &                        1      & 17.34   & \nodata & \nodata & \nodata \\            
J013459.47+303701.9&hot LBVCand     &  \nodata &                        1      & 18.37   & 17.94 &   0.43 & Hartman 210675 \\
J013500.30+304150.9&hot LBVCand     &  \nodata &                        1      & 19.30   & \nodata & \nodata & \nodata \\                                                                
J013509.73+304157.3&Ofpe/WN9        &Romano's Star           & 1,14   & 18.04    & \nodata & \nodata & Hartman 110031 \\
\cutinhead{NGC 6822}
J194503.77-145619.1&hot LBVCand?    &  \nodata &                        1 & 18.24 & \nodata & \nodata & Mennickent Field 2/Star 411  \\
\enddata
\tablerefs{
(1)This paper;
(2) Paper I;
(3) Massey 1998;
(4) Hubble \& Sandage 1953;
(5) King et al.\ 1998;
(6) Massey 2006;
(7) Rosino \& Bianchini 1973;
(8) Keynon \& Gallagher 1985;
(9) Willis et al.\ 1992;
(10) Massey et al.\ 1996;
(11) Corral 1996;
(12) Massey \& Johnson 1998;
(13) Crowther et al.\ 1997;
(14) Viotti et al.\ 2007 and references therein.}
\tablenotetext{a}{     
Cross-IDs:
MCAnnn designations
are from Massey, Conti, \& Armandroff  1987 and are listed in CDS;
UITnnn designations are from Massey et al.\ 1996;
M33WRnnn designations Massey \& Johnson 1998, known to CDS as ``[MJ98] WR nnn";
Bnnn are blue stars from Humphreys \& Sandage 1980;
Hnn and Snnn are from Corral 1996, and known to CDS with the prefix ``[S92b]";
knnn designations are from King et al.\ 1998;
MJ designations are from Massey \& Johnson 1998;
AMnn designations are from Armandroff \& Massey 1985, known to CDS as
   [AM85] M33 nn;
OB69WR2 is from Massey et al.\ 1986}
\tablenotetext{b}{``Other" sources for the photometry are Magnier et al.\ 1992
for M31 and Ivanov et al.\  1993 for M33.}
\tablenotetext{c}{Other variability.  We note if the star is shown in this paper to be
a spectral variable, or cite references which claim that the star is photometrically
variable.  The latter includes the DIRECT series of papers (see text),
Hartman et al.\ 2006, and Mennickent 2006.} 
\tablenotetext{d}{Listed as a Cepheid with a period of 11.0987 days.}
\tablenotetext{e}{Listed as a Cepheid with period  iof 9.985 days.}
\end{deluxetable}
    

\clearpage


\begin{references}
\reference {} Abbott, J. B., Crowther, P. A., Drissen, L, Dessart, L, Martin, P.,
\& Boivin, G. 2004, MNRAS, 350, 552
\reference {} Armandroff, T. E., \& Massey, P. 1985, ApJ, 291, 685
\reference {} Armandroff, T. E., \& Massey, P. 1991, AJ, 102, 927
\reference {} Bianchi, L., Scuderi, S., Massey, P., \& Romaniello, M. 2001,
AJ, 121, 2020
\reference {} Bohannan, B. 1989, in Physics of Luminous Blue Variables,
ed. K. Davidson, A. F. J. Moffat, \& H. J. G. L. M. Lamers (Dordrecht: Kluwer), 35
\reference {} Bohannan, B. 1997, in ASP Conf.\ Ser 120,Luminous Blue Variables: Massive Stars
in Transition, ed. A. Nota \& H. J. G. L. M. Lamers (San Francisco: ASP), 3
\reference {} Bohannan, B., \& Walborn, N. R 1989, PASP, 101, 520
\reference {} Bonanos, A. Z., Stanek, K. Z., Sasselov, D. D., Mochejska, B. J.,
Macri, L. M., \& Kaluzny, J. 2003, AJ, 126, 175 (DIRECT IX)
\reference {} Bresolin, F., Pietrzynski, G., Urbaneja, M. A., Gieren, W., Kudritzki, R.-P.,
Venn, K. A. 2006, ApJ, 648, 1007
\reference {} Cardelli, J. A., Clayton, G. C., \& Mathis, J. S. 1989, ApJ, 345, 245
\reference {} Coluzzi, R. 1993, Bull.\ Inf.\ Centre Donnees Stellaires, 43, 7, \\http://cdsweb.u-strasbg.fr/htbin/Cat?VI/71
\reference {} Conti, P. S. 1997a, in ASP Conf.\ Ser. 120,
Luminous Blue Variables: Massive Stars in Transition,
ed A. Nota \& H. J. H. L. M. Lamers (San Francisco: ASP), 161
\reference {} Conti, P. S. 1997b, in ASP Conf.\ Ser. 120,
Luminous Blue Variables: Massive Stars in Transition,
ed A. Nota \& H. J. H. L. M. Lamers (San Francisco: ASP), 387
\reference {} Corral, L. J. 1996, AJ, 1450
\reference {} Corral, L. J. \& Herrero, A., 2003, Rev. Mex. A\&A, (Serr. Conf.), 16, 265
\reference {} Crowther, P. A., Drissen, L., Abbott, J. B., Royer, P., \& Smartt, S. J.  2003,
A\&A, 404, 483
\reference {} Crowther, P. A., Szeifert, T., Stahl, O., \& Zickgraf, F.-J. 1997, A\&A, 318, 543
\reference {} Drissen, L., Moffat, A. F. J., \& Shara, M. M. 1993, AJ, 105, 1400
\reference {} Hartman, J. D., Bersier, D., Stanek, K. Z., Beaulieu, J.-P.,
Kaluzny, J.,
Marquette, J.-B.,
Stetson, P. B.,
\& 
Schwarzenberg-Czerny, A. 2006, MNRAS, 371, 1405
\reference {} Hubble, E., \& Sandage, A. 1953, ApJ, 118, 353
\reference {} Humphreys, R. M. \& Davidson. K. 1994, PASP, 106, 1025
\reference {} Humphreys, R. M., \& Sandage, A. 1980, ApJS, 44, 319
\reference {} Israelian, G., \& de Groot, M. 1999, Space Sci.\ Rev., 90, 493
\reference {} Ivanov, G. R., Freedman, W. L., \& Madore, B. F. 1993, ApJS, 89, 85
\reference {} Kaluzny, J., Mochejska, B. J., Stanek, K. Z., 
Krockenberger, M., Sasselov, D. D., Tonry, J. L., \& Mateo, M. 1999, AJ, 118, 346
(DIRECT IV)
\reference {} Kaluzny, J., Stanek, K. Z., Krockenberger, M., Sasselov, D. D., 
Tonry, J. L., \& Mateo, M. 1998, AJ, 115, 1016 (DIRECT I)
\reference {} Jacoby, G. 2000, NOAO CCD Mosaic Imager User Manual (Tucson: NOAO), \\ http://www.noao.edu/kpno/mosaic/manual/
\reference {} Jacoby, G. H., Africano, John L., Quigley, Robert J. 1987, PASP, 99, 672
\reference {} Kaluzny, J., Mochejska, B. J., Stanek, K. Z., Krockenberger, M.,
Sasselov, D. D., Tonry, J. L., \& Mateo, M. 1999, AJ, 118, 346 (DIRECT IV)
\reference {} Kenyon, S. J., \& Gallagher, J. S. 1985, ApJ, 290, 542
\reference {} King, N. L., Walterbos, R. A. M., \& Braun, R. 1998, ApJ, 507, 210
\reference {} Maeder, A., Lequeux, J., \& Azzopardi, M. 1980, A\&A 90, L17
\reference {} Magnier, E. A., Lewin, W. H. G., van Paradijs, J., Hasinger, G.,
Jain, A., Pietsch, W., \& ruemper, J. 1992, A\&AS, 96, 379
\reference {} Macri, L. M., Stanek, K. Z., Sasselov, D. D., Krockenberger, M., \& Kaluzny, J. 2001, AJ, 121, 870 (DIRECT VI)
\reference {} Margon, B., 1984, ARA\&A, 22, 507
\reference {} Margon, B., Ford, H. C., Katz, J. I., Kwitter, K. B., Ulrich, R. K., Stone, R. P. S., \& Klemola, A. 1979, ApJ, 230, L41
\reference {} Massey, P. 1998, in Stellar Astrophysics for the Local Group, ed.\ A. Aparicio, A. Herrero, \& F. Sanchez
(Cambridge, Cambridge Univ.\ Press), 95
\reference {} Massey, P. 2000, PASP 112, 144
\reference {} Massey, P. 2003, ARA\&A,41,15
\reference {} Massey, P. 2006, ApJ, 638, L93
\reference {} Massey, P., \& Armandroff, T. E. 1995, AJ, 109, 2470
\reference {} Massey, P., Armandroff, T. E., \& Conti, P. S. 1986, AJ 92, 1303
\reference {} Massey, P., Armandroff, T. E., Pyke, R., Patel, K., \& Wilson, C. D. 1995, AJ, 110, 2715
\reference {} Massey, P., Bianchi L., Hutchings J. B., Stecher T. P., 1996, ApJ, 469, 629
\reference {} Massey, P., \& Conti, P. S. 1983, ApJ 273, 576
\reference {} Massey, P., \& Johnson, O. 1998, ApJ, 505, 793
\reference {} Massey, P., Olsen, K. A. G., Hodge, P. W., Strong, S. B., Jacoby, G. H., Schlingman, W., \& Smith, R. C. 2006, AJ 131, 2478 (Paper I)
\reference {} Massey, P., Olsen, K. A. G., Hodge, P. W., Jacoby, G. H., McNeill, R. T., Smith, R. C. \& Strong S. B.  2007, AJ, 133, 2393 (Paper II)
\reference {} Massey, P., Waterhouse, E., DeGioia-Eastwood, K. 2000, AJ, 119, 2214
\reference {} Mennickent, R. E., Gieren, W., Soszynski, I.,
Pietrzynski, G. 2006, A\&A, 450, 873
\reference {} Mochejska, B. J., Kaluzny, J., Stanek, K. Z.,
Krockenberger, M., Sasselov, D. D. 1999, AJ, 118, 2211 (DIRECT V)
\reference {} Mochejska, B. J., Kaluzny, J., Stanek, K. Z., Sasselov, D. D.,
\& Szentgyorgyi, A. H. 2001a, AJ, 121, 2032 (DIRECT VII)
\reference {} Mochejska, B. J., Kaluzny, J., Stanek, K. Z., Sasselov, D. D.,
\& Szentgyorgyi, A. H. 2001b, AJ, 122, 2477 (DIRECT VIII)
\reference {} Monteverde, M. I., Herrero, A., Lennon, D. J., Kudritzki, R. P. 1996,
A\&A, 312, 24
\reference {} Mould, J., Saha, A., \& Hughes, S. 2004, ApJS, 154, 623
\reference {} Neese, C. L., Armandroff, T. E., \& Massey, P. 1991, in Wolf-Rayet
Stars and Interrelations with Other Massive Stars in Galaxies, 
IAU Symp.\ 143 (Dordrecht: 
Kluwer), 651
\reference {} Oke, J. B. 1974, ApJS, 27, 21
\reference {} Parker, J. W. 1997, in ASP Conf.\ Ser. 120,
Luminous Blue Variables: Massive Stars in Transition,
ed A. Nota \& H. J. H. L. M. Lamers (San Francisco: ASP), 368
\reference {} Romano, G. 1978, A\&A, 67, 291
\reference {} Rosino, L., \& Bianchini, A. 1973, A\&A, 22, 453
\reference {} Rubin, V. C., \& Ford, W. K. 1970, ApJ, 159, 379
\reference {} Shporer, A., \& Mazeh, T. 2006, MNRAS, 370, 1429
\reference {} Smith, L. J., Crowther, P. A., \& Willis, A. J. 1995, A\&A, 302, 830
\reference {} Smith, N. 2007, in Mass Loss from Stars and the Evolution of
Stellar Clusters, ed. A. de Koter, L. J. Smith, \& L. B. F. M. Waters (San Francisco: ASP), in press, astro-ph/0609422
\reference {} Smith, N., \& Owocki, S. P. 2006, ApJ, 645, L45
\reference {} Stahl, O., Wolf, B., Glare, G., Cassatella, A., Krautter, J., Persi, P., \& Ferrari-Toniolo, M. 1983, A\&A, 127, 49
\reference {} Stanek, K. Z., Kaluzny, J., Krockenberger, M., Sasselov, D. D.,
Tonry, J. L., \& Mateo, M. 1998, AJ, 115, 1894 (DIRECT II)
\reference {} Stanek, K. Z., Kaluzny, J., Krockenberger, M., Sasselov, D. D.,
Tonry, J. L., \& Mateo, M. 1999, AJ, 117 (DIRECT III)
\reference {} Szeifert, Th., Humphreys, R. M., Davidson, K., Jones, T. J., Stahl, O., Wolf, B.,
\& Zickgraf, F.-J., 1996, A\&A, 314, 131
\reference {} van den Bergh, S. 2000, The Galaxies of the Local Group (Cambridge: Cambridge Univ. Press)
\reference {} Viotti, R. F., Galleti, S., Gualandi, R., Montagni, F., Polcaro, V. F.,
Rossi, C., \& Norci, L. 2007, A\&A, 464, L53
\reference {} Walborn, N. R. 1977, ApJ, 215, 53
\reference {} Walborn, N. R., \& Fitzpatrick, E. L. 1990, PASP, 102, 379
\reference {} Westerlund, B. E., Azzopardi, M., Breysacher, J., \& Lequeux, J. 1983, A\&A, 123, 159
\reference {} Willis, A. J., Schild, H., \& Smith, L. J. 1992, A\&A, 261, 419
\reference {} Wolf, B., \& Kaufer, A. 1997, in Luminous Blue Variables, ASP Conf.~Ser.~120,
ed. Nota, A., \& Lamers, H. J. G. L. M. (San Francisco, ASP), 26
\reference {} Zickgraf, F.-J., Wolf, B., Stahl, O., Leitherer, C., \& Klare, G. 1985, A\&A 143, 421


\end{references}
\end{document}